\documentclass{emulateapj}

\shorttitle{DENSITY STUDIES IN THE MAGNETIZED ISM}
\shortauthors{BURKHART ET AL.}

\begin{document}

\title{Density Studies of MHD Interstellar Turbulence: Statistical Moments, Correlations and Bispectrum}
\author{Blakesley Burkhart\altaffilmark{1,2}, D. Falceta-Gon\c calves\altaffilmark{1,4}, G. Kowal\altaffilmark{1,3} \& A. Lazarian\altaffilmark{1}}
\altaffiltext{1}{Astronomy Department, University of Wisconsin, Madison, 475 N. Charter St., WI 53711, USA}
\altaffiltext{2}{Department of Physics \& Astronomy, University of Louisville, Louisville, KY 40202, USA}
\altaffiltext{3}{Astronomical Observatory, Jagiellonian University, ul. Orla 171, 30-244 Krak\'ow, Poland}
\altaffiltext{4}{N\' ucleo de Astrof\' isica Te\' orica, Universidade Cruzeiro do Sul - Rua Galv\~ ao Bueno 868, CEP 01506-000, S\~ao Paulo, Brazil}

\begin{abstract} 
 We present a number of statistical tools that show promise for obtaining information on turbulence in molecular clouds and diffuse interstellar medium. For our tests we make use of
three-dimensional $512^3$ compressible MHD isothermal simulations performed for different 
sonic, i.e. ${\cal M}_s\equiv V_L/V_s$, where $V_L$ is the injection velocity, $V_s$ is the sound velocity, and Alfv\'{e}nic ${\cal M}_A\equiv V_L/V_A$, where $V_A$ is the Alfv\'en velocity, Mach numbers. We introduce the bispectrum, a new tool for statistical studies of the interstellar medium which, unlike an ordinary power spectrum of turbulence, preserves the phase information of the stochastic field. We show that the bispectra of the 3D stochastic density field and of column densities, available from observations, are similar. This opens good prospects for studies of molecular clouds and diffuse media with the 
new tool. We use the bispectrum technique to define the role of non-linear wave-wave interactions in the turbulent 
energy cascade.  We also obtained the bispectrum function for density and column densities with varying magnetic field strength. As expected, a strong correlation is obtained for wave modes $k_{1}=k_{2}$ for all models. Larger values of ${\cal M}_{s}$ result in increased correlations for modes with $k_{1}\neq k_{2}$. This effect becomes more evident with increasing magnetic field intensity. We believe that the different MHD wave modes, e.g. Alfv\'{e}n and magneto-acoustic,  which arise in strongly magnetized turbulence, may be responsible for the increased correlations compared to purely hydrodynamical perturbations. In addition to the bispectrum, we calculated the 3rd and 4th statistical moments of density and column density, namely, skewness and kurtosis, respectively. We found a strong dependence of skewness and kurtosis with ${\cal M}_{s}$. In particular, as ${\cal M}_{s}$ increases, so does the Gaussian asymmetry of the density distribution.
 We also studied the correlations of 2D column density with dispersion of velocities and magnetic field, as well as   the correlations of 3D density with magentic and kinetic energy and ${\cal M}_{A}$ for comparison. Our results show that column density is linearly correlated with magnetic field for high ${\cal M}_{s}$. This trend is independent of the turbulent kinetic energy and can be used to characterize inhomogeneities of physical properties in low density clumps in the ISM. 

\end{abstract}

\section{Introduction}
\label{intro}

The interstellar medium (ISM) is known to be highly turbulent but the role of magnetic fields in the distribution of
turbulent eddies, as well as in the energy cascade, is still not completely understood. Numerical 
simulations have proved to be important tools to study turbulent processes in the magnetized ISM \cite[see][]{ball07}. Stars are known to be formed on the denser regions of the ISM. However, the 
formation and survival of these dense cores is not completely understood, mostly because of the complex 
relationship between turbulence, magnetic fields, and self-gravity \cite[see reviews by][]{Lazarian04,evans99,Elmegreen04,mckee07}. In a rather simplistic view, highly supersonic turbulent motions generate shocks that evolve into dense structures, which may be dispersed by turbulence.

This picture may be very different if the magnetic field is taken into account. Observationally, line profiles indeed indicate ISM turbulence is supersonic, but the determination of the magnetic field intensity is much more involved \citep[see][]{falceta03}. In spite of the extensive use of  current techniques, e.g. polarimetry and Zeeman splitting measurements, the actual ratio of the magnetic and turbulent kinetic energies is still a matter of debate \citep[see][]{padoan02,girart06}. 
Magnetic fields may play a key role in the star formation process \citep[see][for review]{mckee02,mac04}, e.g. 
providing clouds with  extra support against gravitational collapse \citep[see][]{ostriker99}. 
Strongly magnetized turbulence presents weaker shocks and decreased densities compared to purely hydrodynamic turbulence. 

Because of the difficulties mentioned above, numerical simulations represent a
unique method to understand the evolution of turbulence and the role of the
magnetic field in the creation of density structures in the ISM.
\cite{vsemadeni97} studied the Larson-type correlations of density and mass with
 cloud radii, using two-dimensional MHD simulations. It was shown that the
topology of density is filamentary and its probability distribution function
(PDF) is, in most cases, log-normal. Further studies
\citep{scalo98,vsemadeni01,ostriker01} confirmed these results (see Beresnyak, Lazarian \& Cho 2005).
Statistical descriptions of turbulence are valuable as they constrain the physical properties of the system. Several techniques involving numerics have been proposed for velocity studies \cite[see][]{Lazarian00,Lazarian04,Lazarian06,Lazarian08} and have been successfully tested in works such as \cite{Lazarian01,Esqui02,padoan06}. These studies are aimed in providing the velocity spectrum of the turbulence. It is important to note that some of the earlier results obtained for velocity spectra studies using velocity centroids, are likely to be in error for high Mach number turbulence in molecular clouds in view of the recent insight into the properties of velocity centroids (see Esquivel \& Lazarian 2005, Esquivel et al. 2007).

In the past, density studies have also provided insight on astrophysical turbulence.  Although density is a less direct measure of turbulence compared to velocity, it is readily available from the observations of column densities. The  index of the density spectrum gets shallow with the sonic Mach number ${\cal M}_s \equiv V_L/V_s$, where $V_L$ and $V_s$ are the injection and sound velocities, respectively, as shown in MHD simulations by Beresnyak, Lazarian \& Cho (2005).  This was also confirmed to be also true for hydrodynamic simulations in the follow-up simulations in Rye \& Kim (2006)\footnote{The first mentioning of the effect can be traced back to Padoan et al. (2004) study.}. The shallow spectra of density, in fact, corresponds to observations (see Lazarian 2008 for a review) of shallow density spectrum of column densities measured for molecular cloud studies. 

The complexity of turbulence calls for the simultaneous use of different statistical measures. 
Recently, Kowal, Lazarian, \& Beresnyak (2007) (henceforth referred to as KLB) tested how different measures, including PDFs, power spectrum, skewness, kurtosis, and higher order She-L\'ev\^eque exponents vary with  ${\cal M}_s$ and ${\cal M}_A$ using an extensive set of MHD simulations. 
They also provided a study of the variations of topology, using {\it genus}, \footnote{Genus was earlier briefly discussed in the context of interstellar medium in Lazarian (1999), Lazarian, Esquivel \& Pogosyan (2001) and Lazarian (2006).} as well as anisotropies of density structures, using correlation functions,
with ${\cal M}_s$ and ${\cal M}_A$. In KLB, emphasis was made on comparing the results available through synthetic observations of column density and
the underlying properties of 3D density. This made this study easily applicable to observations, e.g. to Wisconsin H $\alpha$ Mapper (WHAM) data (see Hill et al. 2008). 

Despite the number of past works, the potential of density studies is far from being exhausted. For instance, when the energy cascade process and the distribution of density structures are analyzed from the power spectra of two-point correlations, phase information is lost. On the other hand the bispectrum, or three-point statistic, measures the magnitude {\it and the phase} of the correlation of signals in Fourier space. As a consequence of being a more informative measure, it can be used to search for nonlinear wave-wave interactions and characterize the turbulent regime. The bispectrum has been widely used in cosmology and gravitational wave studies \citep{fry98,Scoccimarro00,liguori06} and for the characterization of wave-wave interactions in laboratory plasmas \citep{intrator89,tynan01}, and was suggested in \cite{Lazarian99} to be applied to ISM turbulence  
\cite[see also][]{Laz08}. In this paper we apply the bispectrum to density and column density maps for different turbulent regimes in order to test the usefullness of the tool for interstellar studies.

In addition, this work extends the study in KLB in several respects including the use of higher resolution simulations, further investigation of the dependence of Mach number on the skewness and kurtosis of densities, and studies of 3D correlations, which were not attempted in KLB.  For instance, we study the correlations between the physical parameters (velocity dispersion, magnetic field intensity, etc.) with density and column density, and investigate the correlation between velocity dispersion and LOS velocity,  in order to explore their evolution within dense structures and their correlational dependence with turbulent regimes (namely the sonic and Alfv\'enic Mach numbers). We also provided the first interstellar medium related study of the bispectrum. In particular, we study the bispectrum of density and column density for several models of MHD turbulence.  In \S~\ref{sec:numeric} we describe the numerical models of compressible MHD turbulence. In \S~\ref{sec:moments} we discuss the skewness and kurtosis of density and maps of column density. In \S~\ref{sec:correlations} we examine the correlations of Mach numbers,
magnetic and kinetic energies with density and magnetic field and velocity with column density. In \S~\ref{sec:bispectrum} we present an analysis of the bispectrum of compressible MHD turbulence. In \S~ 6 we discuss our results, followed by our conclusions and the summary.

\section{Numerical Simulations}
\label{sec:numeric}

We used a second-order-accurate hybrid essentially nonoscillatory (ENO) scheme \cite[see][]{Cho02} similar to that shown in KLB to solve
the ideal MHD equations in a periodic box,

\begin{eqnarray}
 \frac{\partial \rho}{\partial t} + \nabla \cdot (\rho {\bf v}) = 0, \\
 \frac{\partial \rho {\bf v}}{\partial t} + \nabla \cdot \left[ \rho {\bf v} {\bf v} + \left( p + \frac{B^2}{8 \pi} \right) {\bf I} - \frac{1}{4 \pi}{\bf B}{\bf B} \right] = {\bf f},  \\
 \frac{\partial {\bf B}}{\partial t} - \nabla \times ({\bf v} \times{\bf B}) = 0,
\end{eqnarray}
with zero-divergence condition $\nabla \cdot {\bf B} = 0$, and an isothermal equation of state $p = c_s^2 \rho$, where $\rho$ is density, ${\bf v}$ is velocity, ${\bf B}$ is magnetic field, $p$ is the gas pressure, and $c_s$ is the isothermal speed of sound. On the right-hand side, the source term $\bf{f}$ is a random large-scale driving force (in fact, we drive only the velocity field).   The RMS velocity $\delta V$ is maintained to be approximately unity, so that ${\bf v}$ can be viewed as the velocity measured in units of the RMS velocity of the system and ${\bf B}/\left( 4 \pi \rho \right)^{1/2}$ as the Alfv\'{e}n velocity in the same units. The time $t$ is in units of the large eddy turnover time ($\sim L/\delta V$) and the length in units of $L$, the scale of the energy injection. The magnetic field consists of the uniform background field and a fluctuating field: ${\bf B}= {\bf B}_\mathrm{ext} + {\bf b}$. Initially ${\bf b}=0$.

We drive turbulence solenoidally, at wave scale $k$ equal to about 2.5 (2.5 times smaller than the size of the box). This scale defines the injection scale in our models in Fourier space to minimize the influence of the forcing on the generation of density structures. Density fluctuations are generated later on by the interaction of MHD waves. Density structures in turbulence can be associated with the slow and fast modes \cite[see][]{lithwick01,Cho03,beresnyak05}. We use units in which $V_A=B_\mathrm{ext}/\left( 4 \pi \rho \right)^{1/2}=1$. The values of $B_\mathrm{ext}$ have been chosen to be similar to those observed in the ISM turbulence. The average RMS velocity $\delta V$ in a statistically stationary state is around $1$.

We do not set the viscosity and diffusion explicitly in our models. The scale at which the dissipation starts to act is defined by the numerical diffusivity of the scheme. The ENO-type schemes are considered to be relatively low diffusion ones \cite[see][e.g.]{liu98,levy99}. The numerical diffusion depends not only on the adopted numerical scheme but also on the ``smoothness'' of the solution, so it changes locally in the system. In addition, it is also a time-varying quantity. All these problems make its estimation very difficult and incomparable between different applications. However, the dissipation scales can be estimated approximately from the velocity spectra. In the case of our models we estimated the dissipation scale $k_{\nu}$ at 30 for high resolution.

\begin{figure*}[tbh]
\centering
\includegraphics[scale=.9]{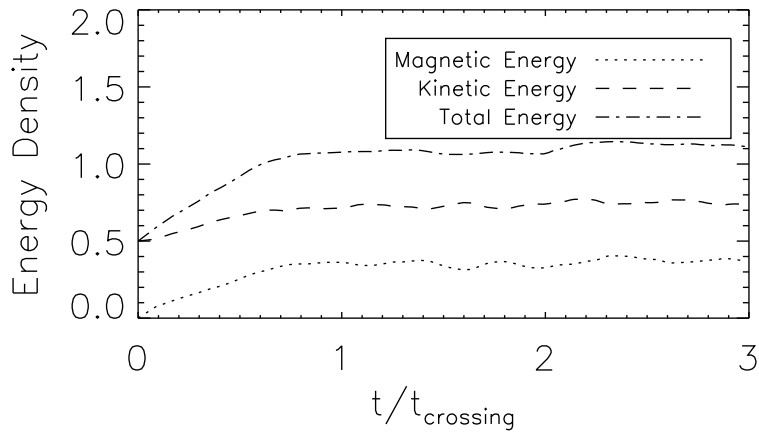}
\caption{Energy density vs. time, showing the evolution of magnetic, kinetic, and total energy for a ${\cal M}_A=2.0$, ${\cal M}_s=0.7$ model. Time is in units of the ``crossing time,'' which is the time it takes the sound speed to travel through the cloud.  It is defined as $t_{crossing}=\frac{L}{c_{max}}$. Here we show that 1 time unit guarantees full development of the turbulent energy cascade. The difference in saturation energies is also seen in incompressible turbulent dynamo simulations, where it is of the order 30\% (Cho et al. 2009)}
\label{fig:saturation}
\end{figure*}

We present 3D numerical experiments of compressible (MHD) turbulence for a range of Mach numbers. We understand the Mach number to be defined as the mean value of the ratio of the absolute value of the local velocity $V$ to the local value of the characteristic speed $c_s$ or $V_A$ (for the sonic and Alfv\'{e}nic Mach number, respectively).We divided our models into two groups corresponding to sub-Alfv\'{e}nic ($B_\mathrm{ext}=1.0$) and super-Alfv\'{e}nic ($B_\mathrm{ext}=0.1$) turbulence. For each group we computed several models with different values of pressure (see Table~\ref{tab:models}). We ran 6 compressible MHD turbulent models, with 512$^3$ resolution, for $t 
\sim 5$ crossing times, to guarantee full development of energy cascade. We show in Figure~\ref{fig:saturation} the full development of global quantities magnetic, kinetic, and total energy. This figure shows that 1 dynamical time unit (time normalized by the crossing time) is sufficient for saturation of energies. Since the saturation level is similar for all models and we solve the isothermal MHD equations, the sonic Mach number is fully determined by the value of isothermal sound speed, which is our control parameter.  The different initial conditions led to three values of sonic Mach number, $\sim
0.7$, $\sim 2$, and $\sim 7$, and two Alfv\'enic Mach numbers, $\sim 0.7$ and
$\sim 2$. The models are listed and described in Table~\ref{tab:models}.

\begin{table*}
\begin{center}
\caption{Description of the simulations - MHD, 512$^3$ \label{tab:models}}
\begin{tabular}{cccccc}
\hline\hline
Model & $P$ & $B_{\rm ext}$ & \multicolumn{1}{c}{${\cal M}_s$\tablenotemark{a}} & ${\cal M}_A$\tablenotemark{b} &Description \\
\tableline
1 &1.00 &1.00 &0.7 &0.7 & subsonic \& sub-Alfv\'enic     \\
2 &0.10 &1.00 &2.0 &0.7 & supersonic \& sub-Alfv\'enic   \\
3 &0.01 &1.00 &7.0 &0.7 & supersonic \& sub-Alfv\'enic   \\
4 &1.00 &0.10 &0.7 &2.0 & subsonic \& super-Alfv\'enic   \\
5 &0.10 &0.10 &2.0 &2.0 & supersonic \& super-Alfv\'enic \\
6 &0.01 &0.10 &7.0 &2.0 & supersonic \& super-Alfv\'enic \\
\hline\hline
\end{tabular}
\begin{itemize}
\item Sonic Mach number is defined as ${\cal M}_s = \langle |{\bf v}|/a \rangle$, where $a$ is the sound speed and the averaging is taken over whole volume
\item Alfv\'{e}nic Mach number is defined as ${\cal M}_A = \langle |{\bf v}|/v_A \rangle$, where $v_A = |{\bf B}|/\sqrt{\rho}$ is the local Alfv\'en speed and the averaging is taken over whole volume
\end{itemize}
\end{center}
\end{table*}

Ultimately, we would like our numerical results to be directly comparable to observations.  Thus, an observer might be curious about what density, temperatures, units, and scalings are appropriate for these models.These are isothermal scale-free simulations.  For the situations when gas can be assumed isothermal, they can easily be rescaled for any parameters (i.e.Temperatures, scales, and densities, see Table~\ref{tab:phase}) of the media concerned.  In Table~\ref{tab:phase} the abbreviations stand for the following six idealized interstellar environments: cold neutral medium (CNM), warm
neutral medium (WNM), warm ionized medium (WIM), molecular cloud (MC), dark cloud (DC), reflection nebula
(RN), and photodissociation region (PDR). We list suggested values for $n_H$ (Hydrogen density), T(k) (gas temperature), $\chi$ (starlight intensity relative to the average starlight background), and other properties of these phases. These properties should be scaled in the models as dictated by the statistics used. An example of these scalings can be found in Hill et al. (2008), who used scaling relationships to set temperature, length, and density of simulated data to match the WIM.  They explored PDFs for the WIM using data taken from the WHAM and synthetic data using these simulations.  They found that scaled models with certain Mach numbers matched the WIM very well. Similar examples of simulation scaling can also be found in Falceta-Gon\c calves et al.(2008). In general, these simulations should be scaled accordingly to the characteristics of the media studied.  These characteristics are listed in Table~\ref{tab:phase} as shown in \cite{Draine98}. As for units, these are defined in terms of the speed of sound, which varies for a given Mach number. For ${\cal M}_{s}$=7.0, $c_{s}$=0.1, for ${\cal M}_{s}$=2.0, $c_{s}$=0.333 and for ${\cal M}_{s}$=0.7, $c_{s}$=1.0.

\begin{table*}
\begin{center}
\caption{Idealized Environments for Interstellar Matter  \label{tab:phase}}
\begin{tabular}{cccccccc}
\hline\hline
Parameter & DC & MC & CNM & WNM & WIM & RN & PDR \\
\tableline
$\textit{n}_{H}\left( cm^{-3} \right) $ & $10^{4}$ & 300. &30 &0.4 & 0.1& $10^3$ & $10^5$ \\
\textit{T}(K) &10. &20. &100. &6000. & 8000.& 100& 300\\
$\chi$ & $10^{4}$ &0.01 &1.0 &1.0 & 1.0& 1000 & 3000 \\
$x_{H} \equiv n\left(H^{+}\right) /\textit{n}_{H}$ & 0 & 0 & 0.0012 & 0.1 & 0.99& 0.001 & 0.0001 \\
$x_{M} \equiv n\left(M^{+}\right) /\textit{n}_{H}$ & $10^{-6}$ &0.0001 & 0.0003 & 0.0003 & 0.001 &0.0002 &0.0002\\
$\textit{y} \equiv 2n(H_{2})/\textit{n}_{H}$ &0.999 &0.99 & 0 & 0 & 0& 0.5& 0.5 \\
\hline\hline
\end{tabular}
\begin{itemize}
From Draine \& Lazarian 1998.
\end{itemize}
\end{center}
\end{table*}

\begin{figure*}[tbh]
\centering
\includegraphics[scale=.6]{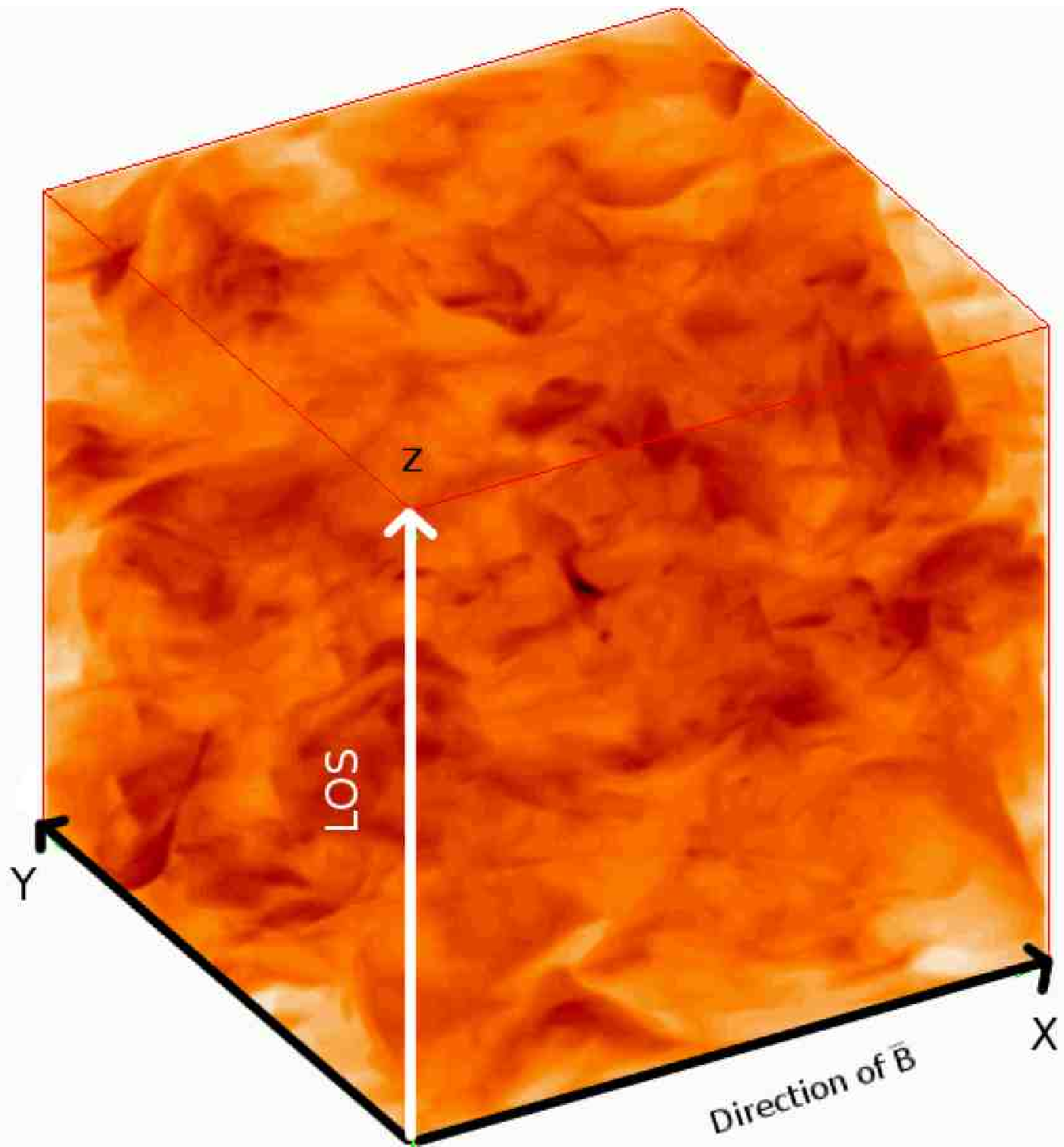}
\caption{Throughout the paper we make mention of directions x, y, and z as well as the line of sight (LOS). This figure shows an example of a subsonic, sub-Alfv\'enic  $512^{3}$ data cube demonstrating one possible position of the line of sight (LOS) and mean magnetic field along the x direction in respect to the  x, y, z axes. Here the LOS is orientated along the z axis, although throughout the paper we will investigate statistics for LOS along all three axes.}
\label{fig:threedplot}
\end{figure*}

For reference, we include a cube visualization in Figure~\ref{fig:threedplot} showing the x, y, z axes, the line of sight (LOS), and the direction of the magnetic field.  We will refer to these directions throughout the paper.  For example, "x column density" or "column density in the x direction"  refers to the density cube being integrated along the x direction to make the 2 dimensional field.  This description is similar for the y and z directions. In the case of Figure~\ref{fig:threedplot}, the line of sight is along the z axis and the magnetic field is oriented along the x axis.

\section{Statistical Moments: Skewness and Kurtosis}
\label{sec:moments}

Skewness and kurtosis are defined by the third and fourth-order statistical moment. Skewness is defined as:

\begin{equation}
\gamma_{\xi} = \frac{1}{N} \sum_{i=1}^N{ \left( \frac{\xi_{i} - \overline{\xi}}{\sigma_{\xi}} \right)^3 }
\label{eq:skew}
\end{equation}
If a distribution is Gaussian, the skewness is zero. Negative skewness indicates the data is skewed in the left direction (the tail is extended to the left) while positive values imply that the distribution is skewed in the right direction (the tail is extended to the right).  

Kurtosis is a measure of whether a quantity has a distribution that is peaked or flattened compared to a normal Gaussian distribution.  Kurtosis is defined in a similar manner to skewness, only is derived from the forth order statistical moment.  If a data set has positive kurtosis then it will have a distinct sharp peak near the mean and have elongated tails.  If a data set has negative kurtosis then it will be flat at the mean. Kurtosis is defined as:

\begin{equation}
\beta_{\xi}=\frac{1}{N}\sum_{i=1}^N \left(\frac{\xi_{i}-\overline{\xi}}{\sigma_{\xi}}\right)^{4}-3
\label{eq:kurt}
\end{equation}

Skewness and kurtosis are statistical measures that can be used to characterize the properties of turbulence.  They have been recently been used to characterize compressible MHD turbulence for simulations and synthetic observations (see KLB). In this work we test their results applying these calculations to cubes of finer resolution ($512^3$) and further investigate how applicable this measure is to observations. 
The interest on these measurements comes from that fact that the skewness and kurtosis of column density and 3D density depends strongly on ${\cal M}_{s}$ (see KLB). Physically, this implies that one can determine ${\cal M}_{s}$ from examining the skewness and/or kurtosis of the observational distributions. Also, by comparing the observed column densities to the synthetic ones and determining the turbulent regime of a given cloud, it is possible to understand the three-dimensional statistics. 

\begin{table*}
\begin{center}
\caption{Skewness and Kurtosis of Density and Column Densities Integrated along X, Y, and Z-direction \label{tab:moments}}
\begin{tabular}{|c|c|cc|cc|cc|cc|} \hline
$M_{s}$ & $M_{A}$ & $\rho$  & & $\Sigma_{X}$  & & $\Sigma_{Y}$ & & $\Sigma_{Z}$ & \\
\hline
 & & Skew & Kurtosis  & Skew & Kurtosis  &  Skew & Kurtosis & Skew & Kurtosis \\
\hline
0.7 & 0.7 & $0.7^{\pm{0.2}}$ & $0.6^{\pm{0.7}}$ & $0.47^{\pm{0.01}}$ & $0.2^{\pm{0.4}}$  & $0.24^{\pm{0.04}}$  & $0.2^{\pm{0.6}}$  & $-0.03^{\pm{0.02}}$ & $0.01^{\pm{0.2}}$  \\
0.7 & 2.0 & $1.0^{\pm{0.2}}$ & $2.0^{\pm{0.7}}$ & $0.37^{\pm{0.01}}$ & $-0.4^{\pm{0.4}}$ & $0.18^{\pm{0.04}}$  & $-0.7^{\pm{0.6}}$  & $-0.06^{\pm{0.02}}$ & $-0.2^{\pm{0.2}}$   \\
2.0 & 0.7 & $3.8^{\pm{0.1}}$ & $20^{\pm{3}}$ & $0.7^{\pm{0.2}}$ & $0.8^{\pm{0.4}}$ & $1.1^{\pm{0.1}}$  & $2.1^{\pm{0.4}}$ & $1.0^{\pm{0.3}}$  &   $2^{\pm{1}}$  \\
2.0 & 2.0 & $4.0^{\pm{0.1}}$ & $30^{\pm{3}}$ & $1.0^{\pm{0.2}}$ & $1.4^{\pm{0.4}}$ & $0.9^{\pm{0.1}}$  & $1.5^{\pm{0.4}}$ & $0.6^{\pm{0.3}}$  &  $0.4^{\pm{1}}$    \\
7.0 & 0.7 & $11^{\pm{0.4}}$ & $200^{\pm{30}}$ & $1.2^{\pm{0.3}}$ & $2^{\pm{2}}$ & $1.4^{\pm{0.1}}$  & $3.6^{\pm{0.9}}$ & $1.4^{\pm{0.1}}$  &  $4^{\pm{1}}$   \\
7.0 & 2.0 & $10.1^{\pm{0.4}}$ & $300^{\pm{30}}$ & $1.6^{\pm{0.3}}$ & $5^{\pm{2}}$ & $1.2^{\pm{0.1}}$  & $2.4^{\pm{0.9}}$  & $1.2^{\pm{0.1}}$  &  $2^{\pm{1}}$      \\
\hline
\end{tabular}
\end{center}
\end{table*}
Statistics are obtained by applying Equations~(\ref{eq:skew}) and~(\ref{eq:kurt}) to data cubes of density and column density. In  Table~\ref{tab:moments} we show the values for the skewness and kurtosis of
density for models with resolution of $512^3$, as well as for the synthetic observations, i.e. the column density
maps ($512^2$), for all models. We applied the moment statistics to 3 different snapshots for each model. Ultimately, we averaged together statistics from all 3 snapshots to obtain the presented results with error bars determined using the standard deviation of the results for each model. These error bars are not the error on the calculated skewness of the models.  Instead they represent the spread of skewness between the snapshots of each model, i.e. the standard deviation. From this data we can see how the skewness and kurtosis of these quantities depend on sonic and Alfv\'en Mach numbers.

The skewness of density and column densities in the x and y directions are strictly positive. We see the same trend in both density and column density in that the asymmetry grows with increasing ${\cal M}_{s}$ for both quantities. As
expected, small ${\cal M}_{s}$ models are more Gaussian ($\gamma \rightarrow
0$), compared to supersonic models. The
presence of shocks, which result in high density structures, is enhanced as
${\cal M}_{s}$ increases and these shocks in supersonic turbulence are the cause for the asymmetry in the density and column density distributions. We also analyze the skewness of averaged data cubes as shown  in Table~\ref{tab:avmoments}.  We average together points in a cube in order to smooth out far outlying data that might be affecting the overall skewness. From Table~\ref{tab:avmoments}, we can see that even after averaging, the asymmetry in the density and column density distributions still increases with increasing ${\cal M}_{s}$, although not as sharply as the non-averaged case, due to the smoothing of outliers.

\begin{table*}
\begin{center}
\caption{Skewness of  Density and Column Densities Integrated along X, Y, and Z-direction averaged with a Gaussian beam of 100 pixels 
\label{tab:avmoments}}
\begin{tabular}{|c|c|c|c|c|c|} \hline
$M_{s}$ & $M_{A}$ & $\gamma$ Av. $\rho$  & $\gamma$ Av. $\Sigma_{X}$  & $\gamma$ Av. $\Sigma_{Y}$ & $\gamma$ Av. $ \Sigma_{Z}$  \\
0.7 & 0.7 & $0.7^{\pm{0.1}}$ & $0.05^{\pm{0.02}}$ & $0.3^{\pm{0.2}}$ & $-0.04^{\pm{0.2}}$ \\
0.7 & 2.0 & $1.0^{\pm{0.3}}$ & $0.02^{\pm{0.2}}$& $0.2^{\pm{0.3}}$ & $-0.1^{\pm{0.2}}$ \\ 
2.0 & 0.7 & $3.1^{\pm{0.1}}$ & $0.63^{\pm{0.04}}$ &  $1.03^{\pm{0.02}}$ &  $0.9^{\pm{0.2}}$ \\
2.0 & 2.0 & $3.1^{\pm{0.2}}$ & $0.8^{\pm{0.4}}$ & $0.9^{\pm{0.2}}$ & $0.5^{\pm{0.1}}$ \\
7.0 & 0.7 & $5.65^{\pm{0.04}}$ & $0.87^{\pm{0.02}}$ & $1.2^{\pm{0.2}}$ & $1.1^{\pm{0.1}}$ \\
7.0 & 2.0 & $5.0^{\pm{0.3}}$ & $1.5^{\pm{0.2}}$ & $1.1^{\pm{0.1}}$ & $1.0^{0.1}$ \\
  
\hline

\hline

\hline
\end{tabular}
\end{center}
\end{table*}

The kurtosis of both density and column density is shown to be also higher for supersonic models, which implies sharper distributions. It is a consequence of the transfer of mass from the average values to the right tail, mostly due to shocks. Again, subsonic models present kurtosis around zero, i.e. similar to Gaussian, in agreement with KLB.  Highly supersonic column density models are systematically more peaked then subsonic models, following the trend noted from the density distributions. 

From this information we can see that the asymmetry of the distribution is greater for cases of high ${\cal M}_{s}$, where shocks produce high density clumps. Just as in KLB, skewness and kurtosis both seem to be relatively unaffected by the strength of the magnetic field.  In this way, we can determine ${\cal M}_{s}$ by examining the skewness of observational column densities. For kurtosis, we see that subsonic models present nearly perfect Gaussian distributions for both densities and column densities. This implies that if density is perturbed weakly, it remains nearly normal. Both measures are strongly dependent on the sonic Mach number, making them useful statistics for observers to characterize interstellar turbulence.

\section{Correlations}
\label{sec:correlations}

\subsection{Correlation of three-dimensional fields}

 It is generally accepted that isothermal simulations, such as the one presented in this paper, are reasonable models of molecular clouds, while non-isothermal simulations better represent the many temperatures of the multiphase ISM. In order to study relationships between global parameters in molecular clouds or ionized gas, correlations are useful.  At the most basic level, correlations can be used to determine a possible relationship between two quantities. Correlation plots can provide information on the dynamical importance of kinetic and magnetic energy in the evolution of density clumps in star forming clouds. 
 In this section we study these correlations, presenting plots of magnetic energy vs. density, and ${\cal M}_{A}$ vs. normalized density. We also discuss the correlation of specific kinetic energy vs. density. In order to make comparisons to observations, plots of magnetic field vs. column density and velocity dispersion vs. column density in directions parallel and perpendicular to the mean magnetic field are included. We also study the correlation between dispersion velocity for constant density vs. non constant density in order to see how density fluctuations affect this observable quantity  in our simulations. Correlations were preformed for the final snapshots of all models.

\begin{figure*}[tbh]
\centering
\includegraphics[scale=.23]{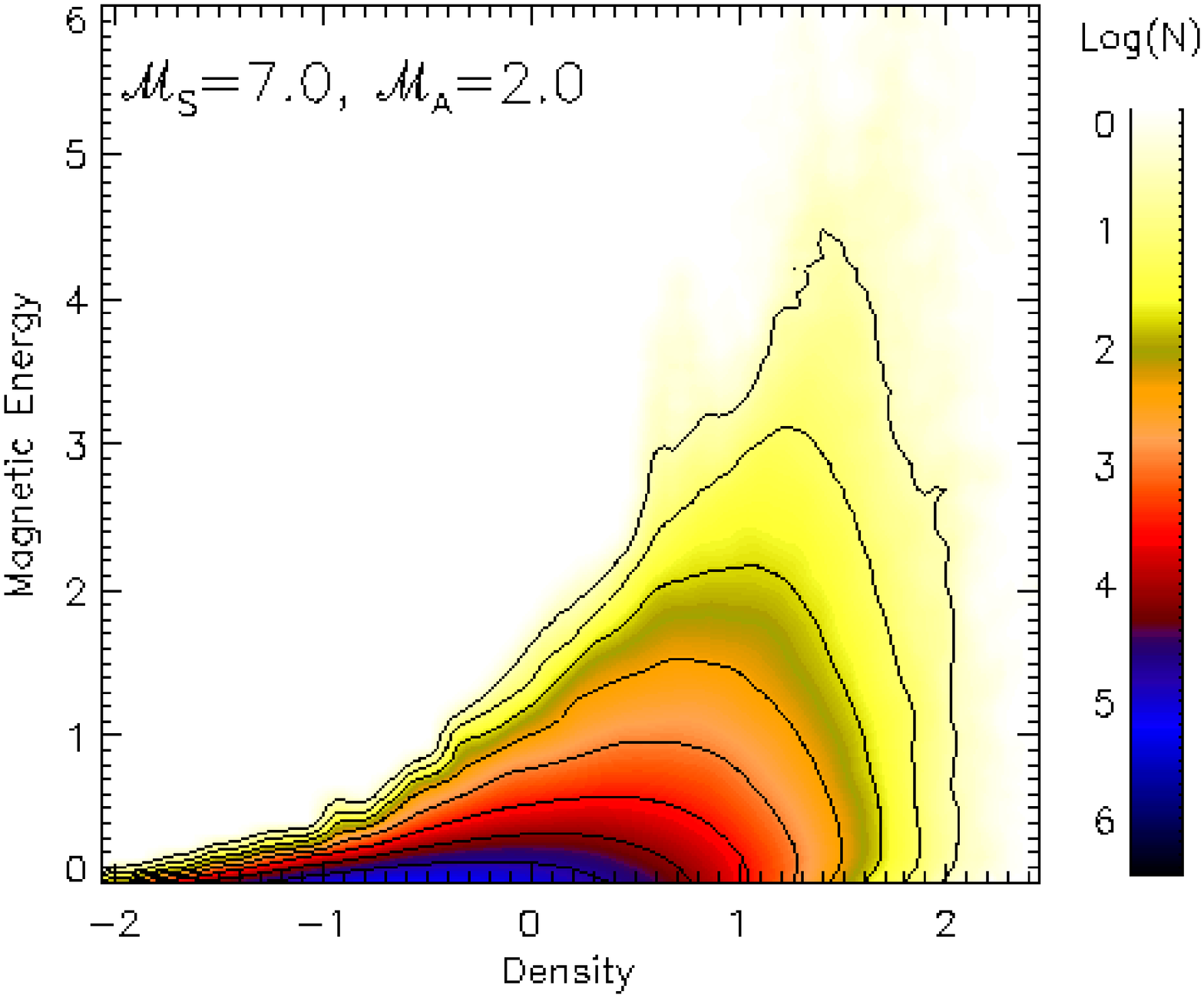} % \includegraphics[scale=.35]{apj/c512b.1p.01/paper/corr_dens_magener.eps}
\includegraphics[scale=.23]{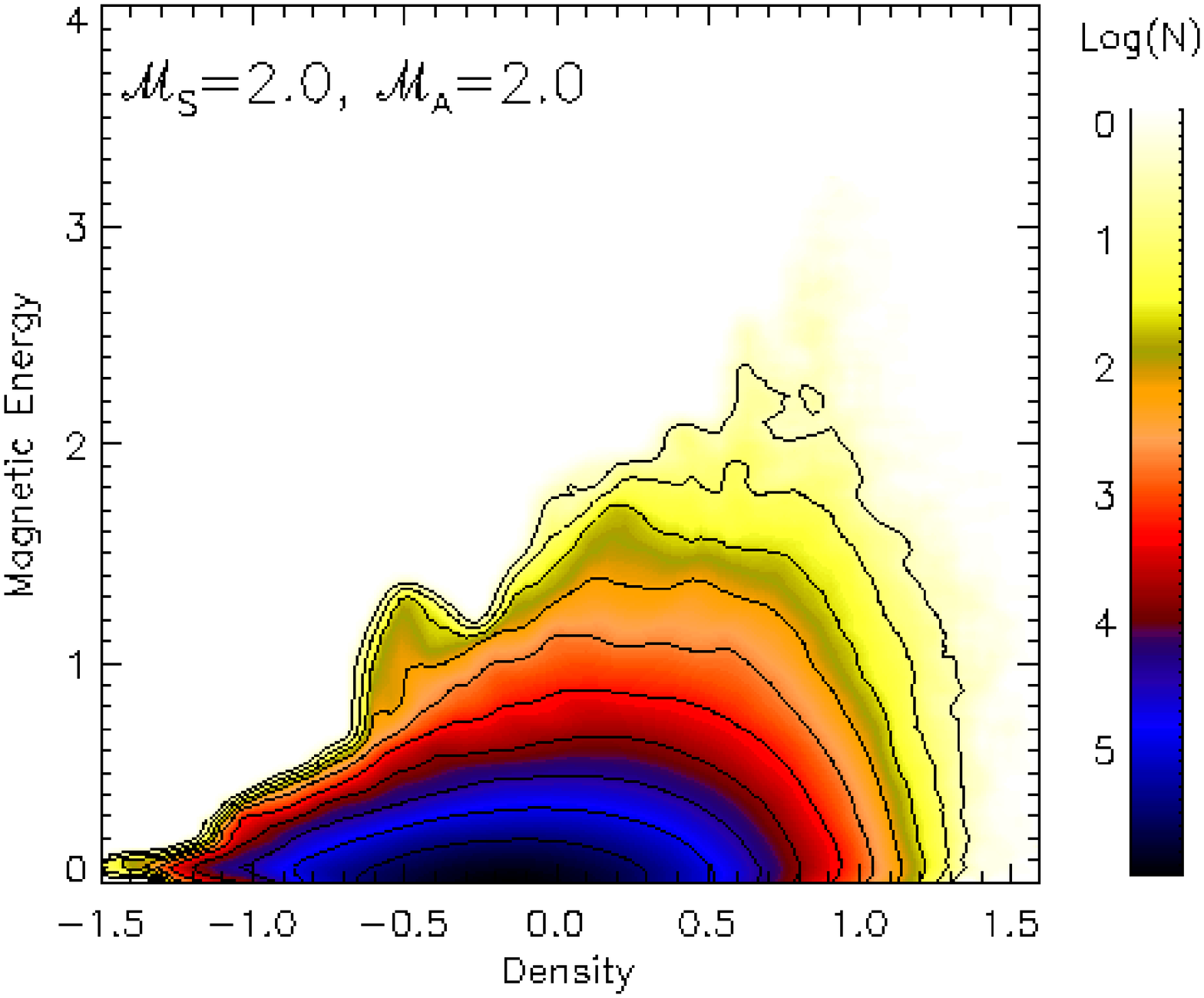} % \includegraphics[scale=.35]{apj/c512b.1p.1/paper/corr_dens_magener.eps}
\includegraphics[scale=.23]{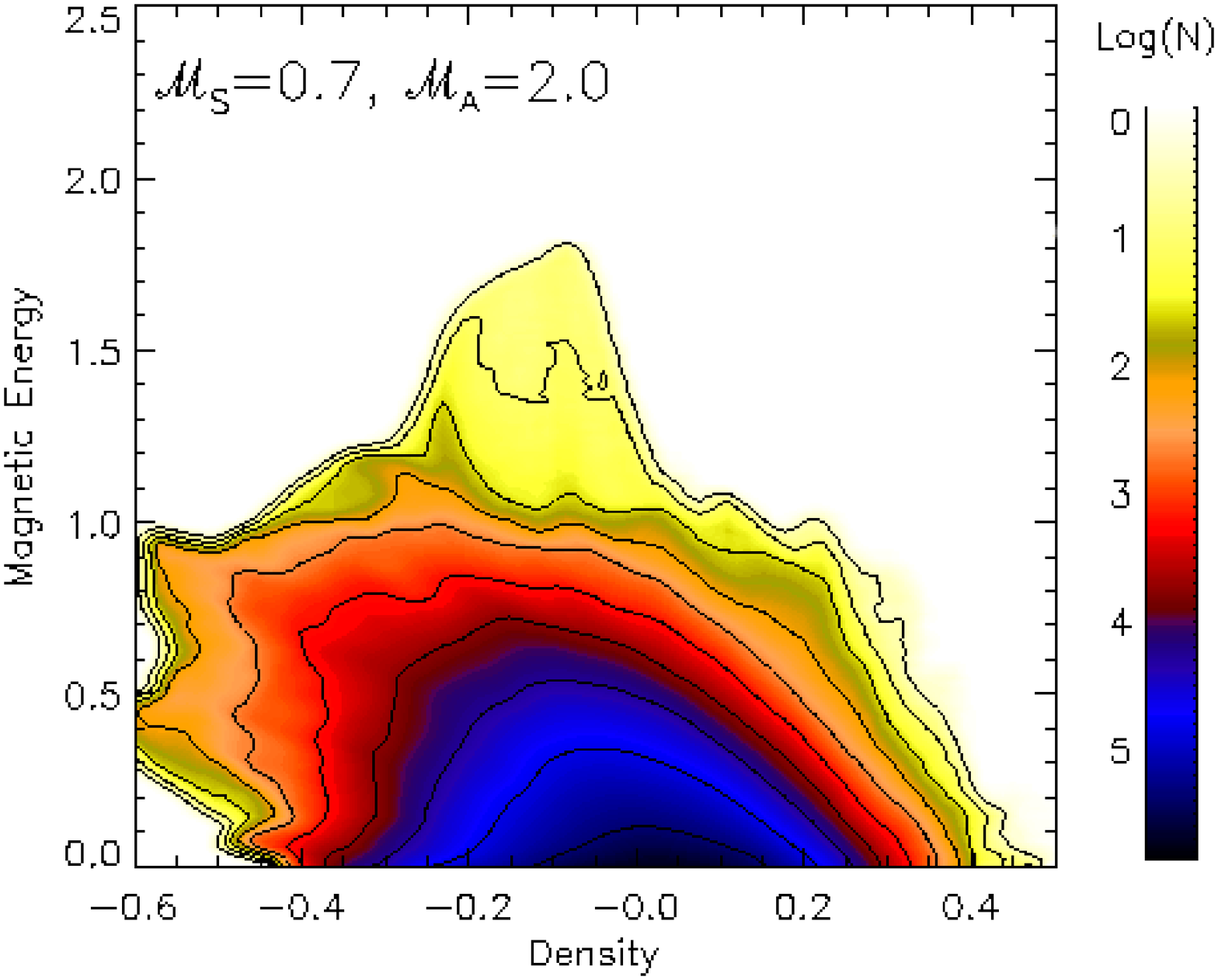} % \includegraphics[scale=.35]{apj/c512b.1p1/paper/corr_dens_magener.eps}
\includegraphics[scale=.3]{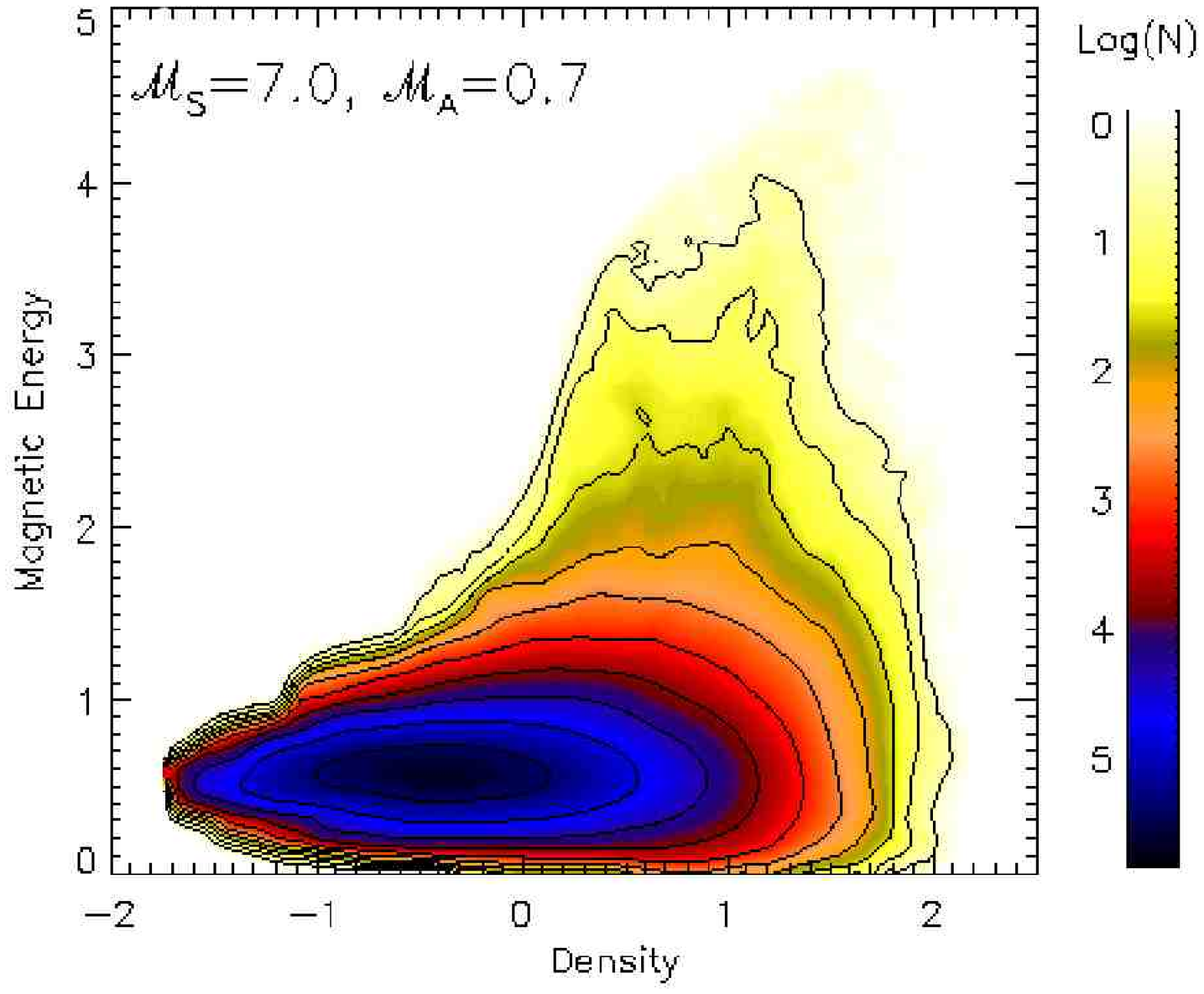} % \includegraphics[scale=.35]{apj/c512b1p.01/paper/corr_dens_magener.eps}
\includegraphics[scale=.3]{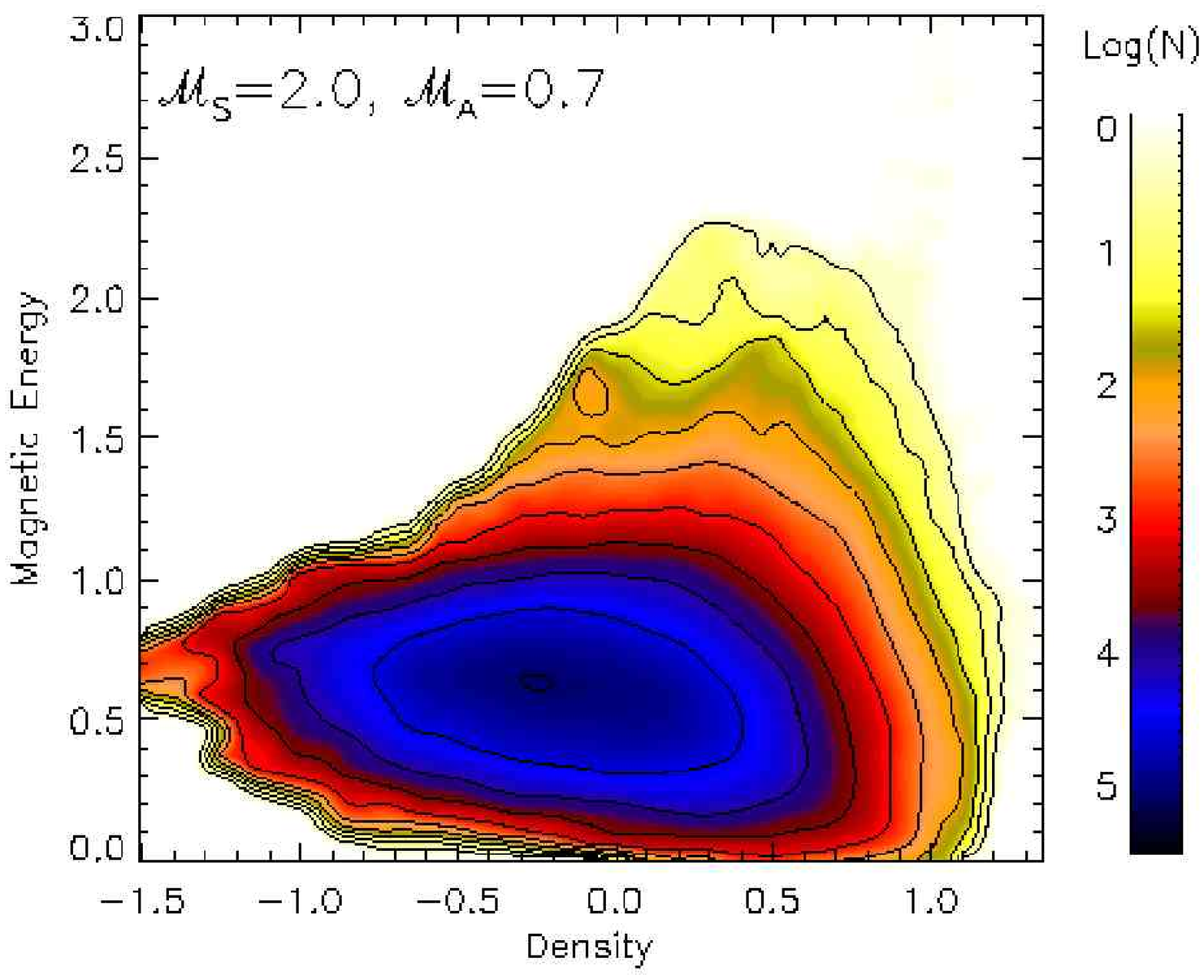} % \includegraphics[scale=.35]{apj/c512b1p.1/paper/corr_dens_magener.eps}
\includegraphics[scale=.3]{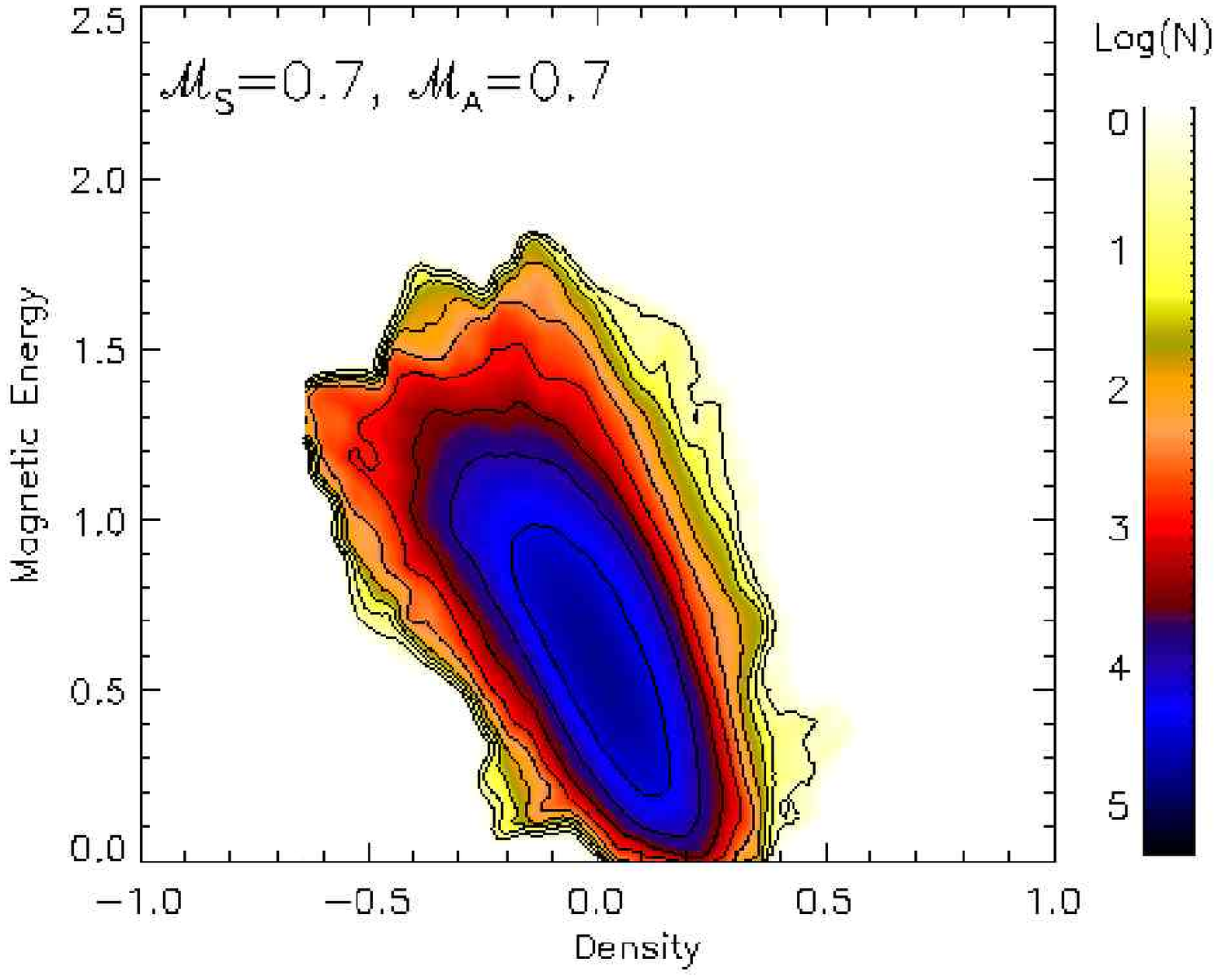} % \includegraphics[scale=.35]{apj/c512b1p1/paper/corr_dens_magener.eps}
\caption{The 3D correlation of magnetic energy vs. the logarithm of normalized density. The first row consists of
super-Alfv\'{e}nic cases while the bottom row is sub-Alfv\'{e}nic. Images are
ordered from left to right as supersonic to subsonic. Blue contours indicate
regions of high data counts while red and yellow have lower counts.}
\label{fig:dens_emag}
\end{figure*}

The plots of magnetic energy vs. log density are shown in 
Figure~\ref{fig:dens_emag}. For supersonic cases we see that the magnetic energy
is correlated with density, namely, denser regions contain stronger magnetic
fields, which is due to the compressibility of the gas and magnetic fields being frozen in.
This causes the magnetic field to follow the flow of plasma if its tension is negligible. The compressed regions are dense enough to distort the magnetic field lines, enhance the magnetic field intensity, and effectively trap the magnetic energy due to the frozen-in condition. The magnetic energy in the super-Alfv\'{e}nic model also reaches peaks higher than those in sub-Alfv\'{e}nic cases due to a larger magnetic pressure in the latter case. For subsonic cases we see a very different trend as a result of incompressible turbulent flows. Due to lack of shocks trapping the magnetic field in the density clumps, the magnetic energy is anti-correlated to density for sub-Alfv\'enic turbulence, which shows a narrow PDF peaked at the mean density. For super Alfv\'enic cases, the distribution is more spread and no relationship is obtained.

\begin{figure*}
\centering
\includegraphics[scale=.3]{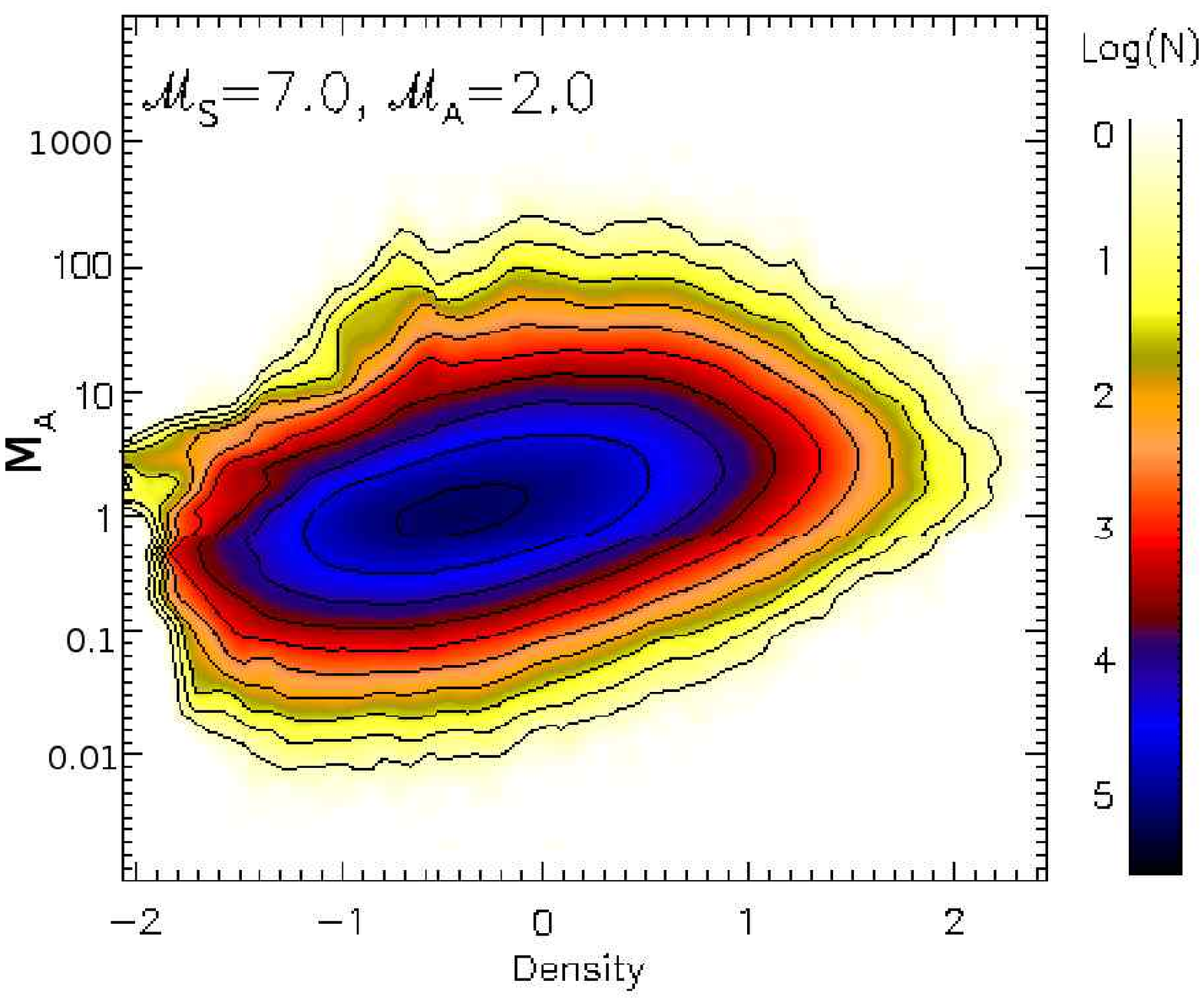} % \includegraphics[scale=.35]{apj/c512b.1p.01/paper/corr_dens_machalf}
\includegraphics[scale=.3]{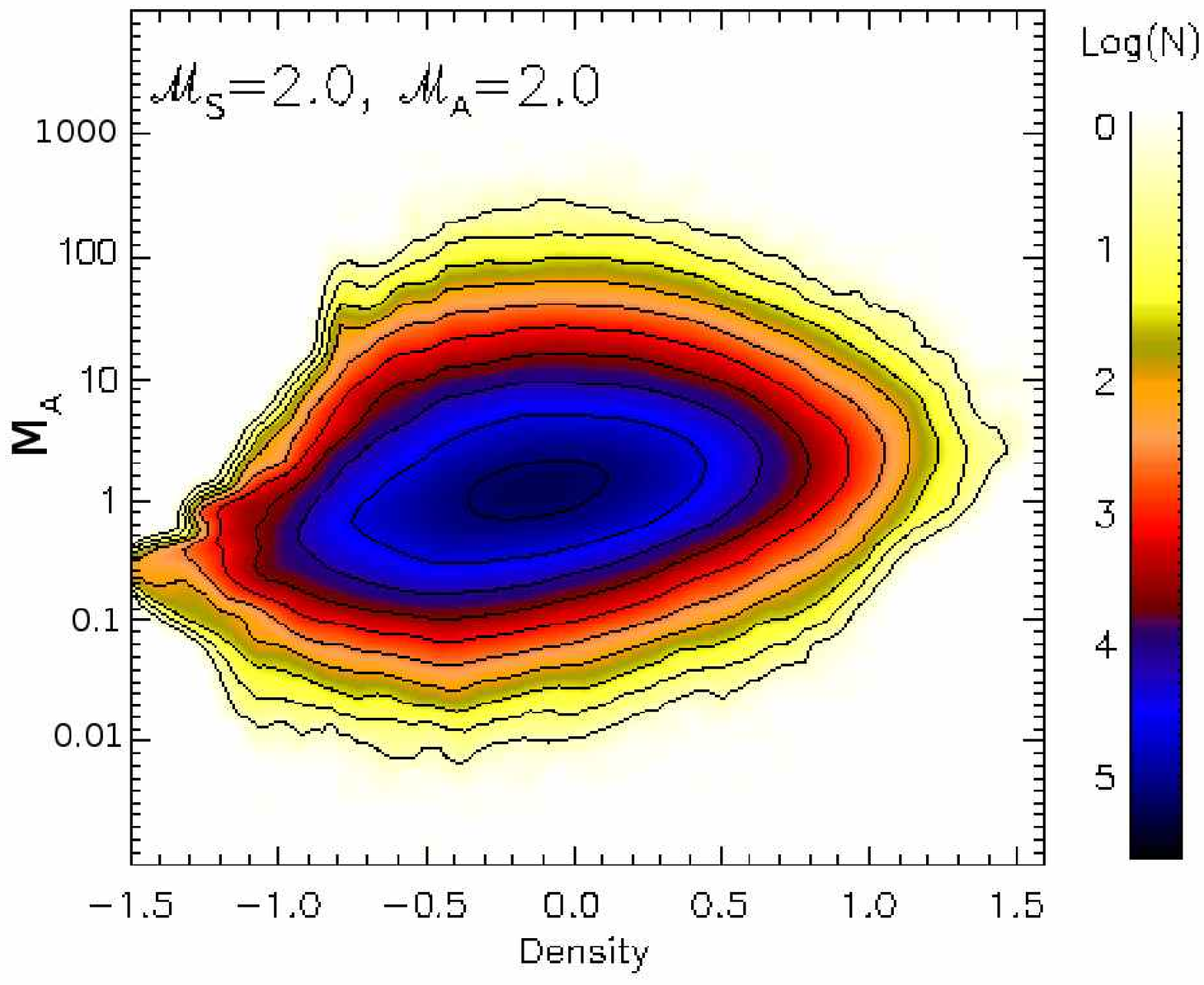} % \includegraphics[scale=.35]{apj/c512b.1p.1/paper/corr_dens_machalf}
\includegraphics[scale=.3]{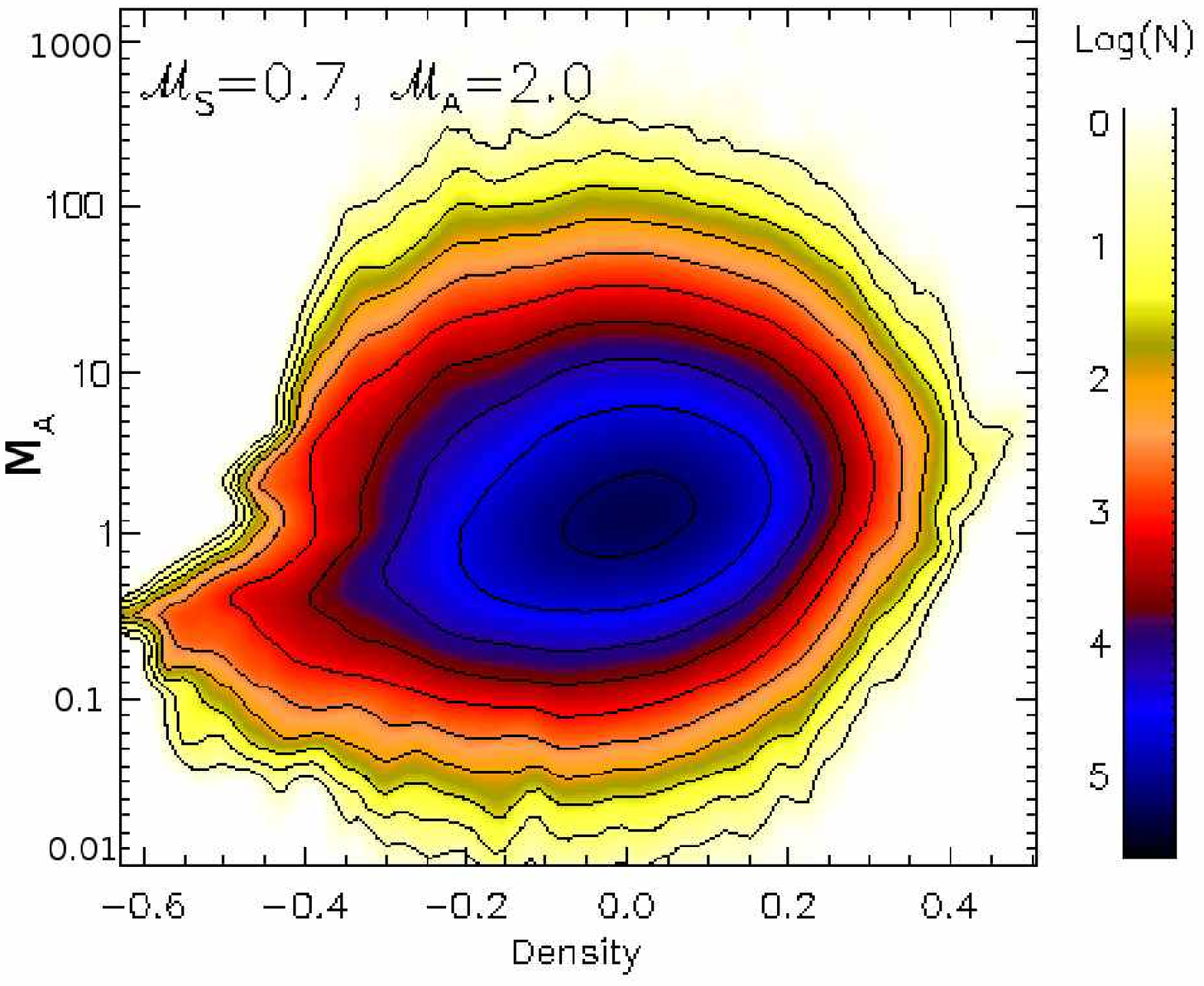} % \includegraphics[scale=.35]{apj/c512b.1p1/paper/corr_dens_machalf}
\includegraphics[scale=.3]{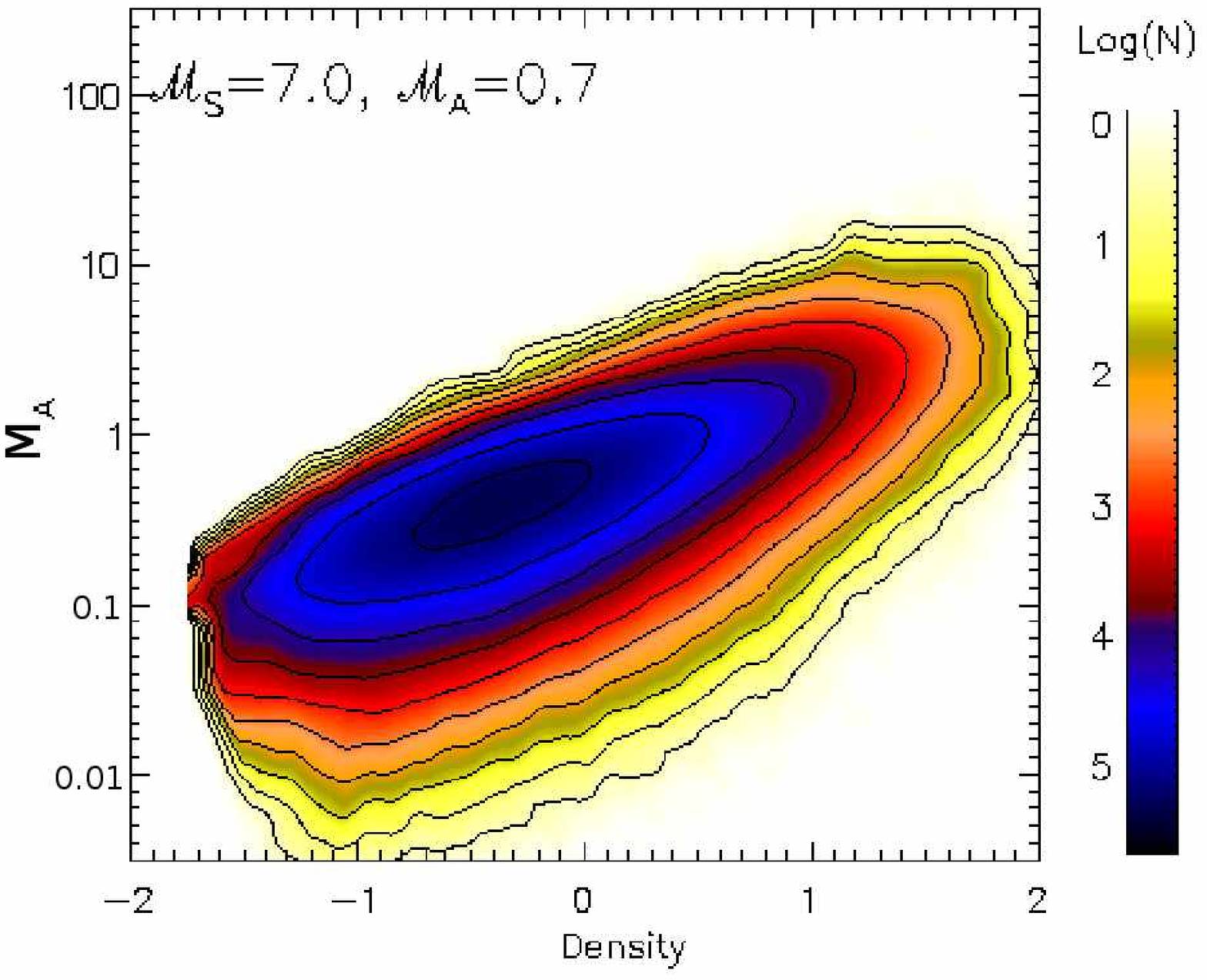} % \includegraphics[scale=.35]{apj/c512b1p.01/paper/corr_dens_machalf}
\includegraphics[scale=.3]{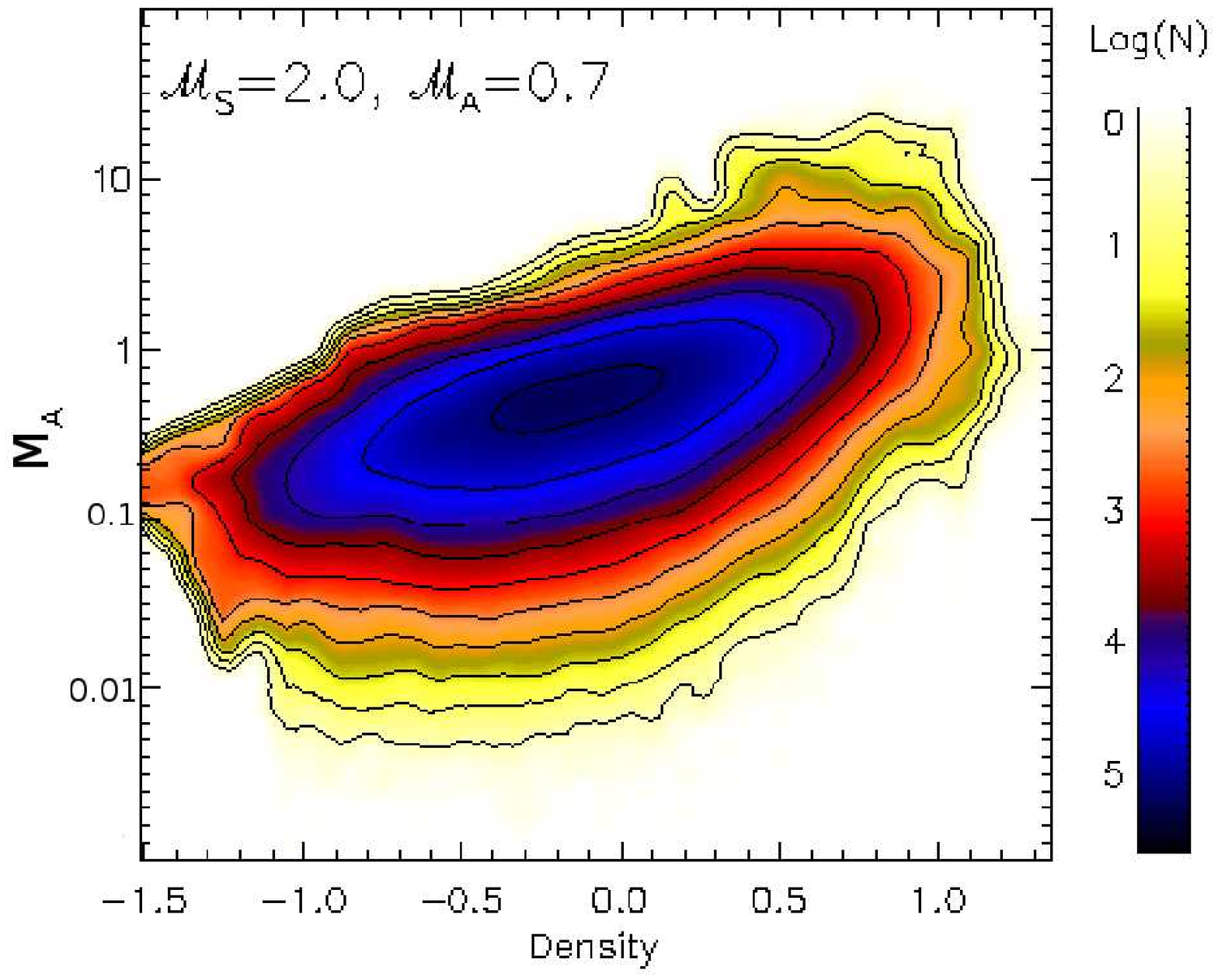} % \includegraphics[scale=.35]{apj/c512b1p.1/paper/corr_dens_machalf}
\includegraphics[scale=.3]{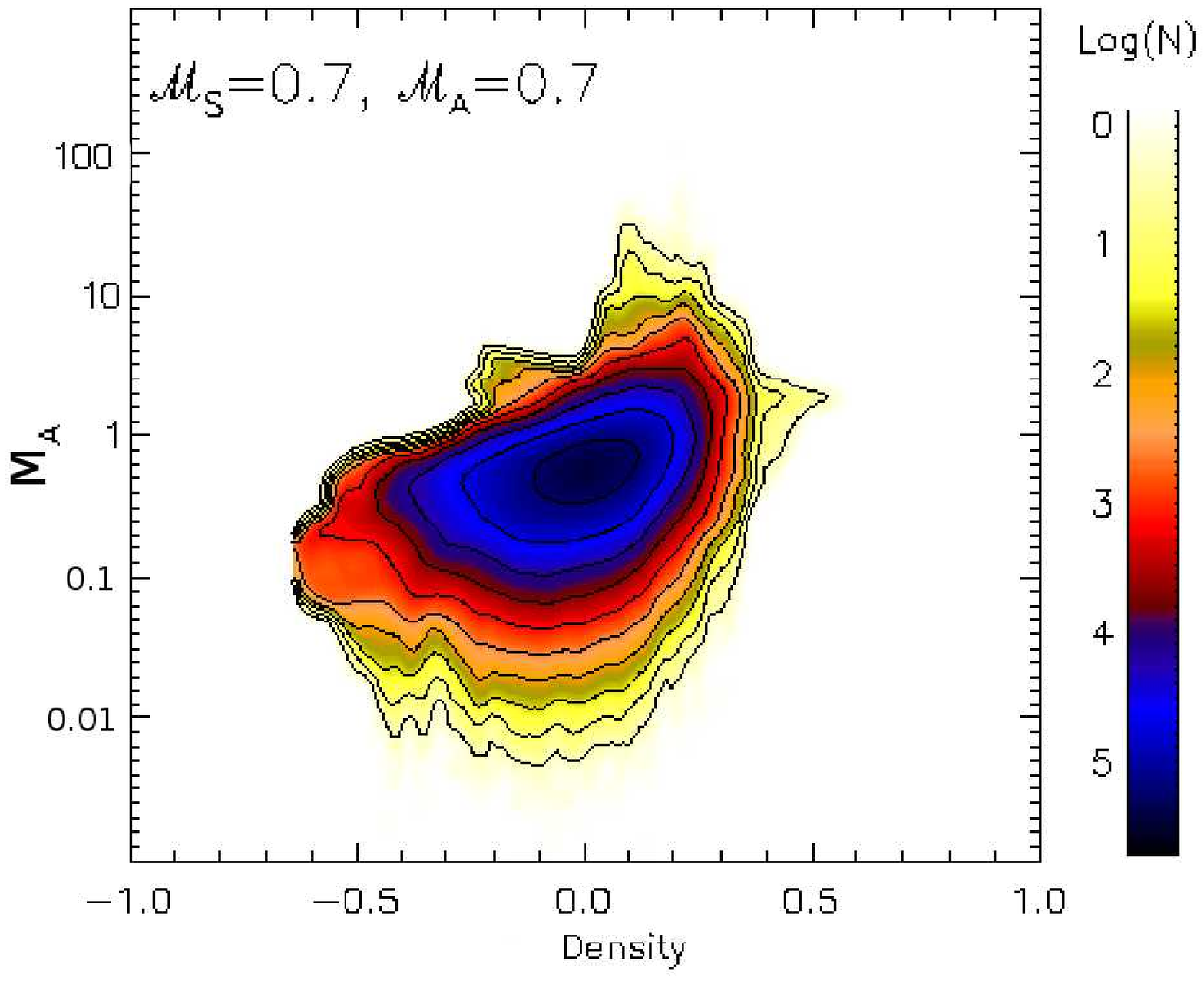} % \includegraphics[scale=.35]{apj/c512b1p1/paper/corr_dens_machalf}
\caption{The 3D correlation of the local Alfv\'enic Mach number ($M_{A}$) vs. normalized density for a log log scale. The first row consists of  globally super-Alfv\'enic cases while the bottom row is globally sub-Alfv\'enic. The mean Alfv\'enic number is always nearly the global value for the model. Images are ordered left to right as supersonic to sub-sonic. Blue contours indicate regions of high data counts while red and yellow have lower counts. We show the actual values of $M_{A}$ for ease of reading.  \label{fig:dens_malf}}
\end{figure*}

We examined the correlation of squared velocity (i.e. the specific kinetic energy (${\cal E}_k=V^2$) vs. the
logarithm of density, but do not include plots since no clear relationship is seen.  We observe a decrease in velocity for higher density, however here we are also interested in how density and the kinetic energy density (i.e.  $E_{k}=\rho* V^{2}$ vs. $\rho$) relate to one another.  For specific kinetic energy ${\cal E}_k$ vs. $\rho$, the distribution tends to isotropically cluster around the mean value of density. Therefore, while the specific kinetic energy might decrease with larger densities, the  kinetic energy density increases in a near linear relationship for supersonic models. Subsonic models have minimum kinetic energies at values of log-density that vary less from the mean density then do supersonic models (-0.6 to 0.4 for subsonic and -2 to 2 for supersonic). Compressibility in the subsonic models plays less of a role than in supersonic ones, and thus they reach their minimum kinetic energy with clumps that are less dense then supersonic models. Because density clumps are not prevalent in subsonic models, we see no clear relationship between density and kinetic energy, with the highest values for $E_{k}$ being around the mean of density and then falling off.

Apart from the global Alfv\'en Mach number, which is defined as ${\cal M}_{A}=\frac{V_{L}}{V_{A}}$ where $V_{L}$ is the velocity flow \footnote{In this section, we denote global Alfv\'en Mach number with a calligraphic ${\cal M}_{A}$, versus the local Alfv\'en Mach number which is denoted by $M_A$}, which is computed by taking the magnitude of velocity $V_L=\sqrt{v_{x}^{2}+v_{y}^{2}+v_{z}^{2}}$, and $V_{A}$ (the Alfv\'en speed), one can also determine the local Alfv\'en Mach number, which is what observers can potentially measure \footnote{The estimate of the local Mach number is possible, for instance, using the $B_{z}$ component of magnetic field available through Zeeman splitting and the velocity dispersion of clouds available through spectroscopic measurements.}. ${\cal M}_{A}$ depends on both the mean magnetic field and density $\rho$, which can result in differences between the global ${\cal M}_{A}$ of a cloud and a density clump within the cloud. In order to explore the relationship between density and $M_{A}$ we present here the correlation of the  log Alfv\'{e}nic Mach number versus the log of $\frac{\rho}{\rho_{0}}$, shown in Figure~\ref{fig:dens_malf}. Although we use a log-log scale, the actual numbers of $M_A$ are used for ease of reading. We obtained a general trend of $M_{A}$ slightly increasing with density, especially for sub Alfv\'enic cases. This is because ${\cal M}_{A}$ has dependence on density: ${\cal M}_{A}=\frac{V_{L}}{v_{A}}=\frac{V_{L}\sqrt{4\pi \rho_{0}}}{B}$.  Cases with globally sub-Alfv\'enic turbulence show local $M_{A}$ continually increasing with density until local values become super-Alfv\'enic. Cases that are globally super-Alfv\'enic reach local values that are higher then $M_{A}=2.0$ which peak at the mean density. However, values of density larger then the mean yield decreasing values of $M_{A}$ for the globally super-Alfv\'enic models, although they continue to stay locally super-Alfv\'enic. In essence, we find that globally sub-Alfv\'enic models become locally super-Alfv\'enic with increasing density while the globally super-Alfv\'enic models remain so for all values of density studied. This can best be explained by ${\cal M}_{A}$'s dependence on B and the relationship between B and $\rho$.  From Figure~\ref{fig:dens_emag}, it is noticeable that the magnetic energy does not grow as effectively with density in sub-Alfv\'enic cases as it does in super-Alfv\'enic ones. Therefore, even in globally sub-Alfv\'{e}nic models, the very dense regions may be super-Alfv\'{e}nic, in spite of the global regime of the cloud, since density clumps do not effectively trap the higher  magnetic field. However, most density clumps lie around  $M_{A} = 0.7$ and only the highest density clumps are able to become super-Alfv\'enic. Similarly for ${\cal M}_{A}$=2.0, past the mean density the  Alfv\'enic Mach number begins to decrease.  This is because the magnetic field increases as the density of clumps increase, causing the clumps to become more magnetically dominated and thus causing $M_{A}$ to decline. As the models progress from subsonic to supersonic, we see that these effect becomes more pronounced.

\subsection{Synthetic Observational Correlations}

\begin{figure*}
\centering
\includegraphics[scale=.25]{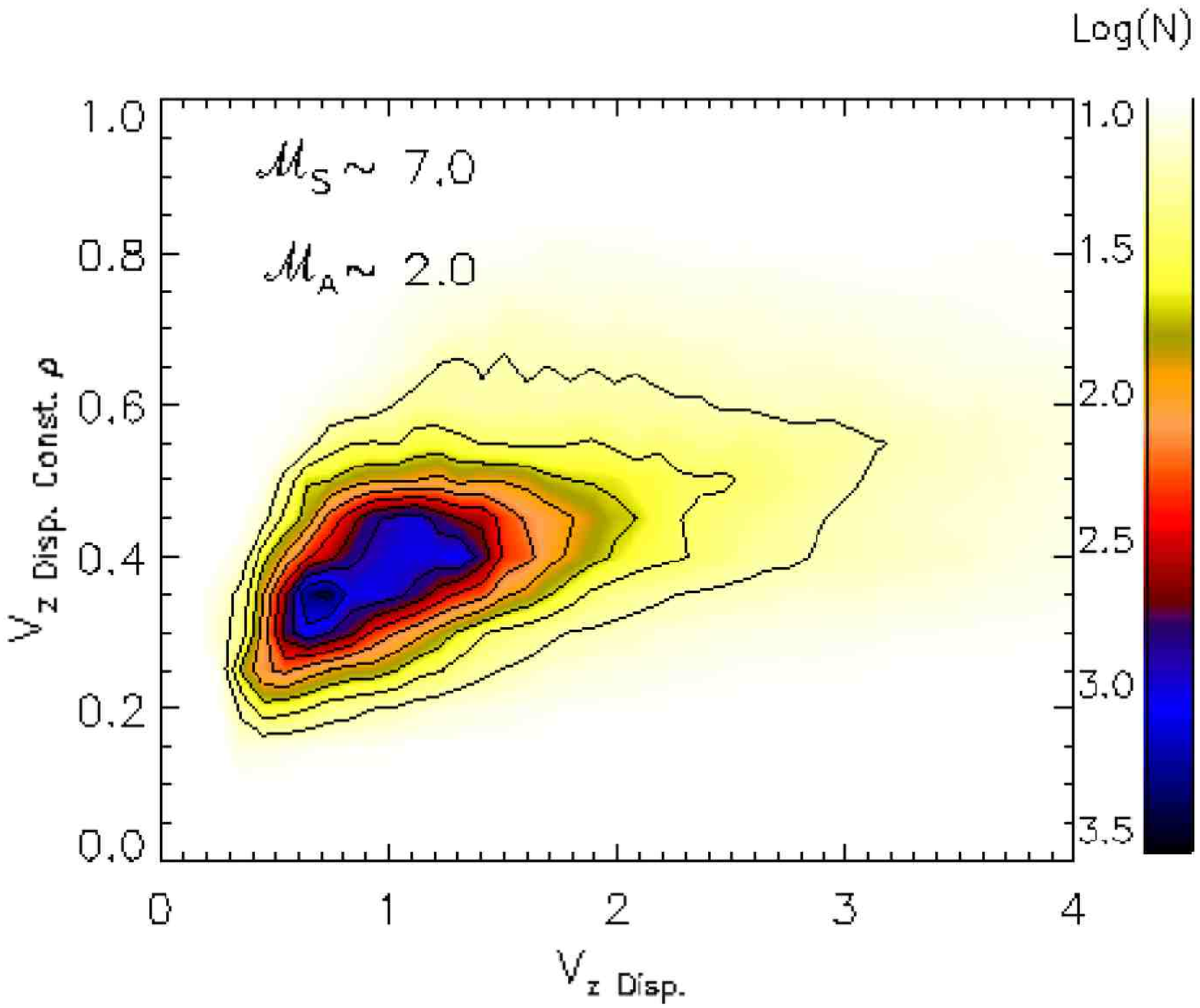} % \includegraphics[scale=.35]{apj/c512b.1p.01/paper/corr_coldens_vel_z}
\includegraphics[scale=.25]{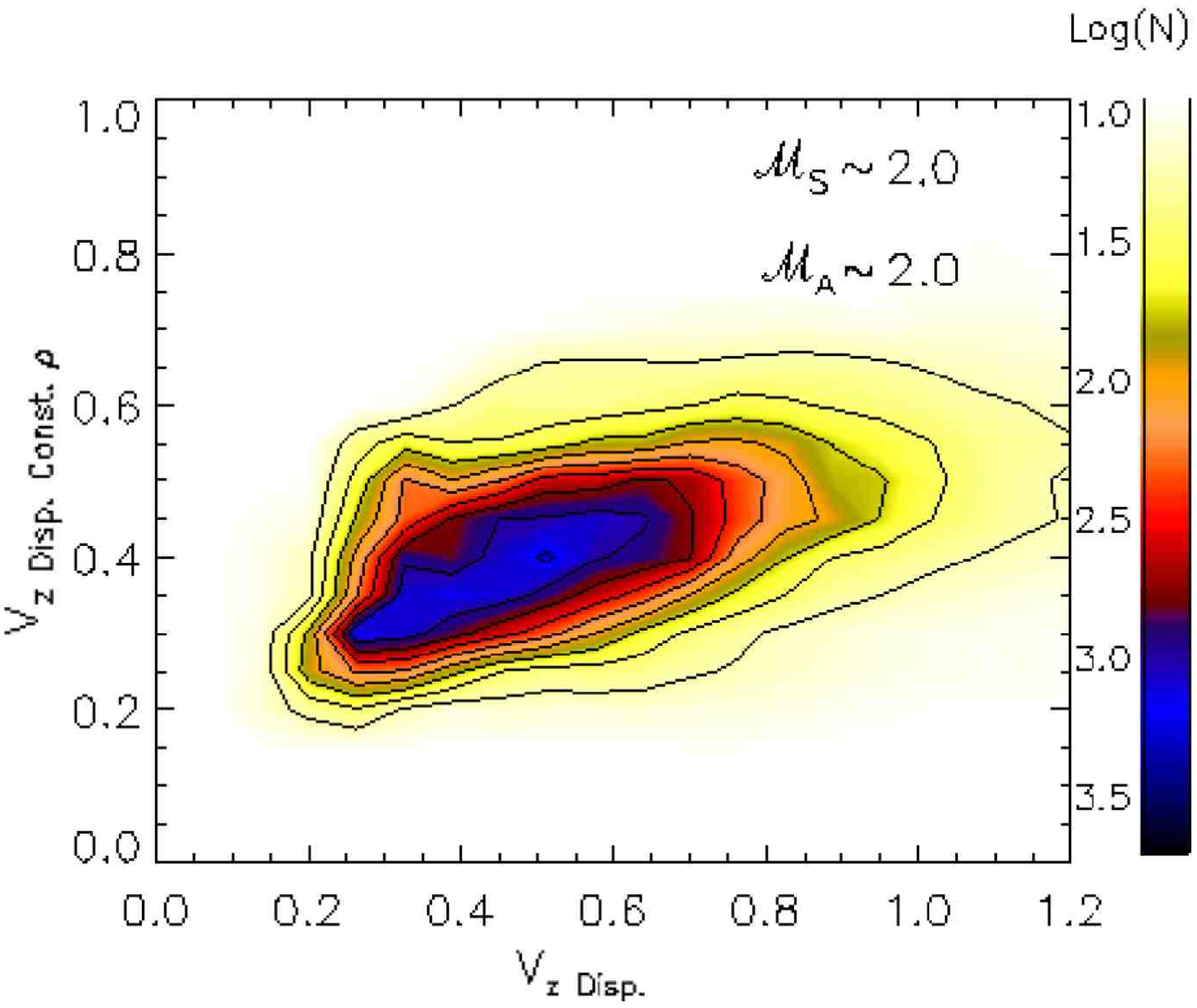} % \includegraphics[scale=.35]{apj/c512b.1p.1/paper/corr_coldens_vel_z}
\includegraphics[scale=.25]{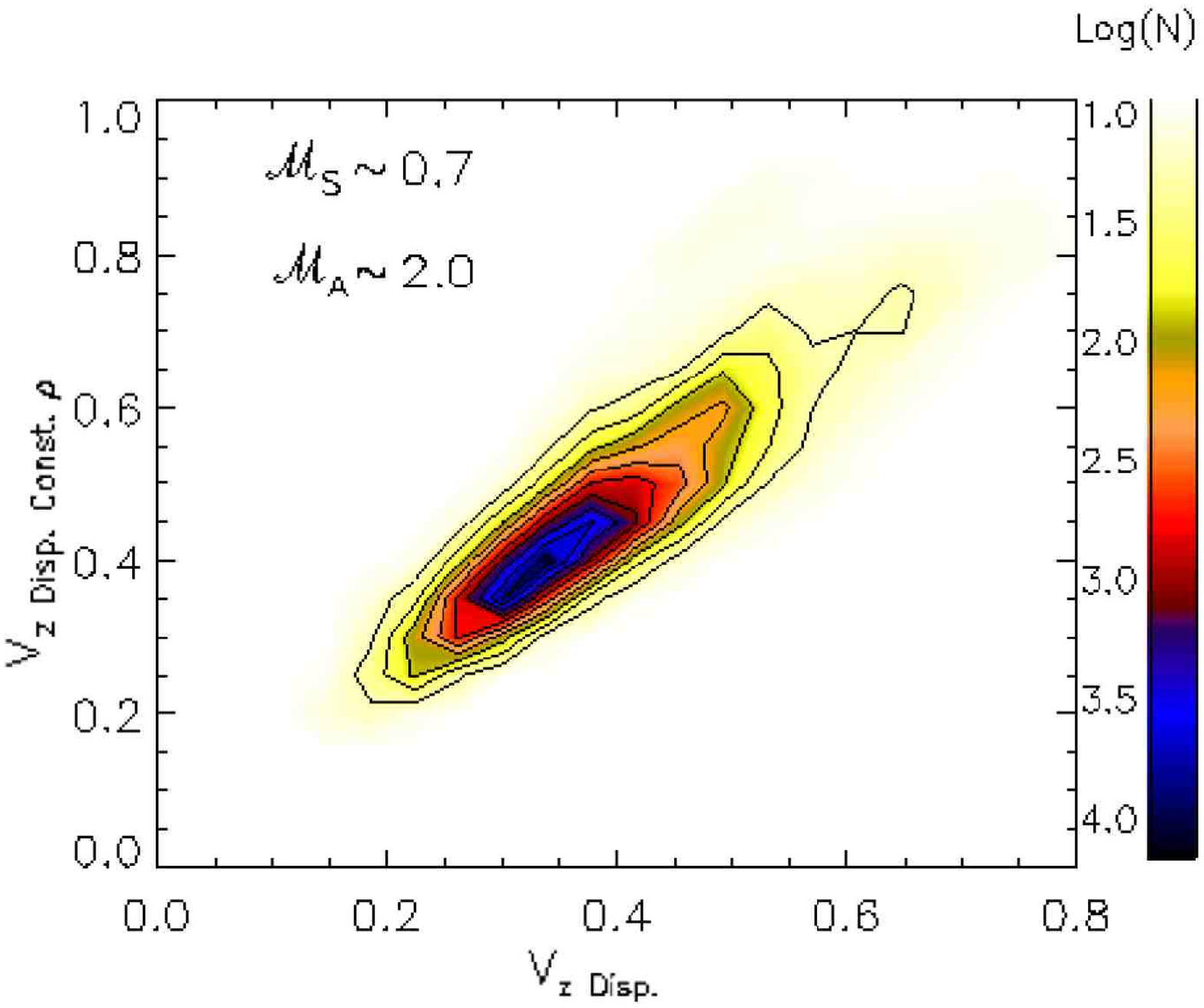} % \includegraphics[scale=.35]{apj/c512b.1p1/paper/corr_coldens_vel_z}
\includegraphics[scale=.3]{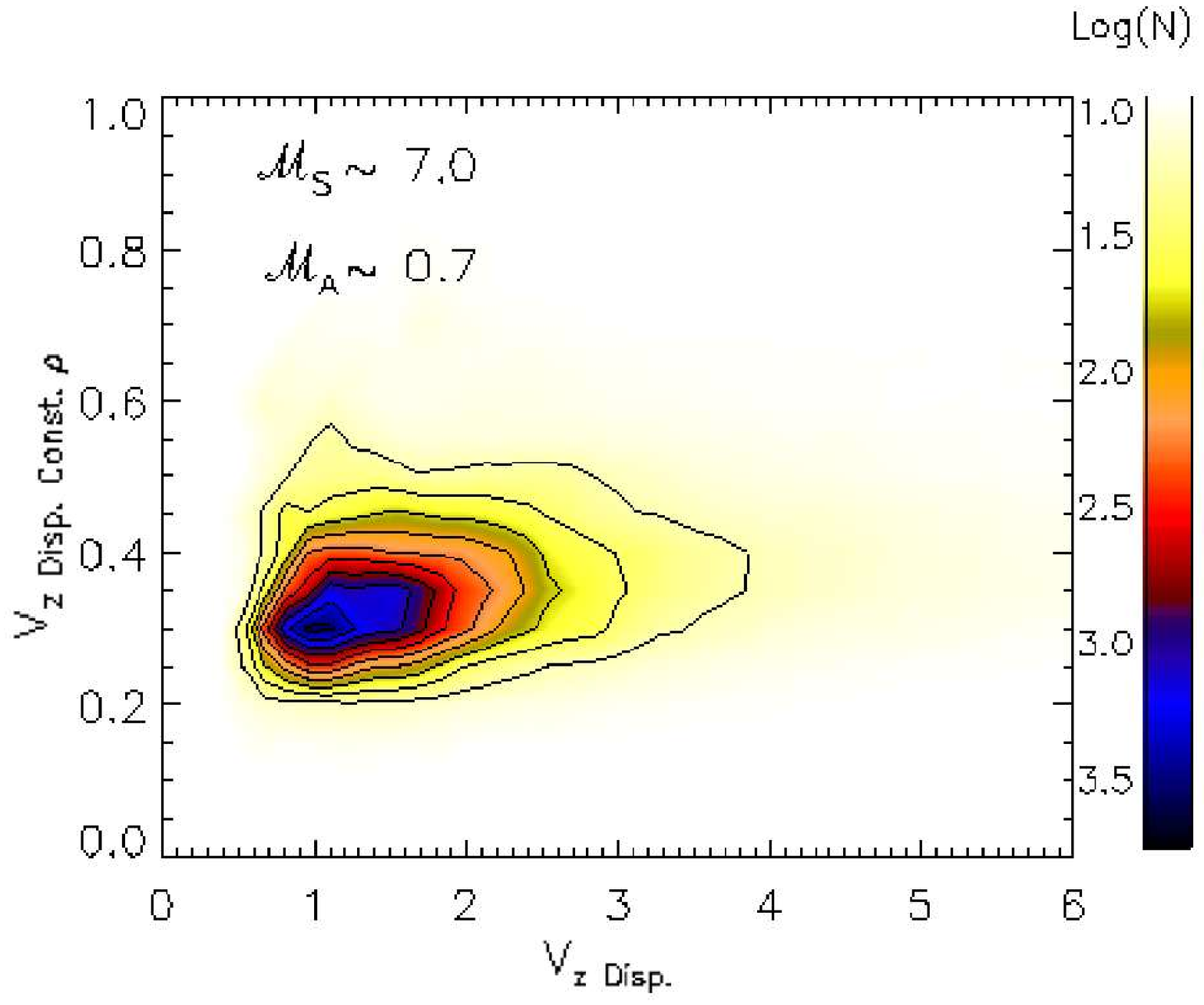} % \includegraphics[scale=.35]{apj/c512b1p.01/paper/corr_coldens_vel_z}
\includegraphics[scale=.3]{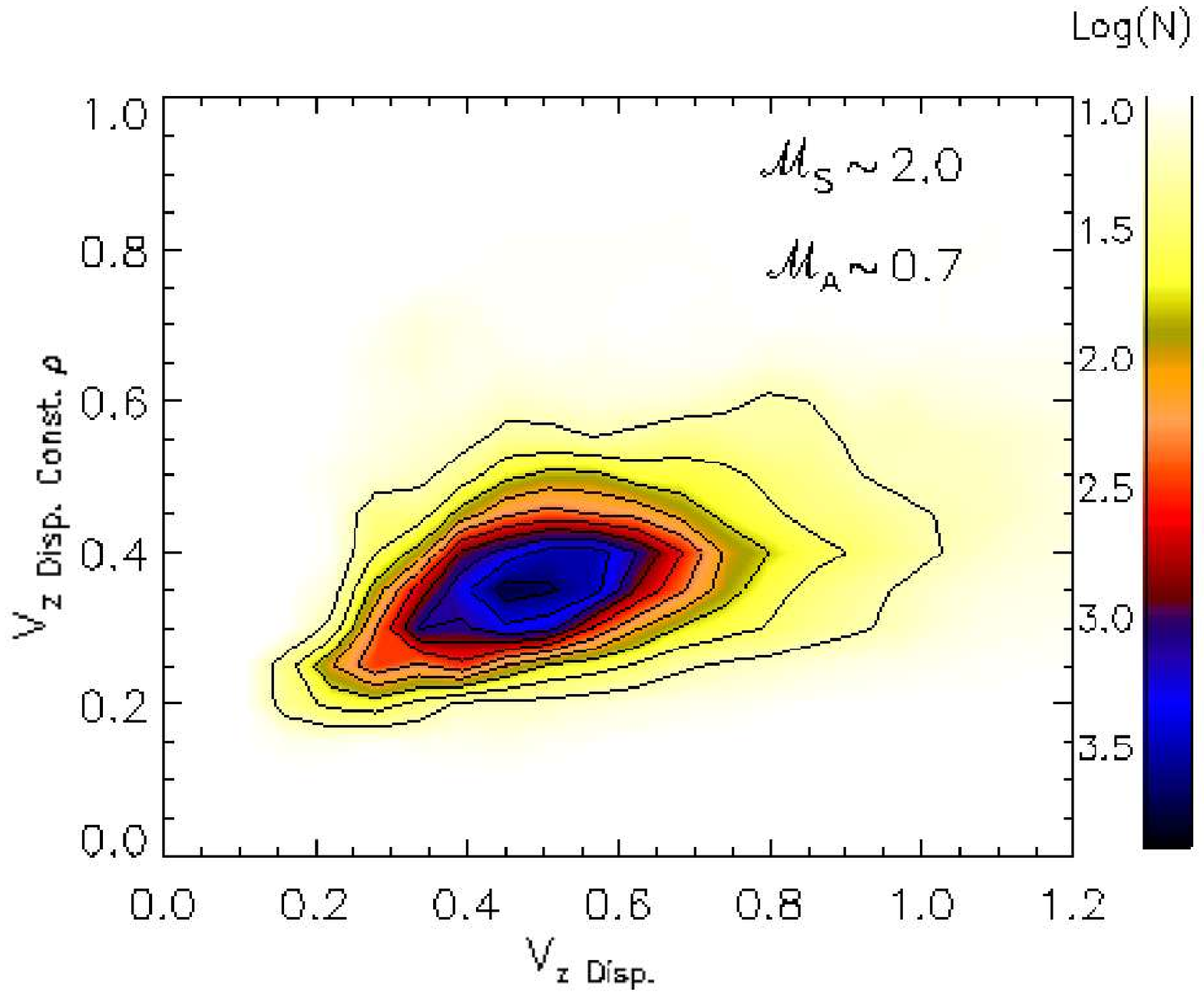} % \includegraphics[scale=.35]{apj/c512b1p.1/paper/corr_coldens_vel_z}
\includegraphics[scale=.3]{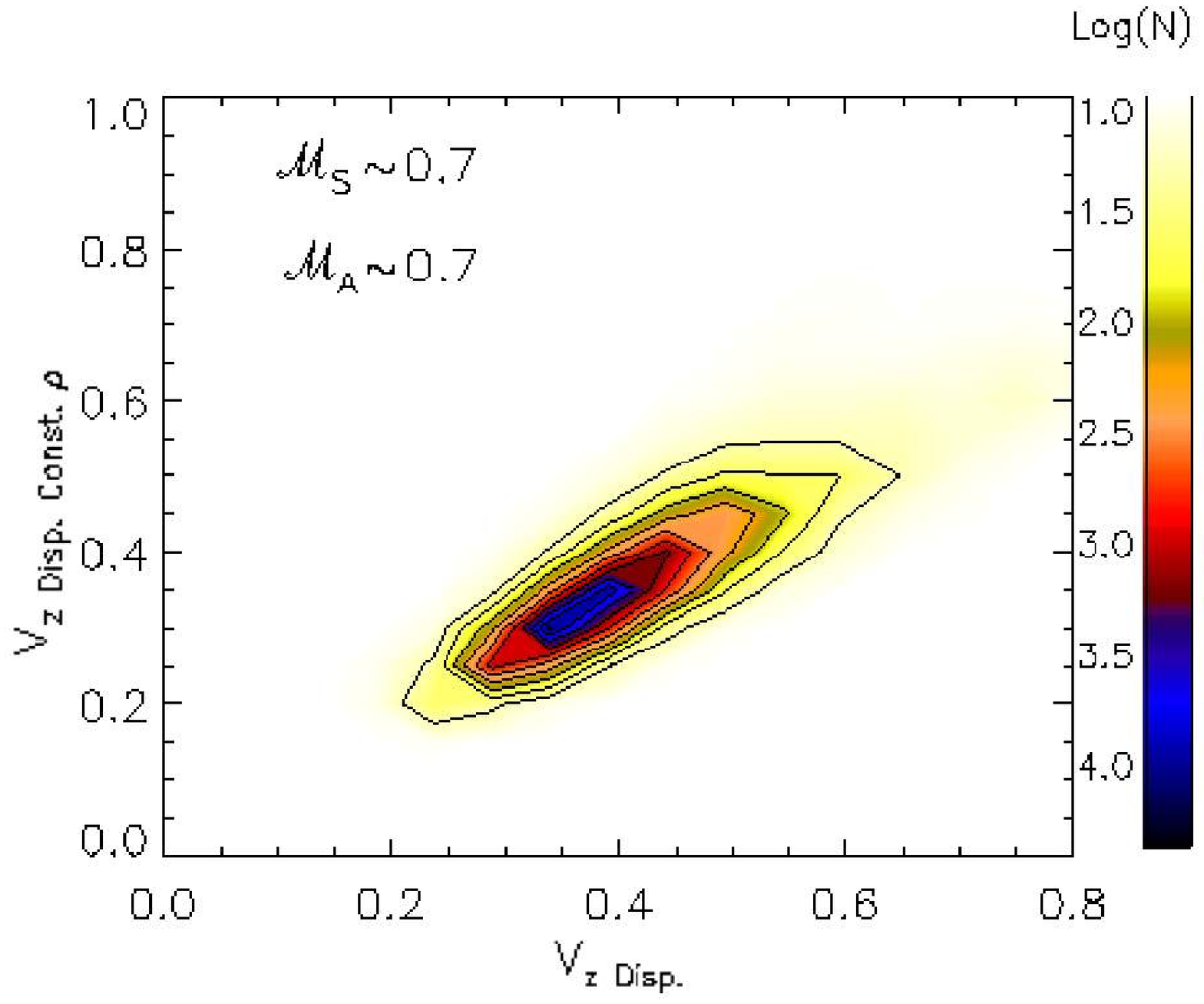} % \includegraphics[scale=.35]{apj/c512b1p1/paper/corr_coldens_vel_z}
\caption{The 2D correlation of dispersion velocity for constant $\rho$  vs. the dispersion velocity along the Z direction (i.e. mean  magnetic field $\perp$ to the LOS).    The first row consists of super-Alfv\'enic models while the bottom row is sub-Alfv\'enic. Images are ordered left to right as supersonic to subsonic. Blue contours indicate regions of high data counts while red and yellow have lower counts. Supersonic cases show evidence of nonlinear density fluctuations while subsonic cases are linear. \label{fig:vdisp}}
\end{figure*}

In order to make our studies applicable to observational analysis, we provide
correlational studies using column densities. We performed this analysis
for magnetic field and velocity dispersion in order to gain a better understanding of how
these quantities affect turbulent gas. Interest in these measures is due to the fact that the magnetic field
intensity parallel to the line of sight (LOS) is directly compared to Zeeman
measurements, and perpendicular components may be estimated using polarimetric maps \cite[see][]{falceta08}.
Regarding the velocity field, we focused on correlating column density to dispersion velocity along the LOS as obtained from line profiles. We also compare in Figure~\ref{fig:vdisp} the correlation of theoretical velocity dispersion for constant density with actual velocity dispersion available from synthetic observations. The goal of this is to study how density fluctuations affects the observables. In Figure~\ref{fig:threedplot}, we present a cubic depiction of what is meant by the directions x, y, z and the LOS. We can place an observer at different points along the plane of the cube to look vertically through with parallel sightlines. This visualization utilizes every cell in the simulated cube. However, in reality we only have one direction of viewing molecular clouds. The directions presented here are all measurable and mearly used to get a variety of column density samples along different lines of integration.  However, a cloud with a very specific  viewing geometry might require integration along a sightline above or below a certain latitude \cite[see][]{Hill08} and we do not specifically consider all these cases here.

\begin{figure*}[tbh]
\centering
\includegraphics[scale=.3]{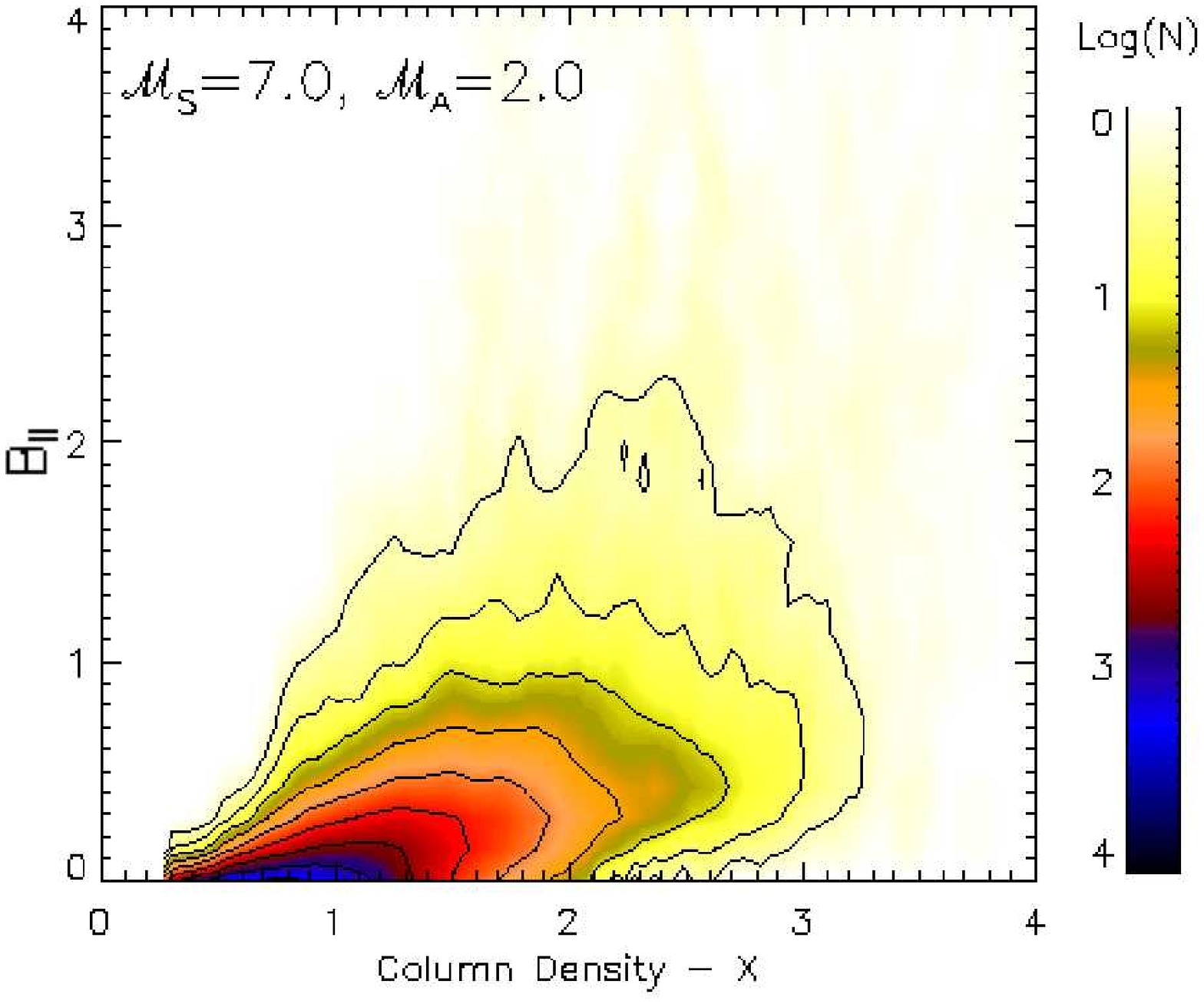} % \includegraphics[scale=.35]{apj/c512b.1p.01/paper/corr_coldens_magpar_x}
\includegraphics[scale=.3]{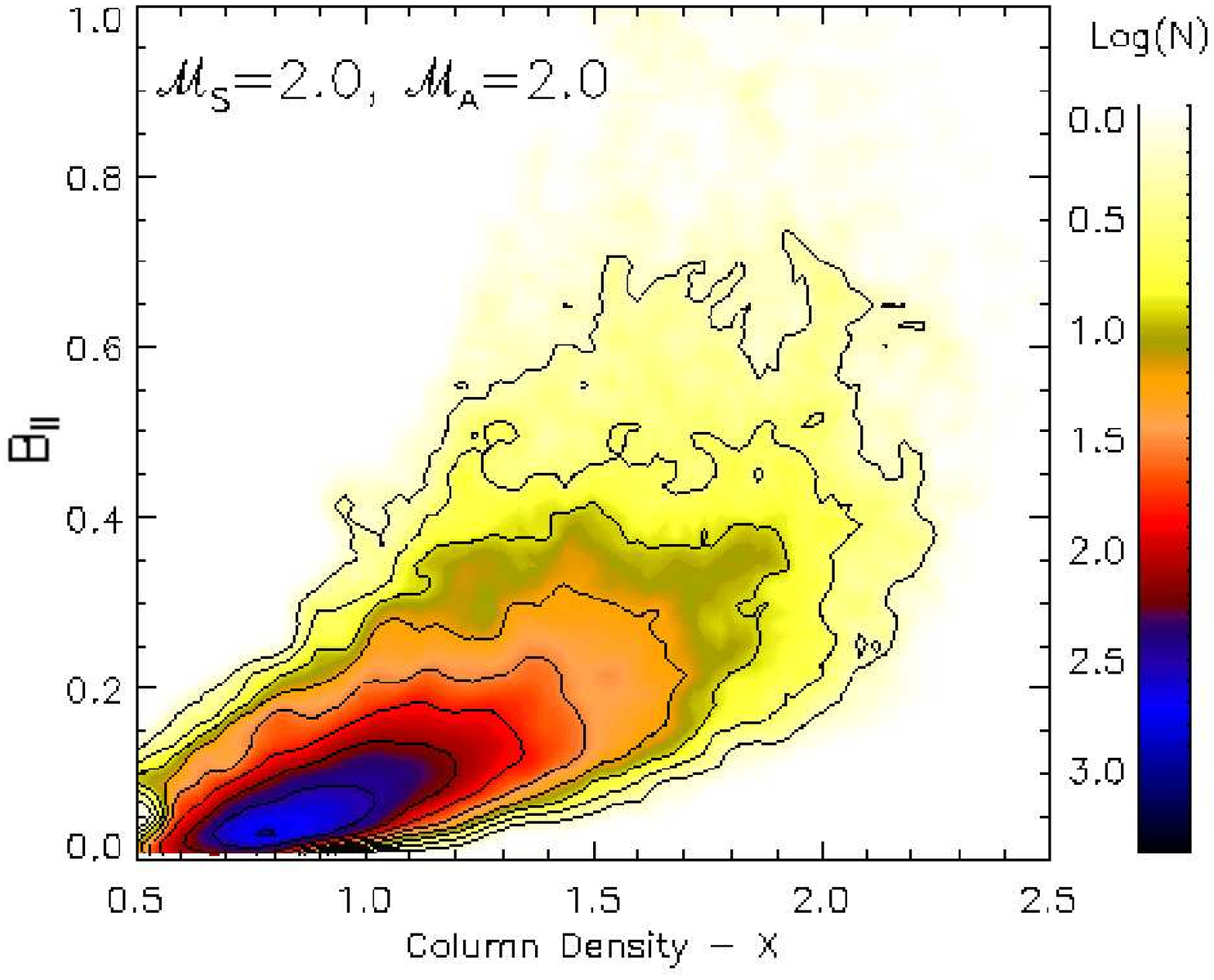} % \includegraphics[scale=.35]{apj/c512b.1p.1/paper/corr_coldens_magpar_x}
\includegraphics[scale=.3]{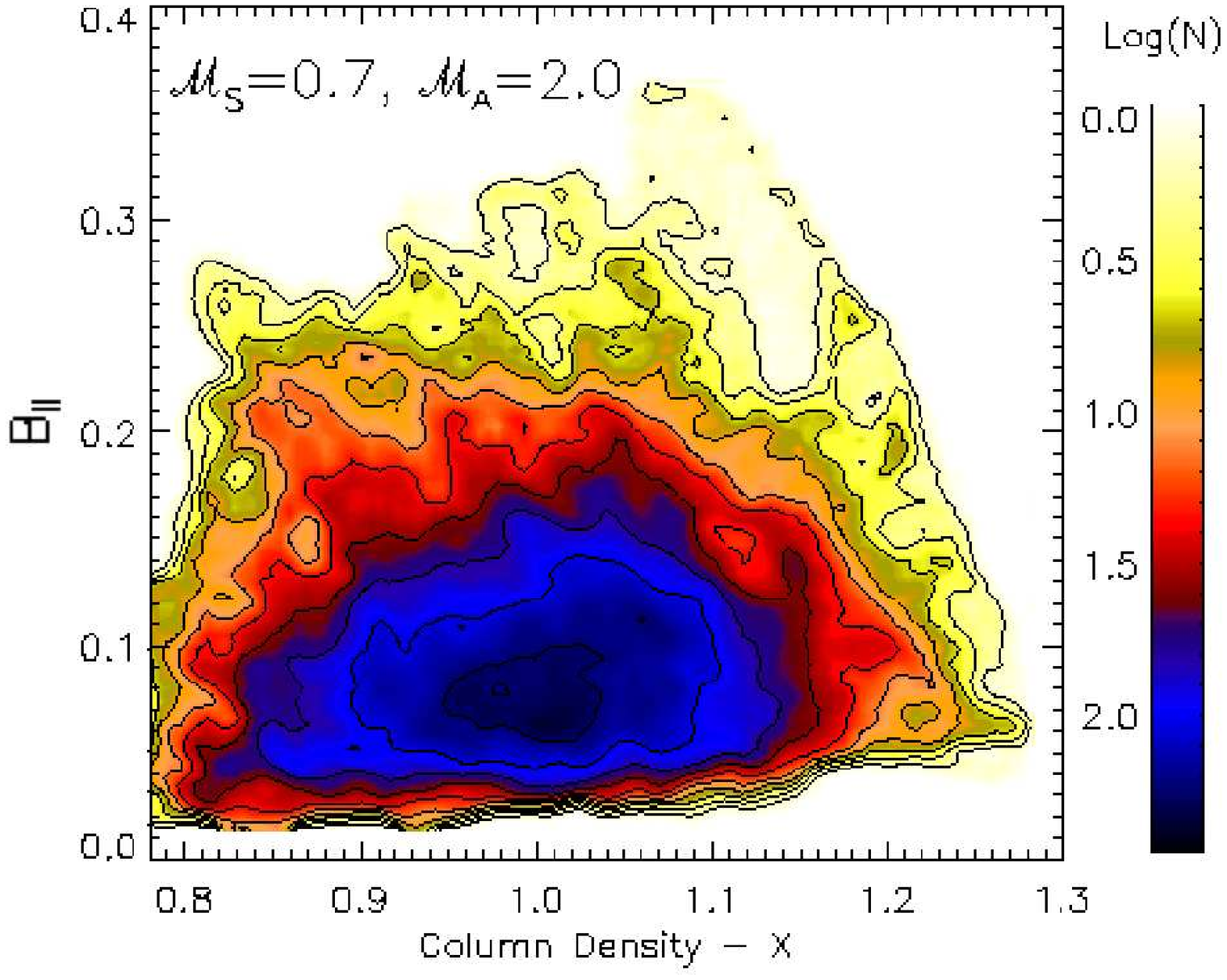} % \includegraphics[scale=.35]{apj/c512b.1p1/paper/corr_coldens_magpar_x}
\includegraphics[scale=.3]{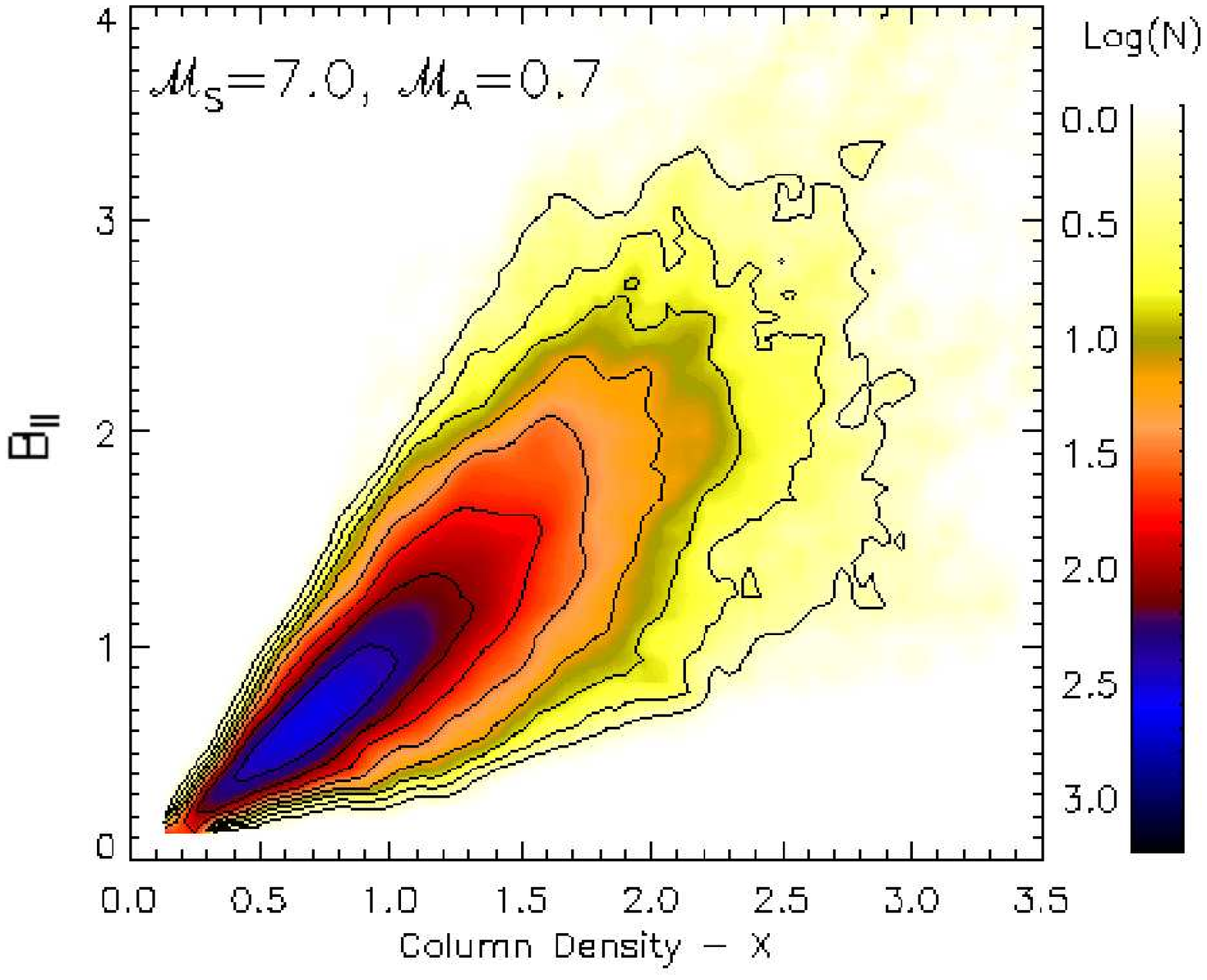} % \includegraphics[scale=.35]{apj/c512b1p.01/paper/corr_coldens_magpar_x}
\includegraphics[scale=.3]{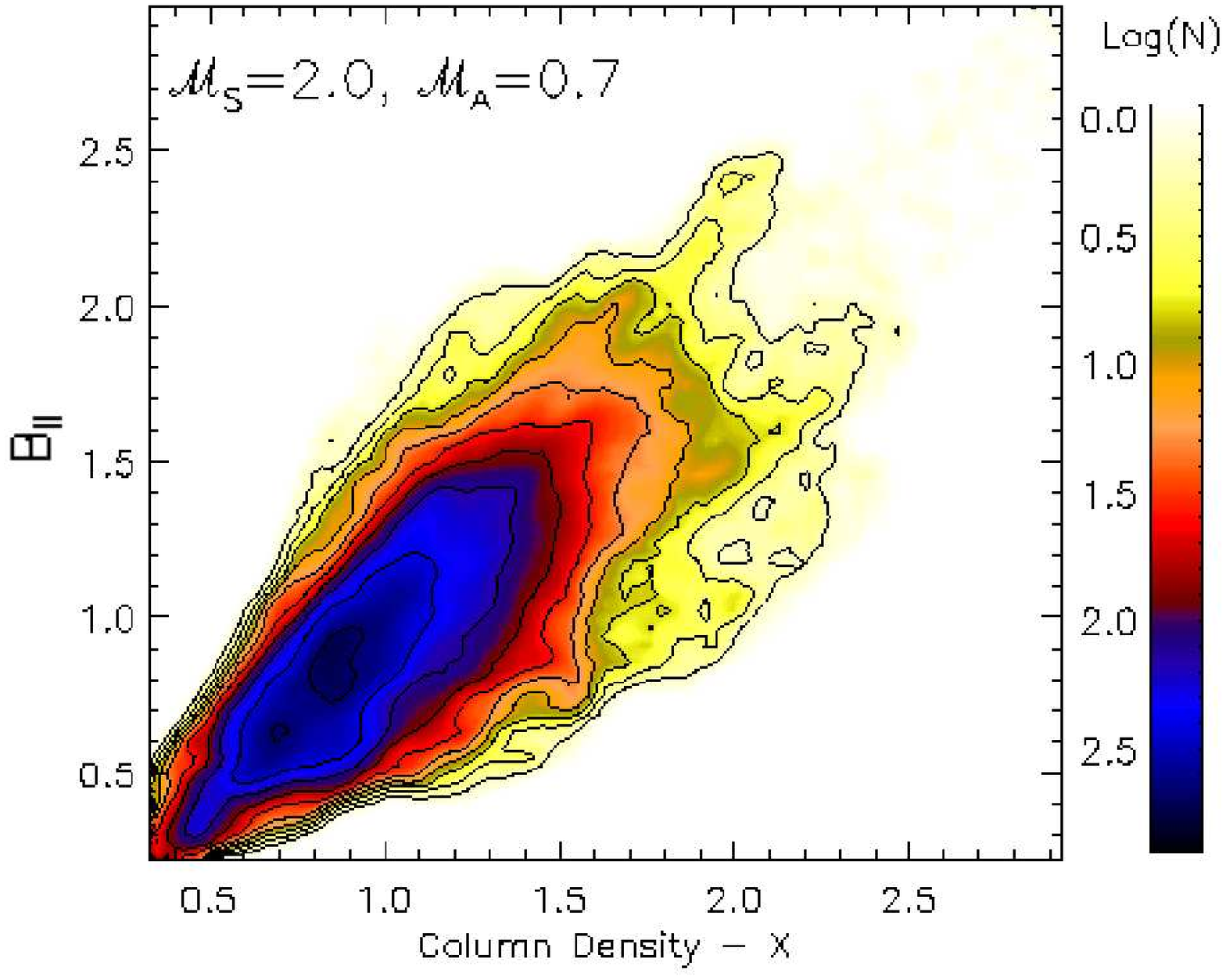} % \includegraphics[scale=.35]{apj/c512b1p.1/paper/corr_coldens_magpar_x}
\includegraphics[scale=.3]{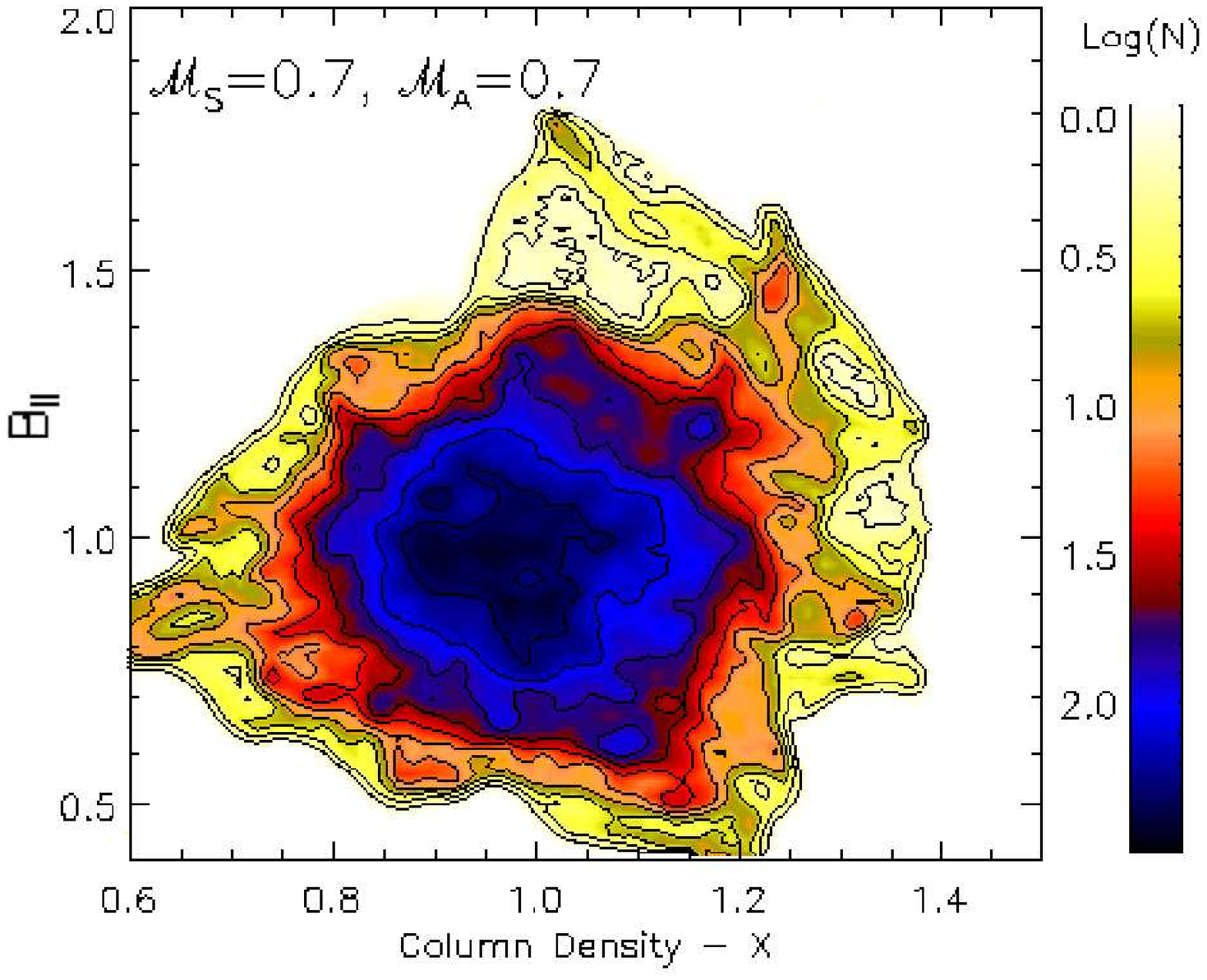} % \includegraphics[scale=.35]{apj/c512b1p1/paper/corr_coldens_magpar_x}
\caption{The 2D correlation of the integrated magnetic field component parallel to the line of sight vs. column density in the x direction. The first row consists of super-Alfv\'enic models while the bottom row is sub-Alfv\'enic. Images are ordered left to right as supersonic to subsonic. Blue contours indicate regions of high data counts while red and yellow have lower data counts. \label{fig:cden_bpar_x}}
\end{figure*}

In Figure~\ref{fig:cden_bpar_x} we show the correlation of column density along the 
x-direction and the magnetic field component parallel to the LOS, i.e. parallel to
$B_{\rm ext}$.   This correlation may be compared, for instance, directly to
Zeeman splitting measurements. Similarly to the 3D correlation of Figure~\ref{fig:dens_emag}, column density and magnetic field along the LOS increase together for supersonic models due to density clumps trapping the magnetic field. As expected, sub-Alfv\'{e}nic models present steeper correlations due to higher mean magnetic field intensity. Correlations are near linear for column density values around the mean density for supersonic sub-Alfv\'enic models.   This is due to the stronger field becoming entangled with high density clumps. In super Alfv\'enic cases, the magnetic field is not as strong and hence  density clumps will increase, yet the magnetic energy avaliable for clumps to trap will level off.  Subsonic models showed no correlation due to incompressible densities.

\begin{figure*}
\centering
\includegraphics[scale=.3]{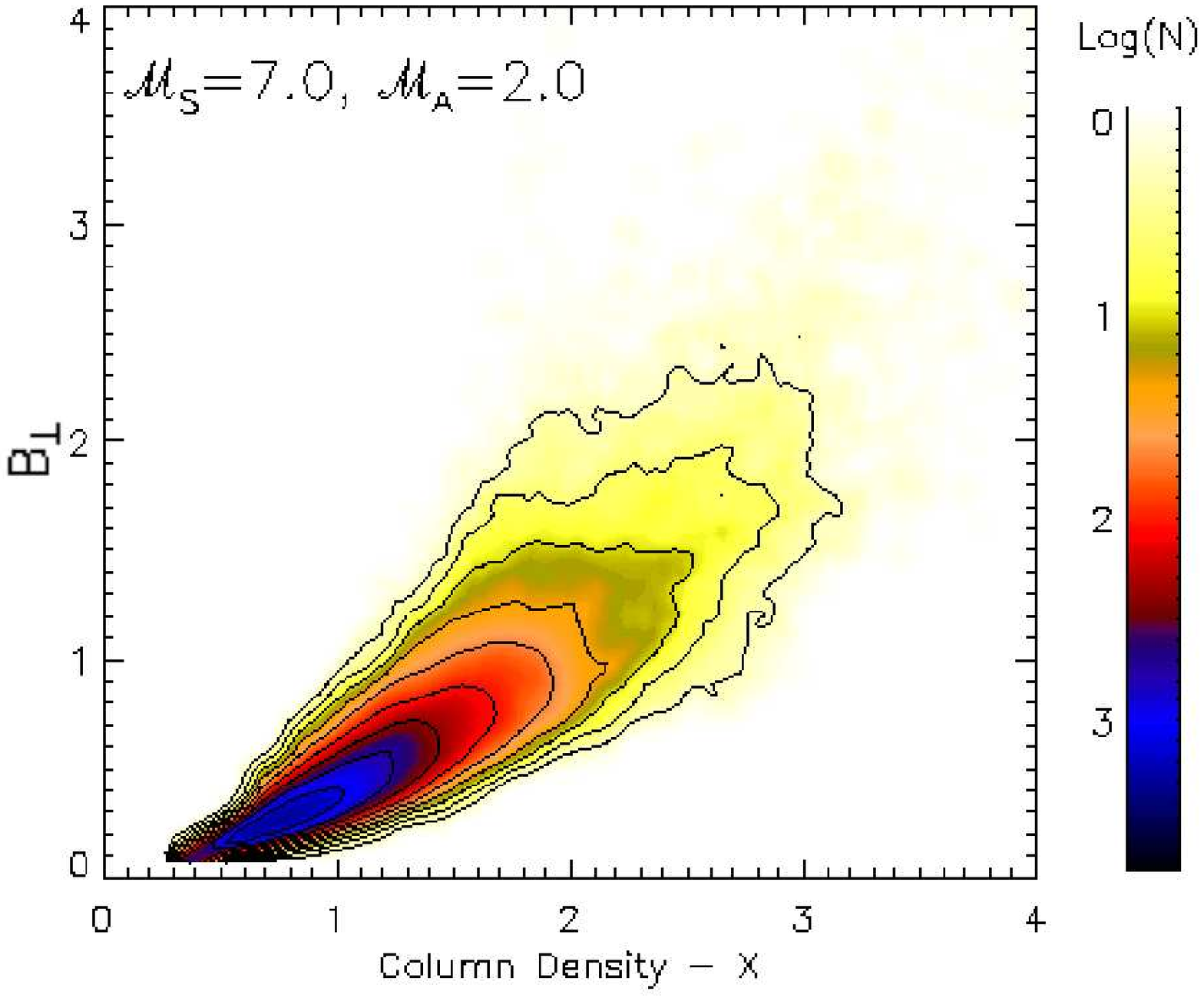} % \includegraphics[scale=.35]{apj/c512b.1p.01/paper/corr_coldens_magper_x}
\includegraphics[scale=.3]{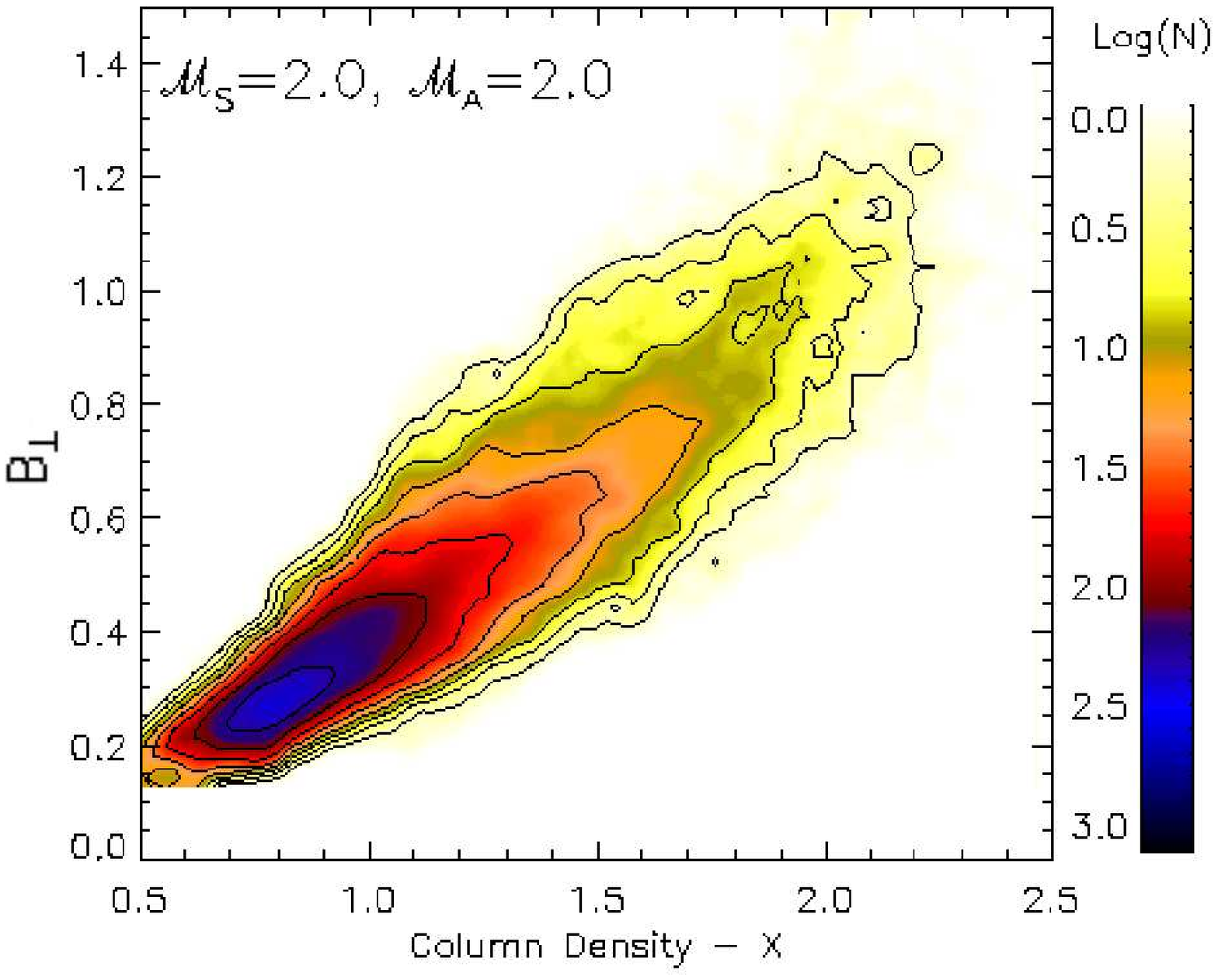} % \includegraphics[scale=.35]{apj/c512b.1p.1/paper/corr_coldens_magper_x}
\includegraphics[scale=.3]{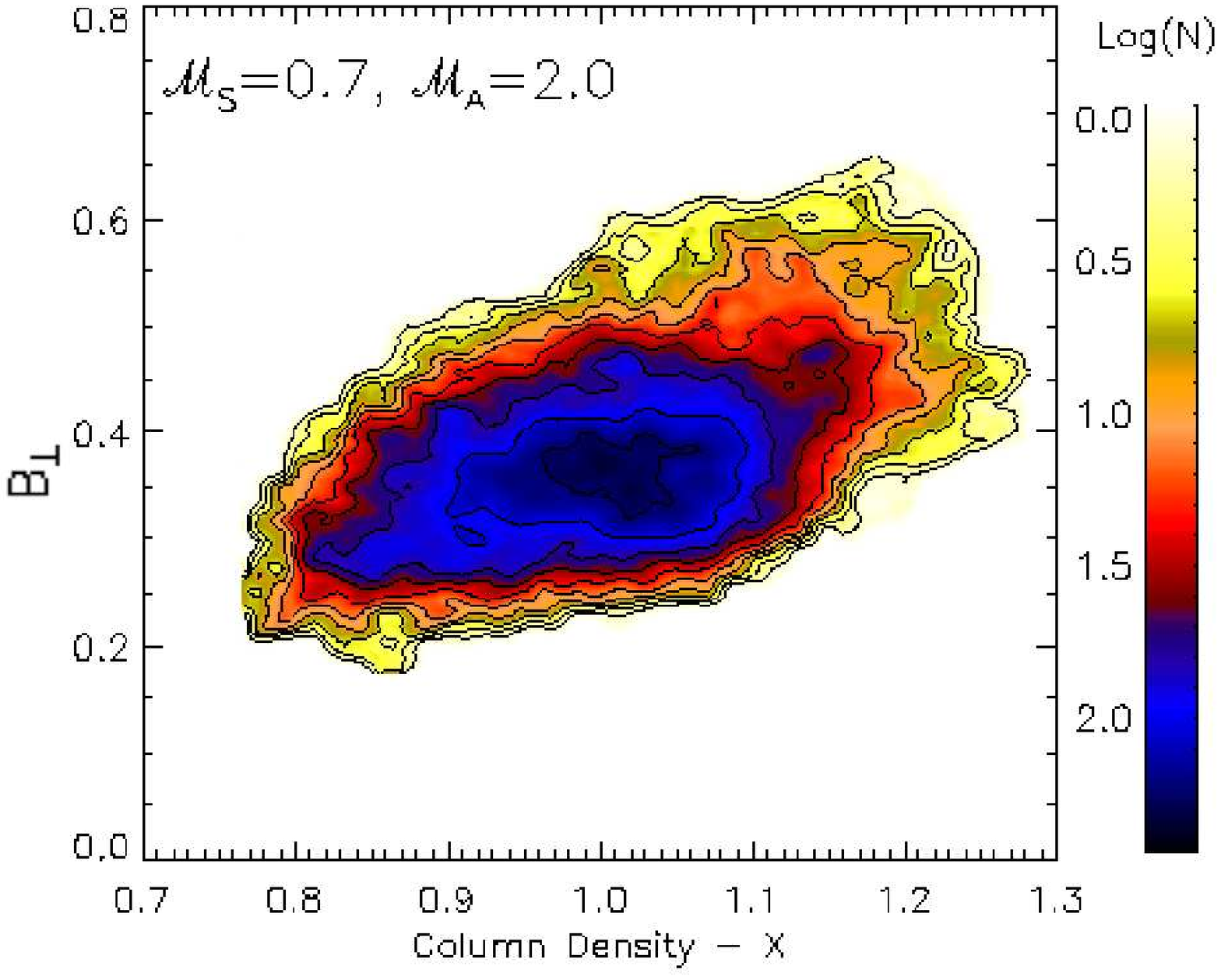} % \includegraphics[scale=.35]{apj/c512b.1p1/paper/corr_coldens_magper_x}
\includegraphics[scale=.3]{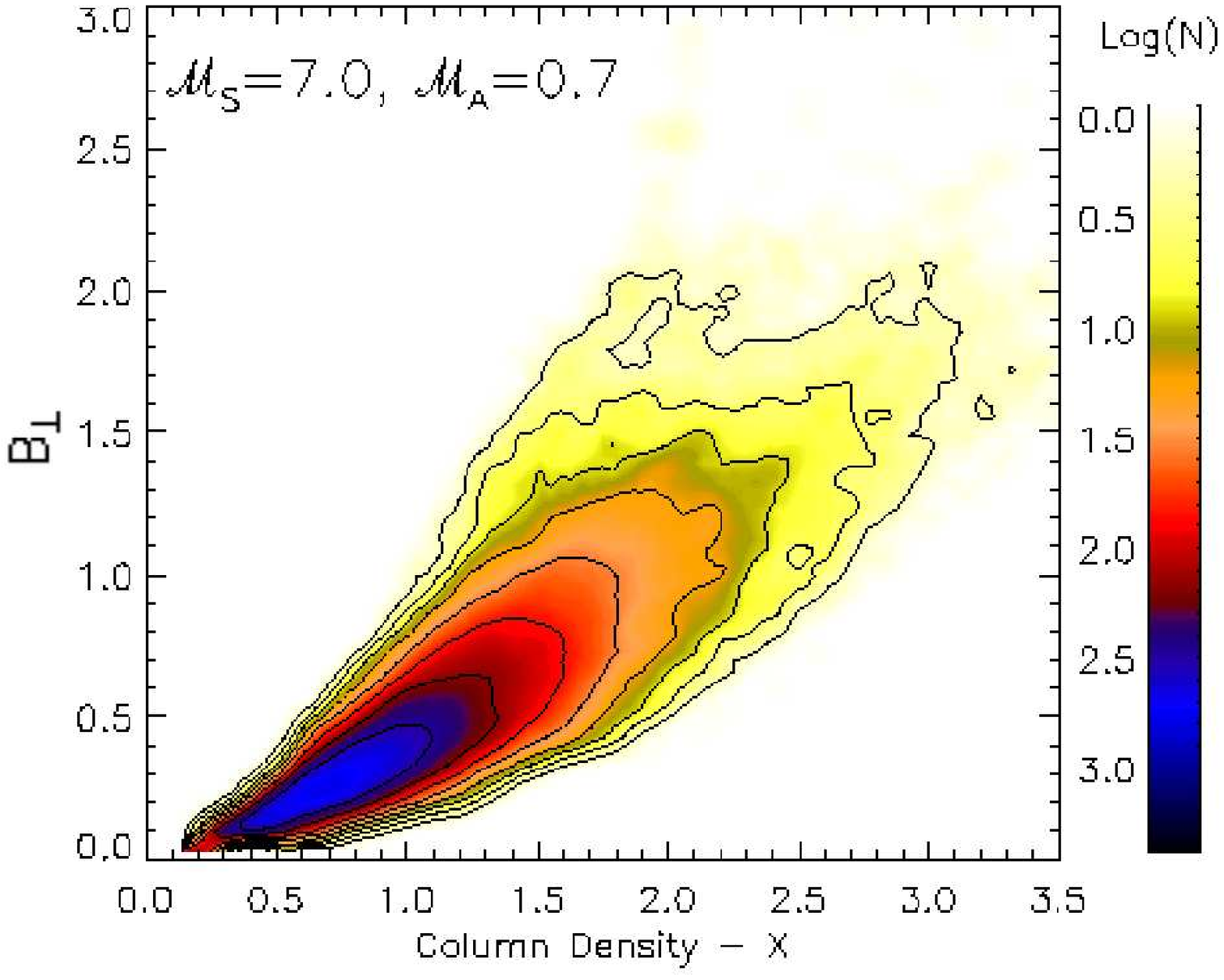} % \includegraphics[scale=.35]{apj/c512b1p.01/paper/corr_coldens_magper_x}
\includegraphics[scale=.3]{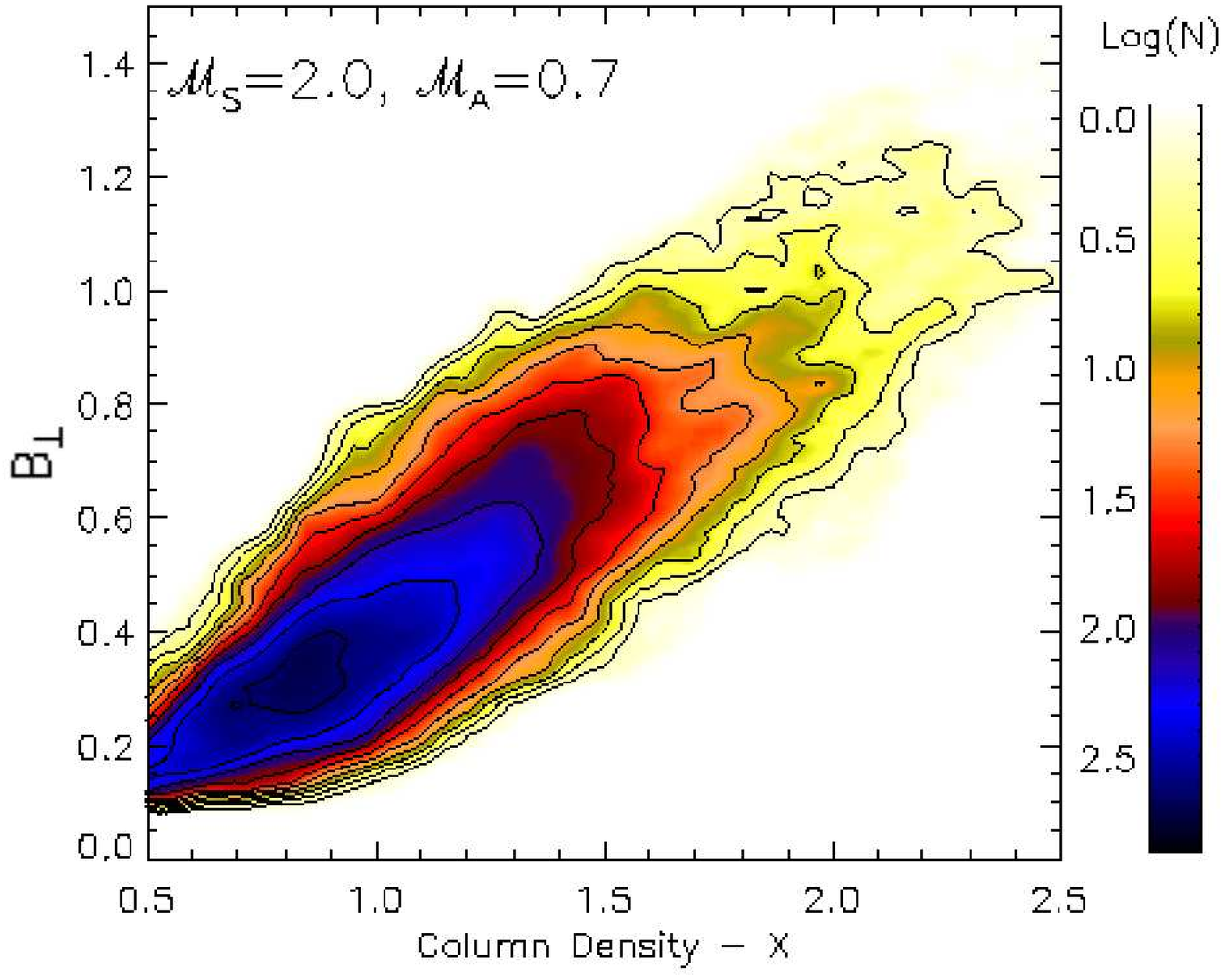} % \includegraphics[scale=.35]{apj/c512b1p.1/paper/corr_coldens_magper_x}
\includegraphics[scale=.3]{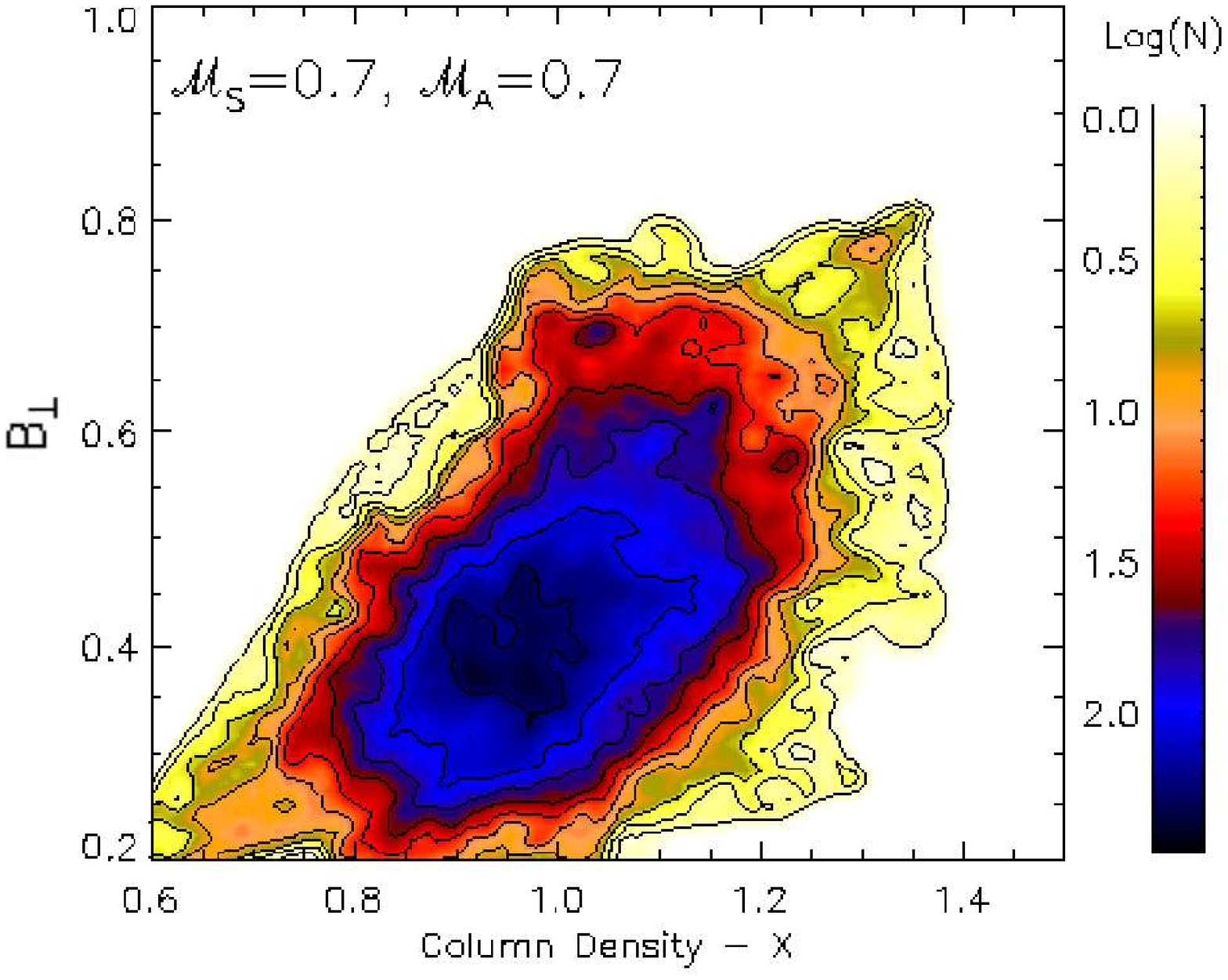} % \includegraphics[scale=.35]{apj/c512b1p1/paper/corr_coldens_magper_x}
\caption{The 2D correlation of the integrated magnetic field component perpendicular to the line of sight vs. column density in the x direction. The first row consists of  super-Alfv\'enic models while the bottom row is sub-Alfv\'enic. Images are ordered left to right as supersonic to subsonic. Blue contours indicate regions of high data count while red and yellow have lower counts. \label{fig:cden_bper_x}}
\end{figure*}

A behavior similar to Figure~\ref{fig:cden_bpar_x} was obtained for the correlation of column density along
x-direction and the magnetic field component perpendicular to the line of sight\footnote{The magnetic field $\perp$ to the line of sight can be estimated with the Chandrasechar-Fermi technique.}, i.e. LOS $\perp B_{\rm ext}$, shown in Figure~\ref{fig:cden_bper_x}.  The subsonic models  show no correlation between column density and magnetic field. Supersonic
cases are able to reach higher field strength for higher densities due to the
entanglement of the field lines with density clumps, which is similar to both the 3D correlation of magentic energy vs. $\rho$ in  Figure~\ref{fig:dens_emag} and the 2D correlation in Figure ~\ref{fig:cden_bpar_x}. The difference between Figures~\ref{fig:cden_bpar_x} and~\ref{fig:cden_bper_x} is the orientation of the field with respect to the plane of projected column density. When the field is perpendicular to this plane we see a higher degree of linearity  then in the parallel case.   This is because density clumps have more freedom when the compressions are perpendicular to the field lines and thus clumps are able to better trap magentic energy. When the field is parallel to the plane, plasmas tend to flow along field lines and cannot trap them. Correlations are stronger for even subsonic cases when the column density is perpendicular to the field and the correlation coefficients of Table~\ref{tab:coefficients2}  confirm this.   We must keep in mind that while the field is oriented parallel and perpendicular to the column density integrated along the x direction for Figures \ref{fig:cden_bpar_x} and~\ref{fig:cden_bper_x} respectively, the mean volume magnetic field is parallel to the x direction in our simulations.

\begin{figure*}
\centering
\includegraphics[scale=.3]{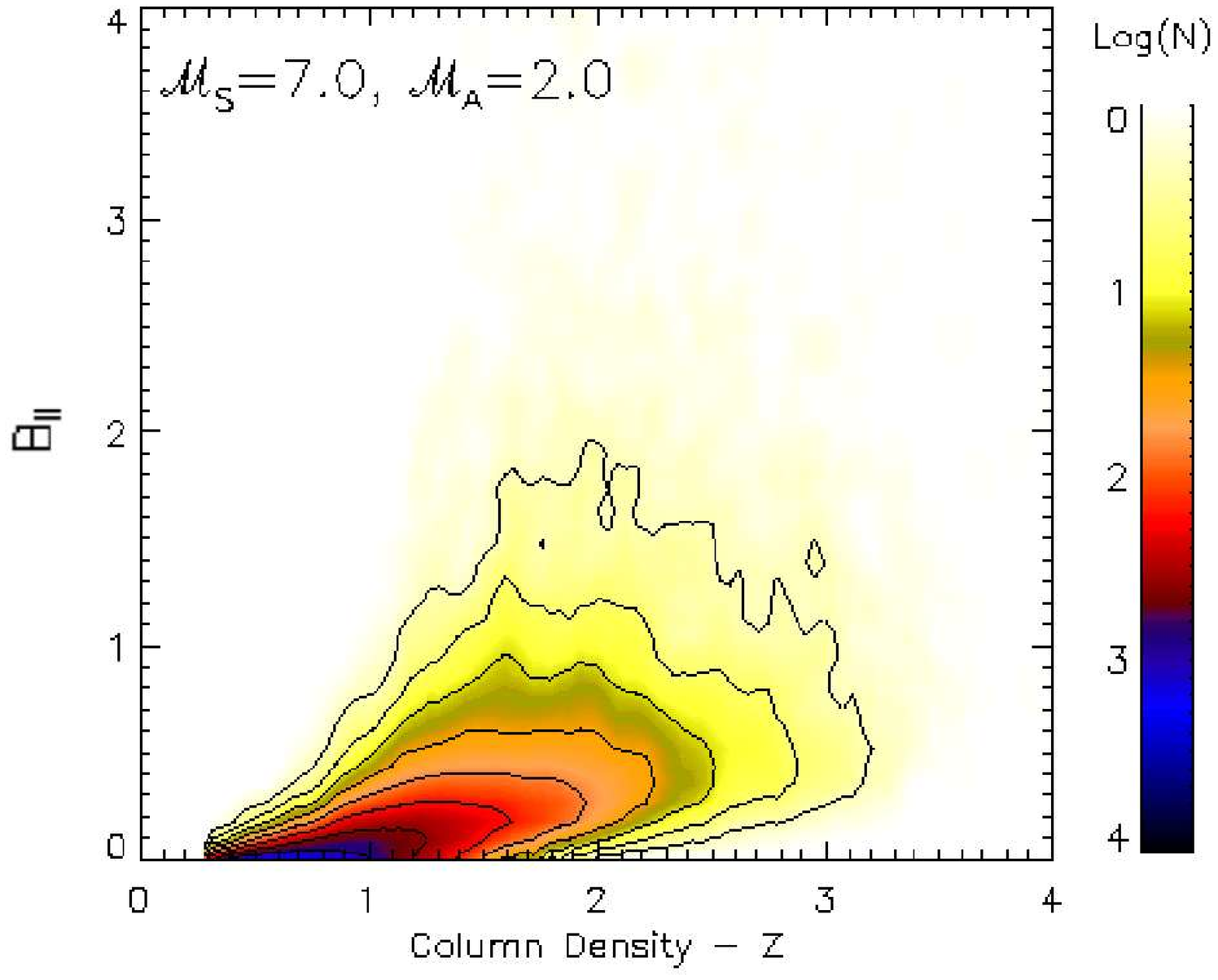} % \includegraphics[scale=.35]{apj/c512b.1p.01/paper/corr_coldens_magpar_z}
\includegraphics[scale=.3]{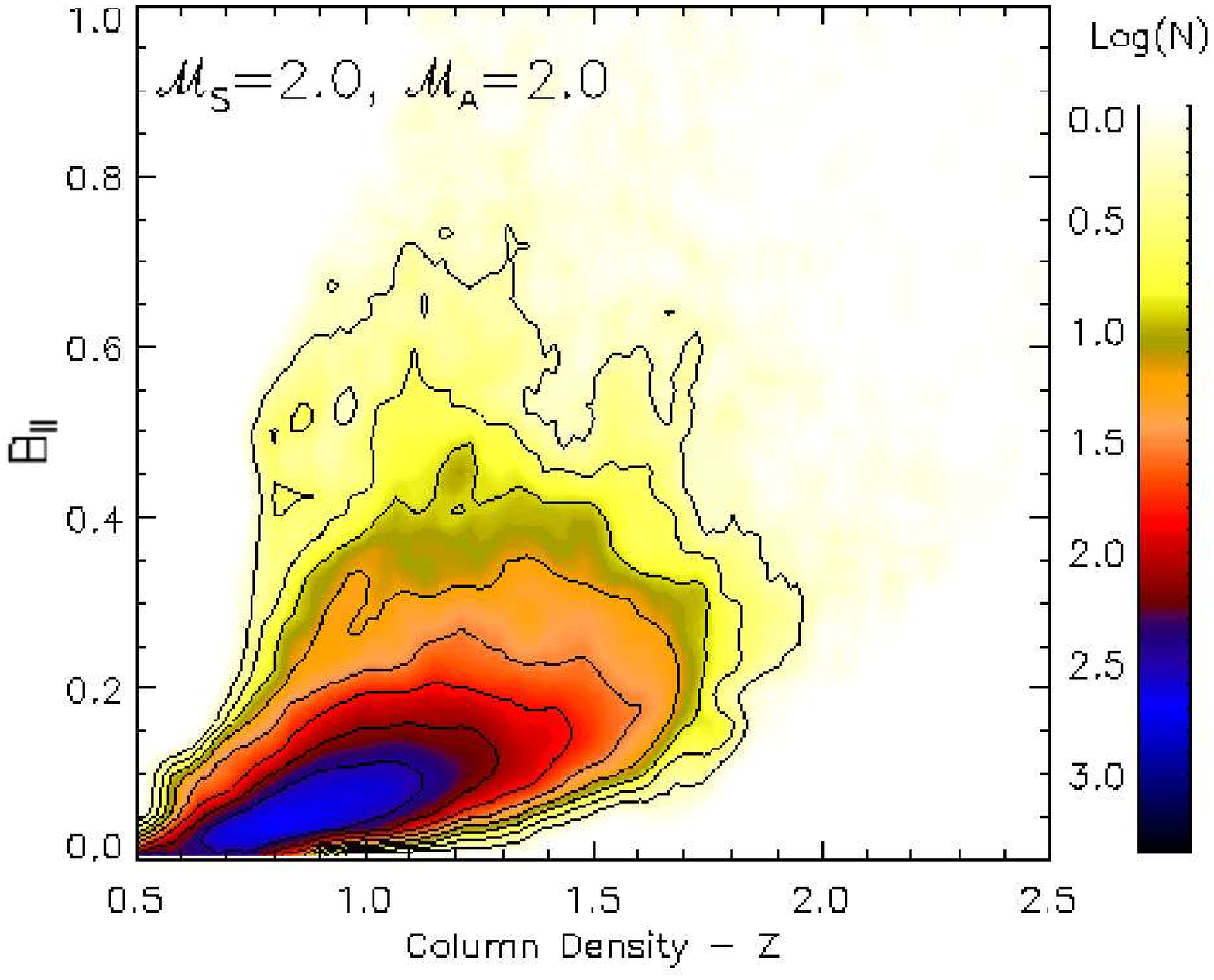} % \includegraphics[scale=.35]{apj/c512b.1p.1/paper/corr_coldens_magpar_z}
\includegraphics[scale=.3]{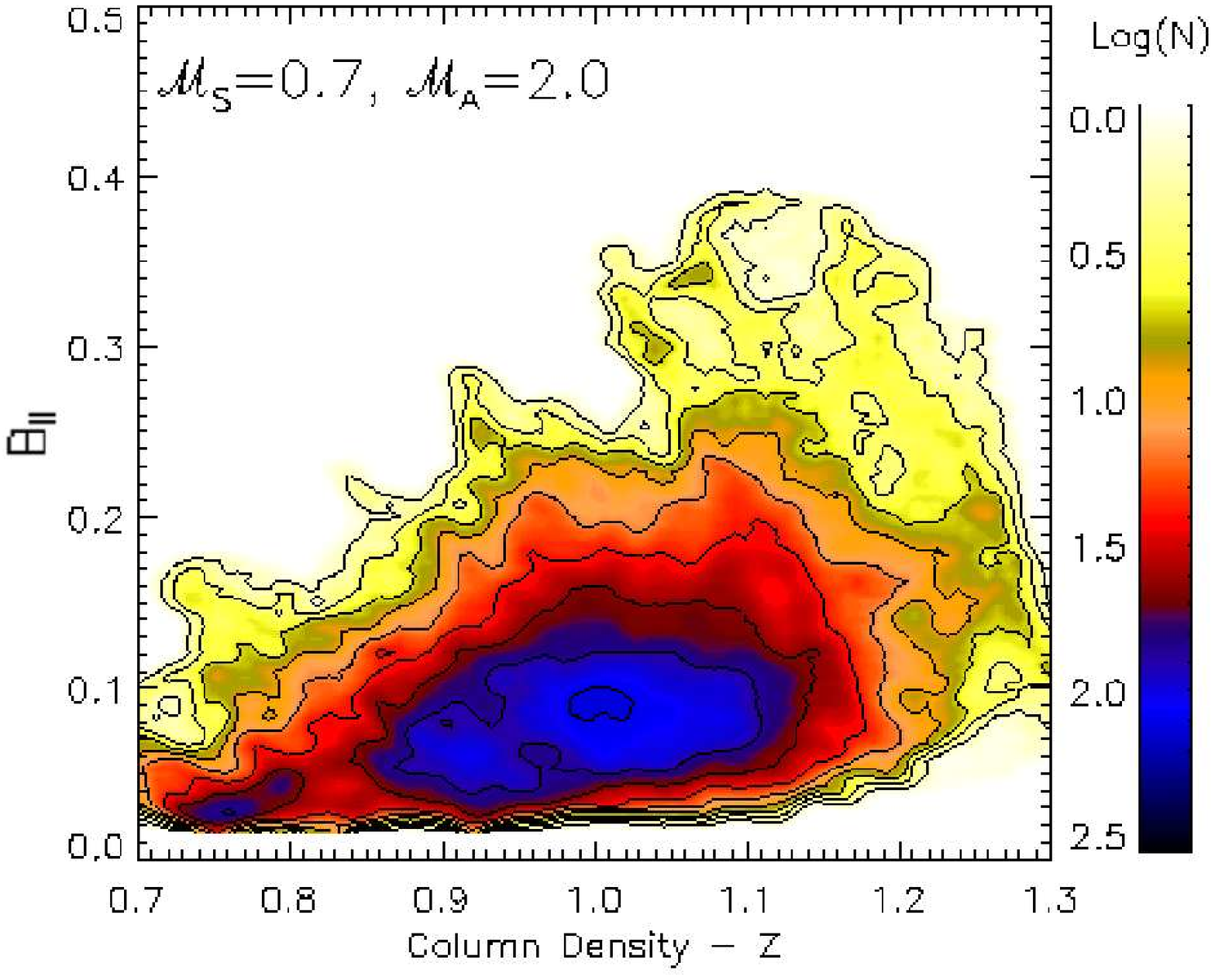} % \includegraphics[scale=.35]{apj/c512b.1p1/paper/corr_coldens_magpar_z}
\includegraphics[scale=.3]{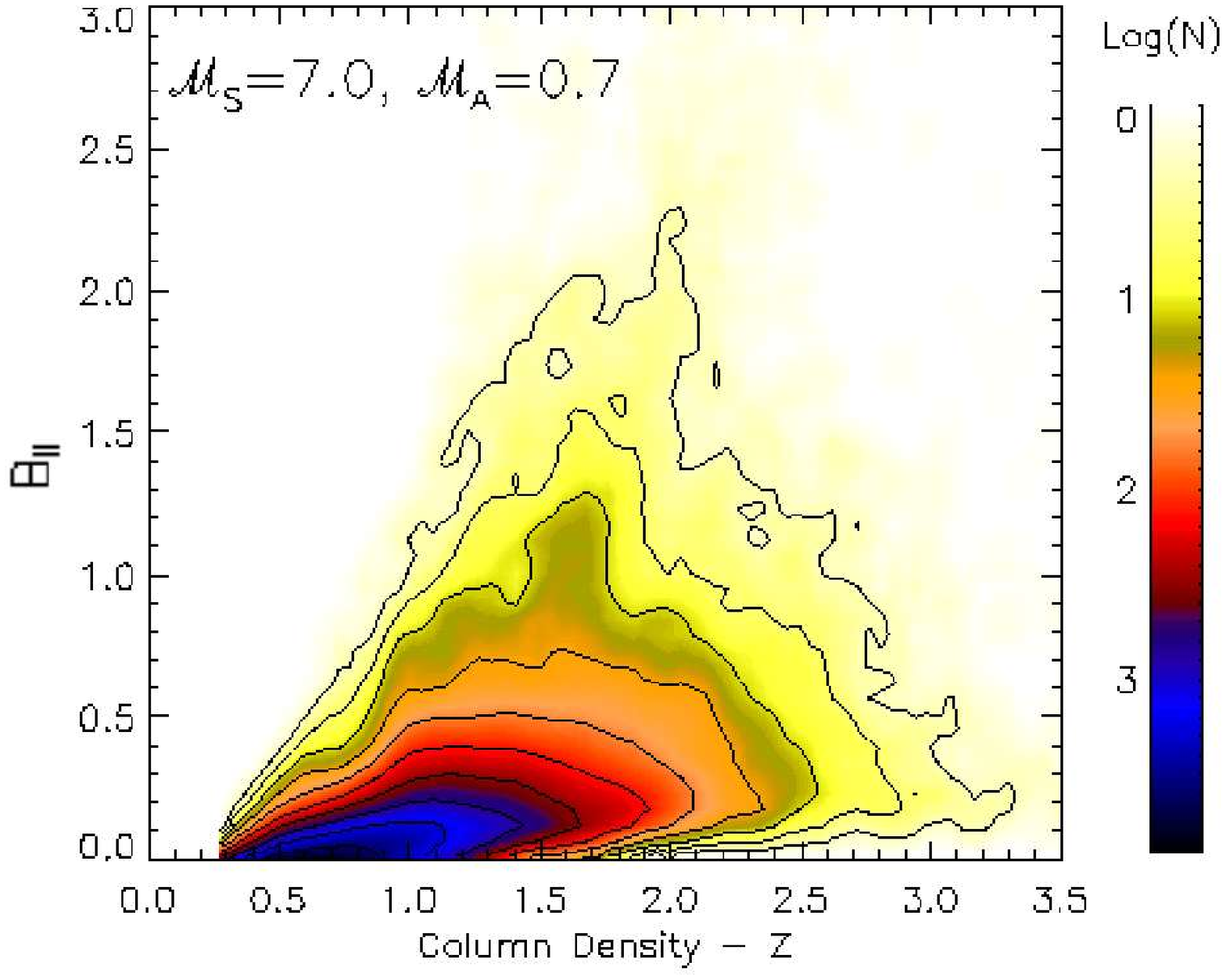} % \includegraphics[scale=.35]{apj/c512b1p.01/paper/corr_coldens_magpar_z}
\includegraphics[scale=.3]{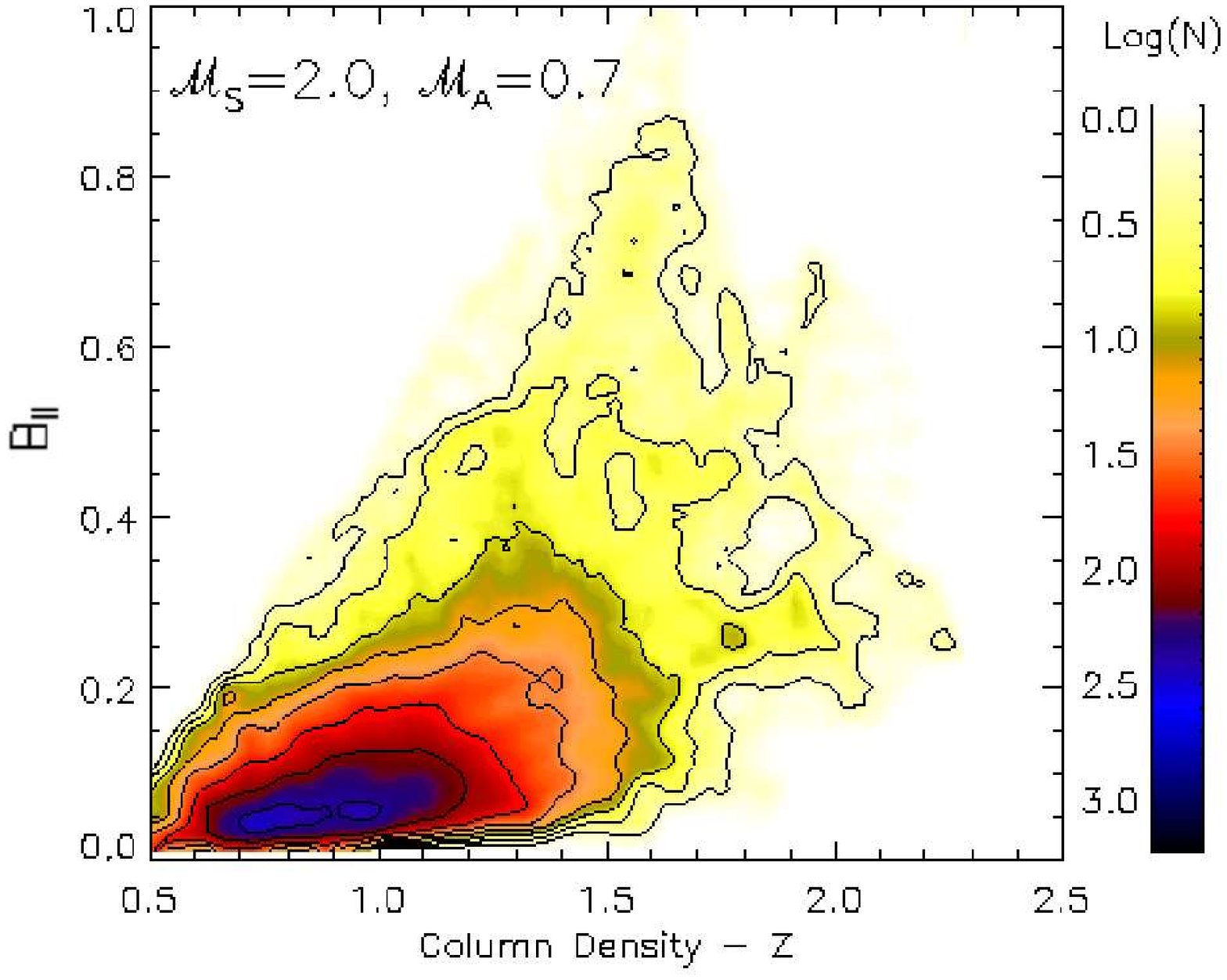} % \includegraphics[scale=.35]{apj/c512b1p.1/paper/corr_coldens_magpar_z}
\includegraphics[scale=.3]{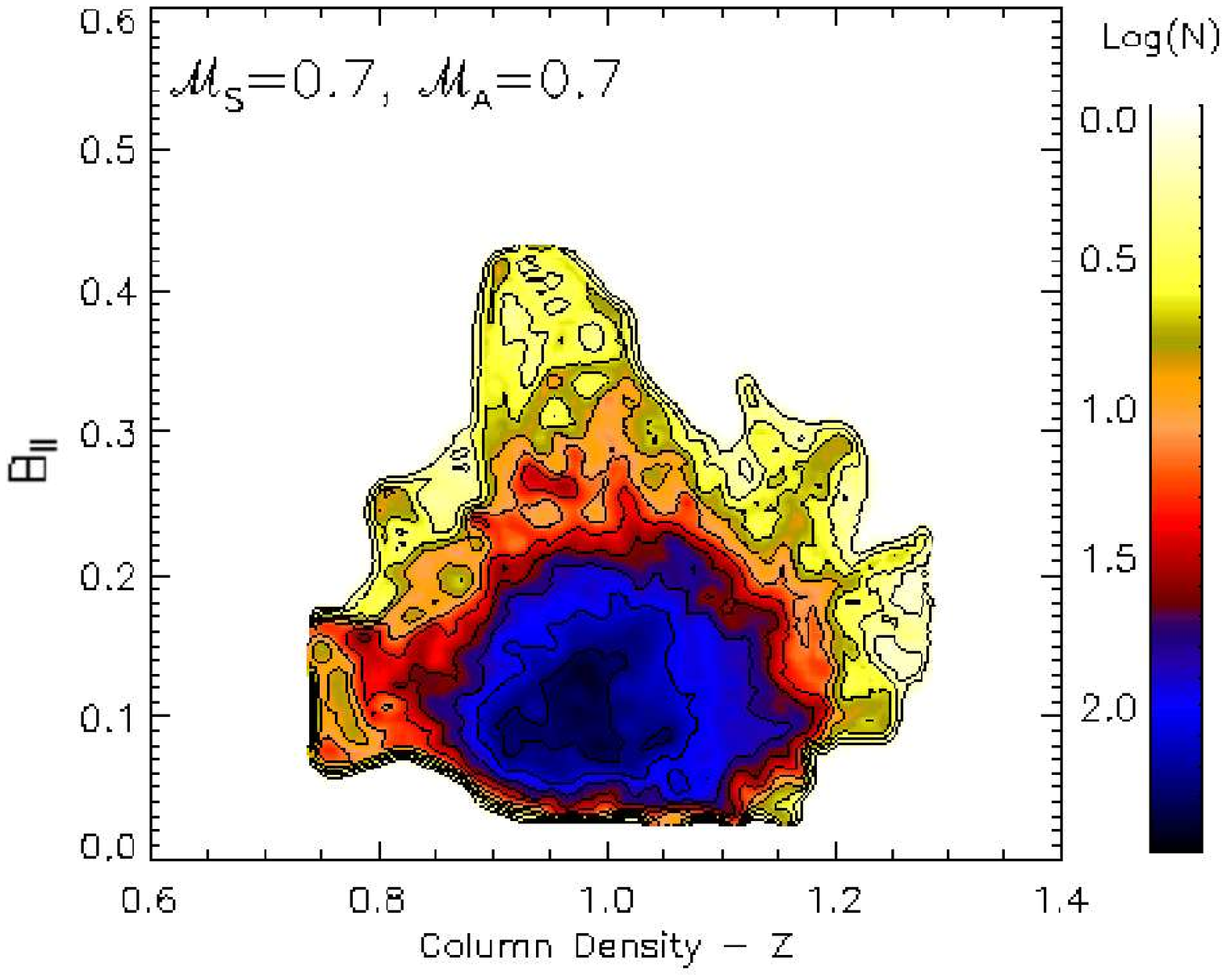} % \includegraphics[scale=.35]{apj/c512b1p1/paper/corr_coldens_magpar_z}
\caption{The 2D correlation of the integrated magnetic field component parallel to the line of sight vs. column density in the z direction.  The first row consists of super-Alfv\'enic models while the bottom row is sub-Alfv\'enic. Images are ordered left to right as supersonic to subsonic. Blue contours indicate regions of high data counts while red and yellow have lower counts. \label{fig:cden_bpar_z}}
\end{figure*}

\begin{figure*}
\centering
\includegraphics[scale=.3]{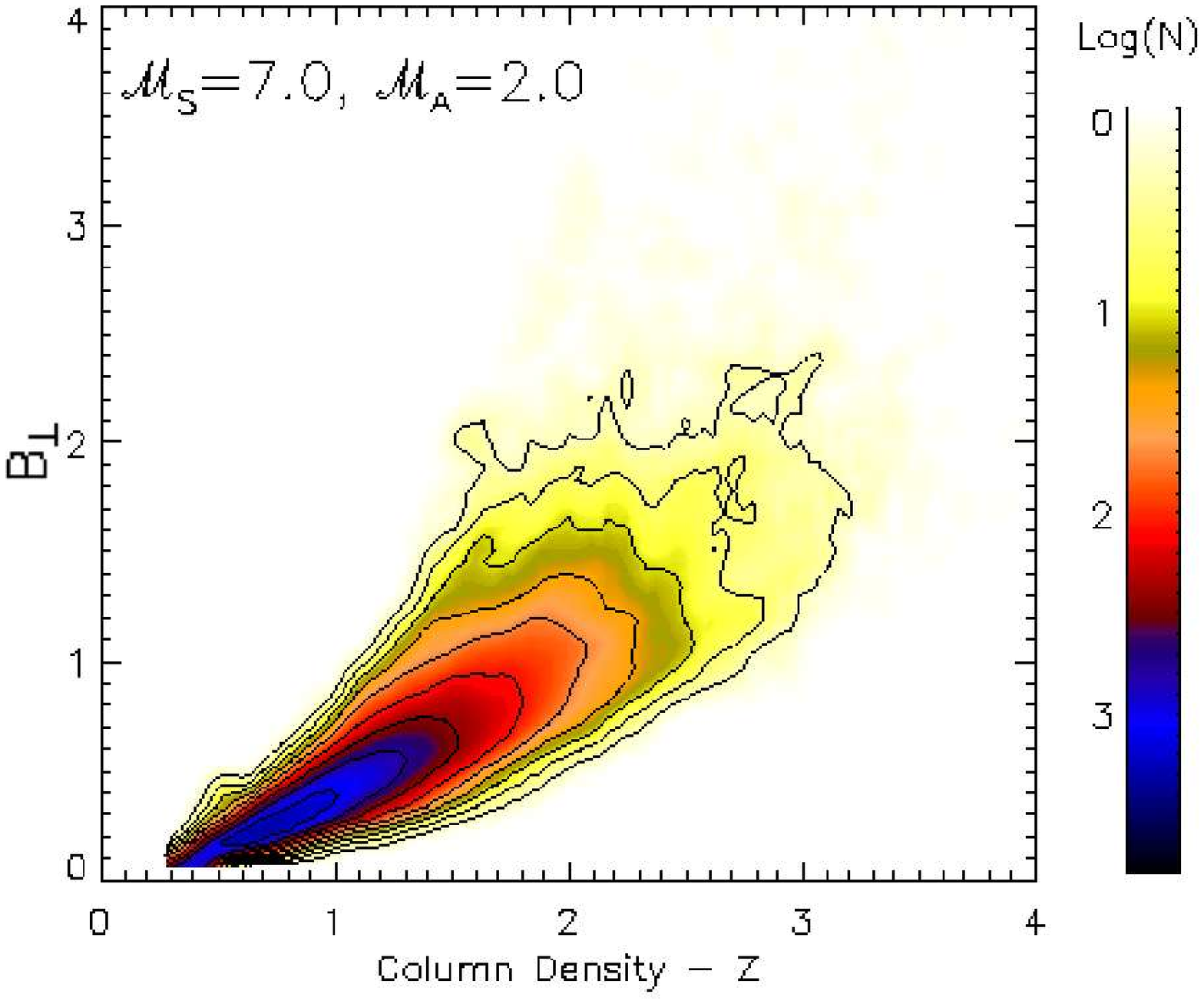} % \includegraphics[scale=.35]{apj/c512b.1p.01/paper/corr_coldens_magper_z}
\includegraphics[scale=.3]{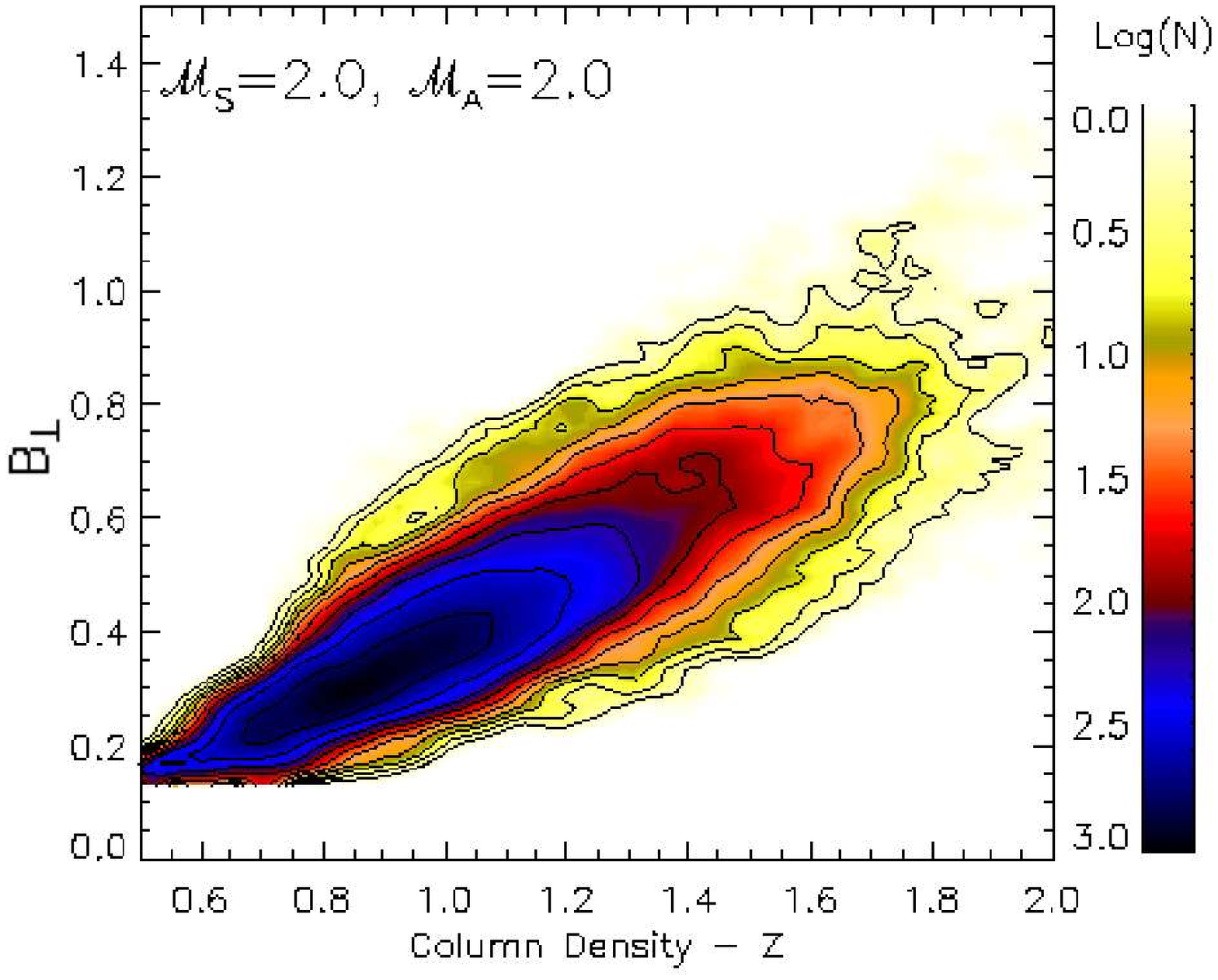} % \includegraphics[scale=.35]{apj/c512b.1p.1/paper/corr_coldens_magper_z}
\includegraphics[scale=.3]{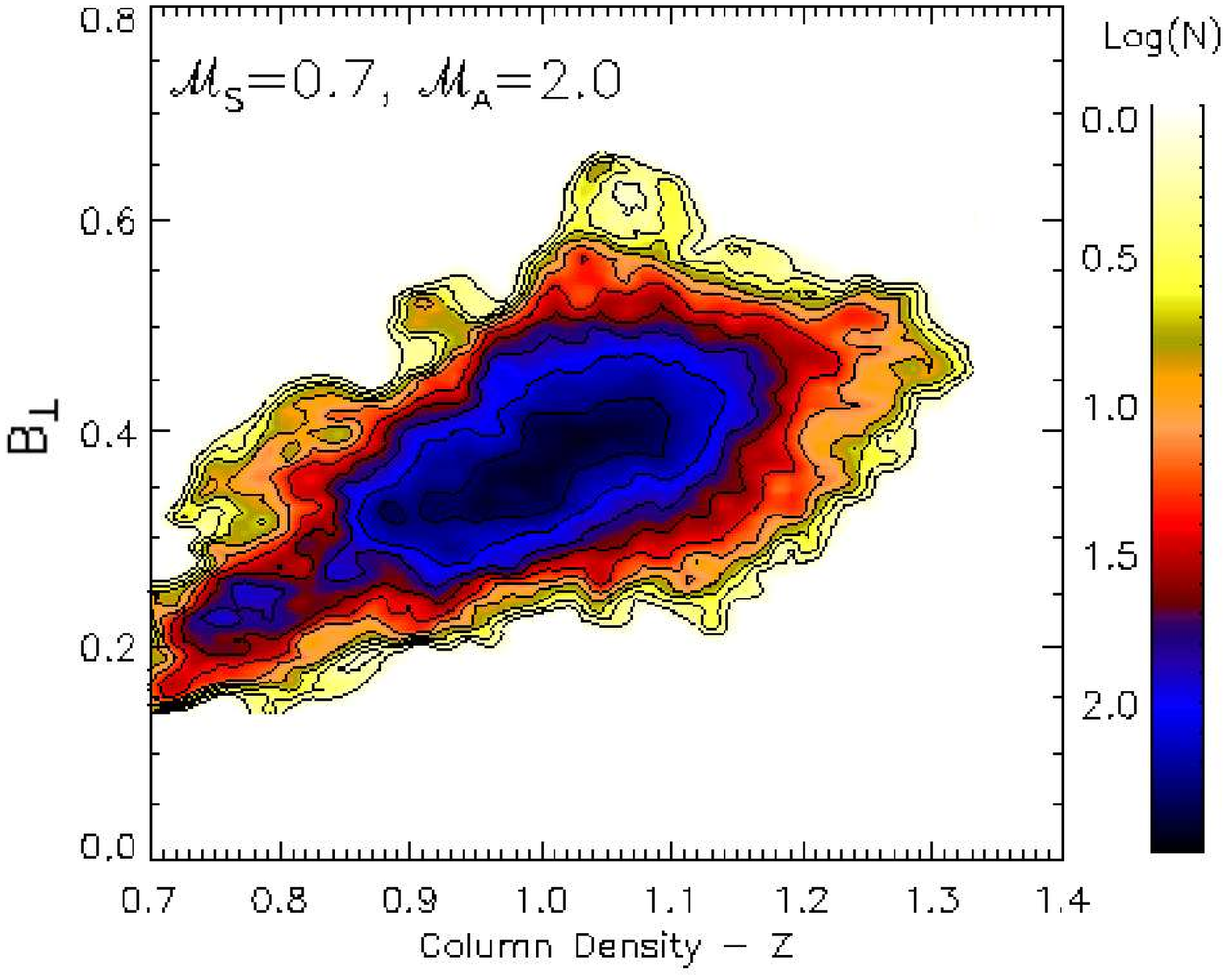} % \includegraphics[scale=.35]{apj/c512b.1p1/paper/corr_coldens_magper_z}
\includegraphics[scale=.3]{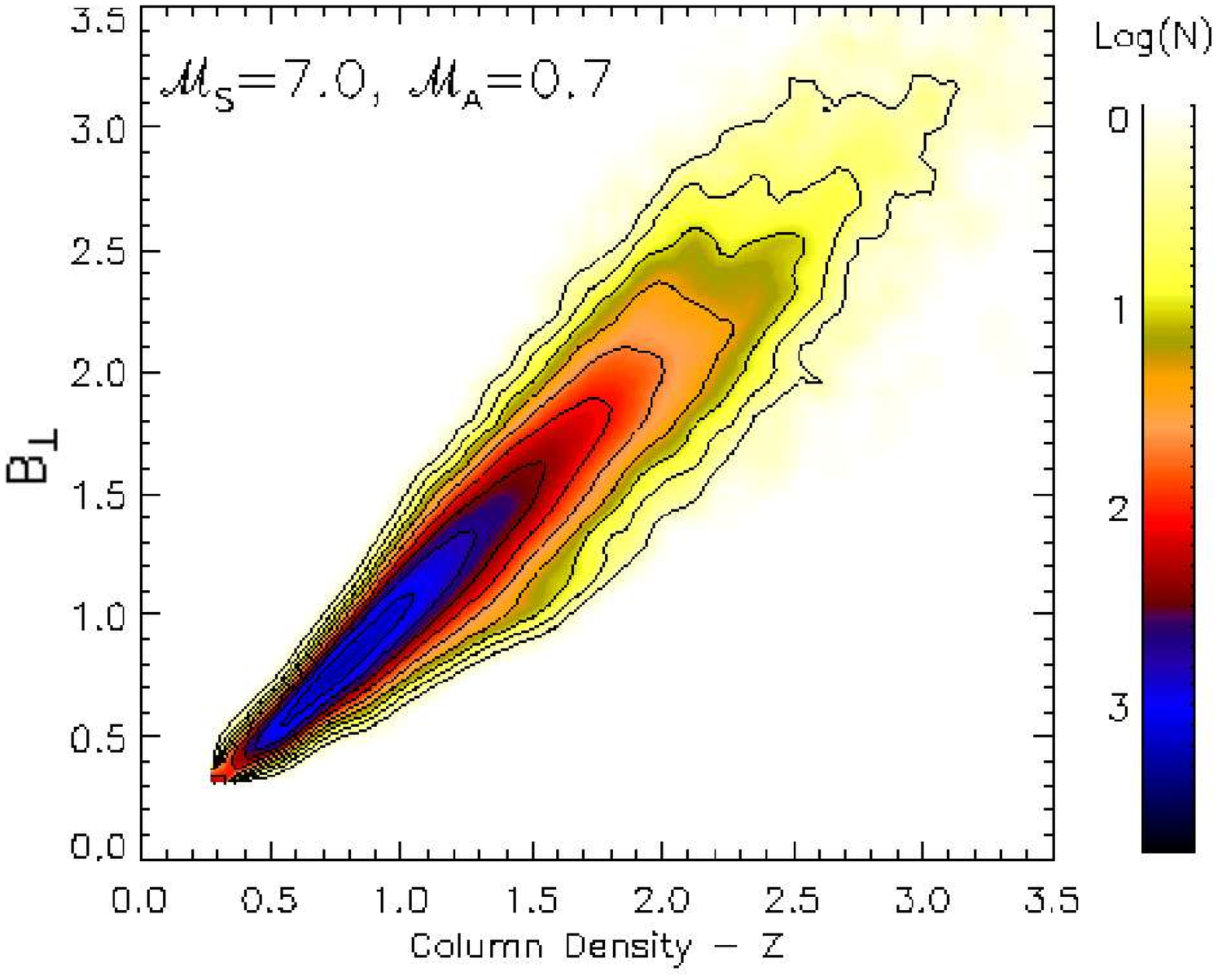} % \includegraphics[scale=.35]{apj/c512b1p.01/paper/corr_coldens_magper_z}
\includegraphics[scale=.24]{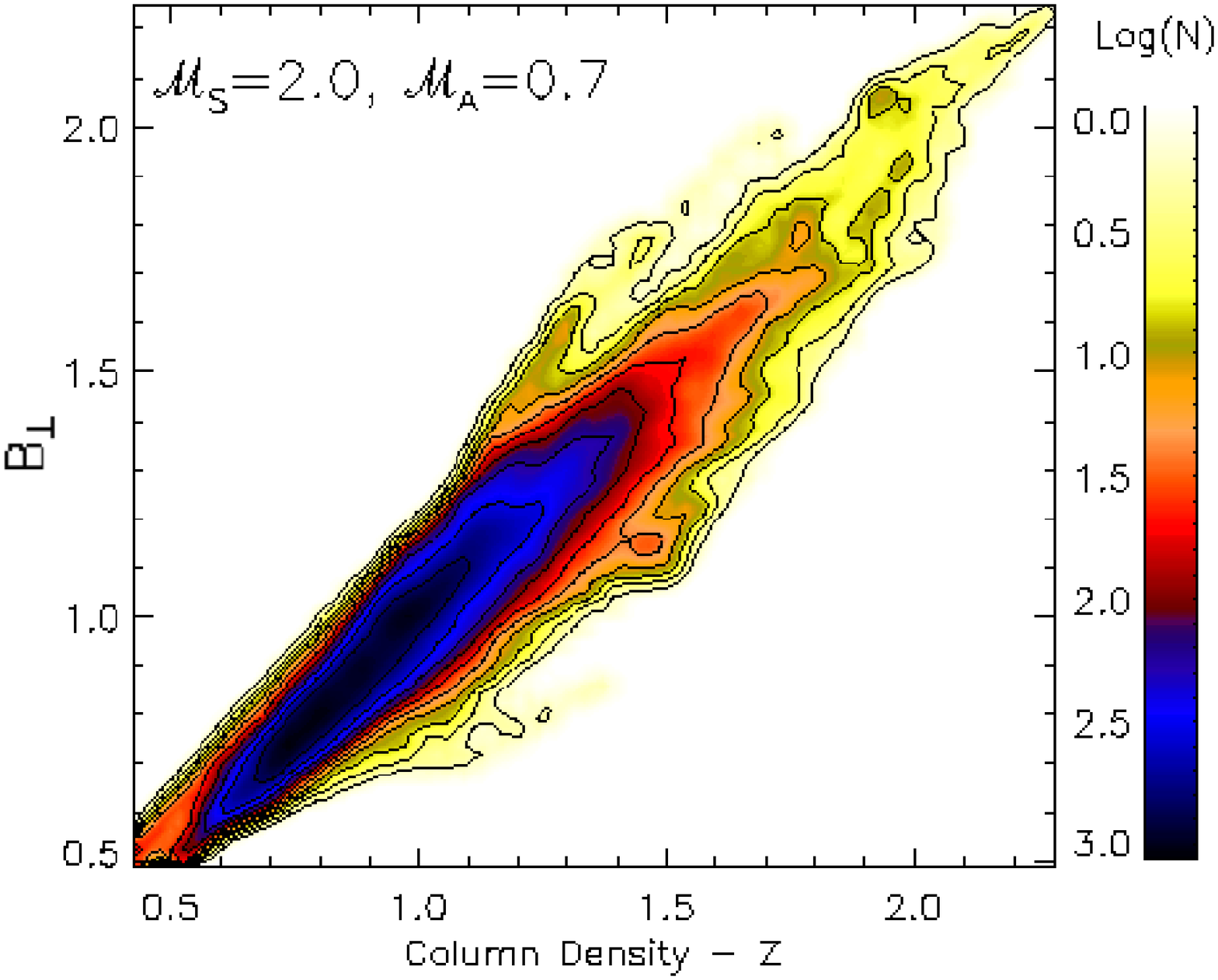} % \includegraphics[scale=.35]{apj/c512b1p.1/paper/corr_coldens_magper_z}
\includegraphics[scale=.3]{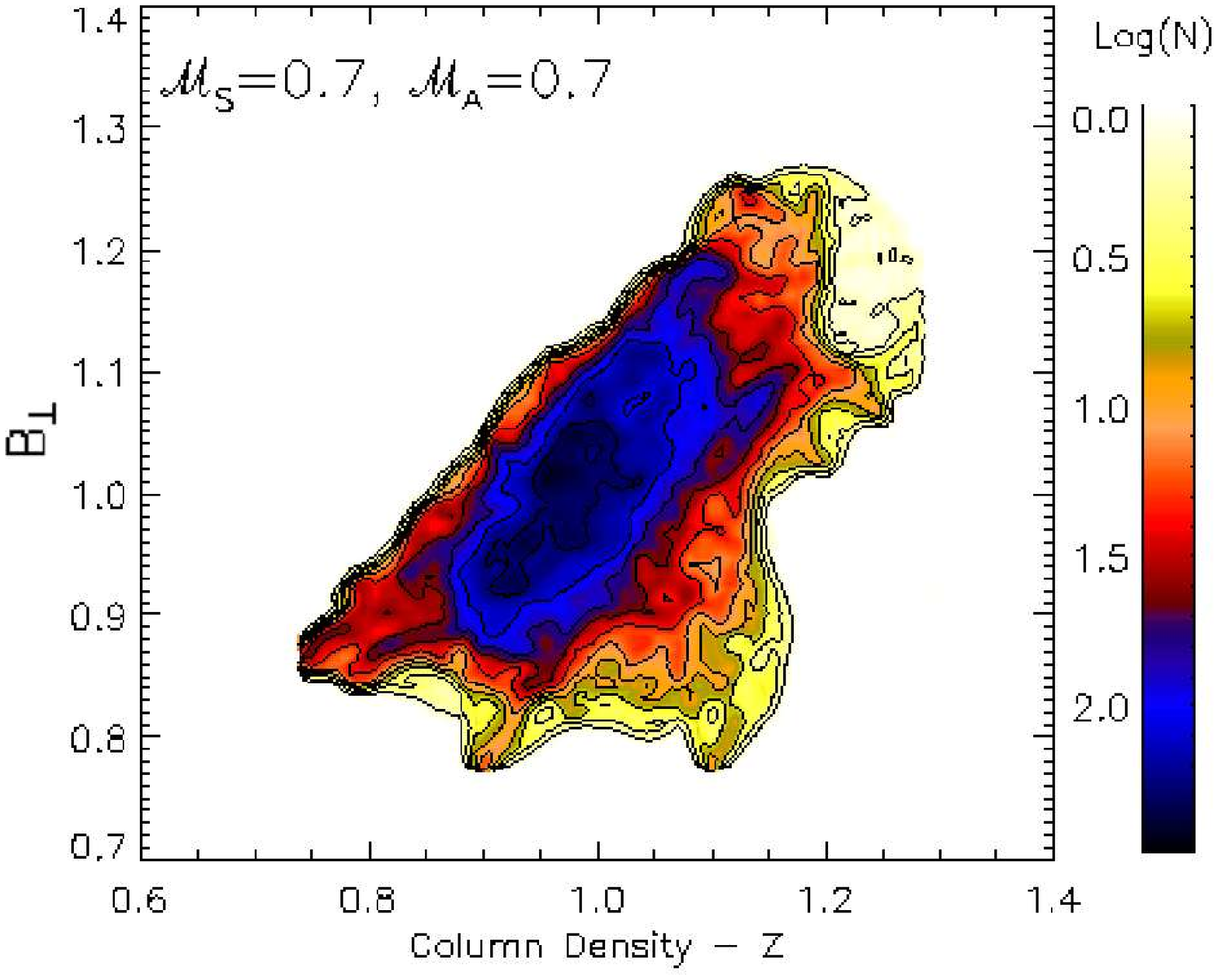} % \includegraphics[scale=.35]{apj/c512b1p1/paper/corr_coldens_magper_z}
\caption{The 2D correlation of the integrated magnetic field component perpendicular to the line of sight vs. column density in the z direction. The first row consists of super-Alfv\'enic models while the bottom row is sub-Alfv\'enic. Images are ordered left to right as supersonic to subsonic. Blue contours indicate regions of high data counts while red and yellow have lower counts.\label{fig:cden_bper_z}}
\end{figure*}

Figure~\ref{fig:cden_bpar_z} shows the correlation of column density along
z-direction and the magnetic field component parallel to the line of sight.
Figure~\ref{fig:cden_bper_z} shows the correlation of column density along
z-direction and the magnetic field component perpendicular to the line of sight.
 Both correlations are obtained along the z direction, i.e. perpendicular to $B_0$. Figure~\ref{fig:cden_bpar_z} is similar to
 Figure~\ref{fig:cden_bpar_x} for models with ${\cal M}_{A} = 2.0$. and Figure~\ref{fig:cden_bper_z} is similar to and shows steeper correlations then \ref{fig:cden_bper_x} for all models. This occurs because the
line of sight is chosen to be $\perp B_{0}$ and the observed magnetic
field is simply the random/perturbed component. We must keep in mind that the
density structures in MHD turbulence are, in general, filamentary and anisotropic
regarding the magnetic field orientation.

 As a consequence, column density
maps along x and z-directions will be different. This effect is more prominent
on sub-Alfv\'{e}nic models, as the magnetic field lines are slightly changed due to
turbulent motions. We visually see that a strong magnetic field compresses matter more when densities are perpendicular. This is clearly shown since the plots in Figure ~\ref{fig:cden_bper_z} are the most linear of all four B vs. column density correlations. Also, despite the orientation of the field, it is  noticeable from Figures~\ref{fig:cden_bpar_x}-\ref{fig:cden_bper_z} that supersonic models present smaller dispersion over the column density-magnetic field space.

\begin{figure*}
\centering
\includegraphics[scale=.3]{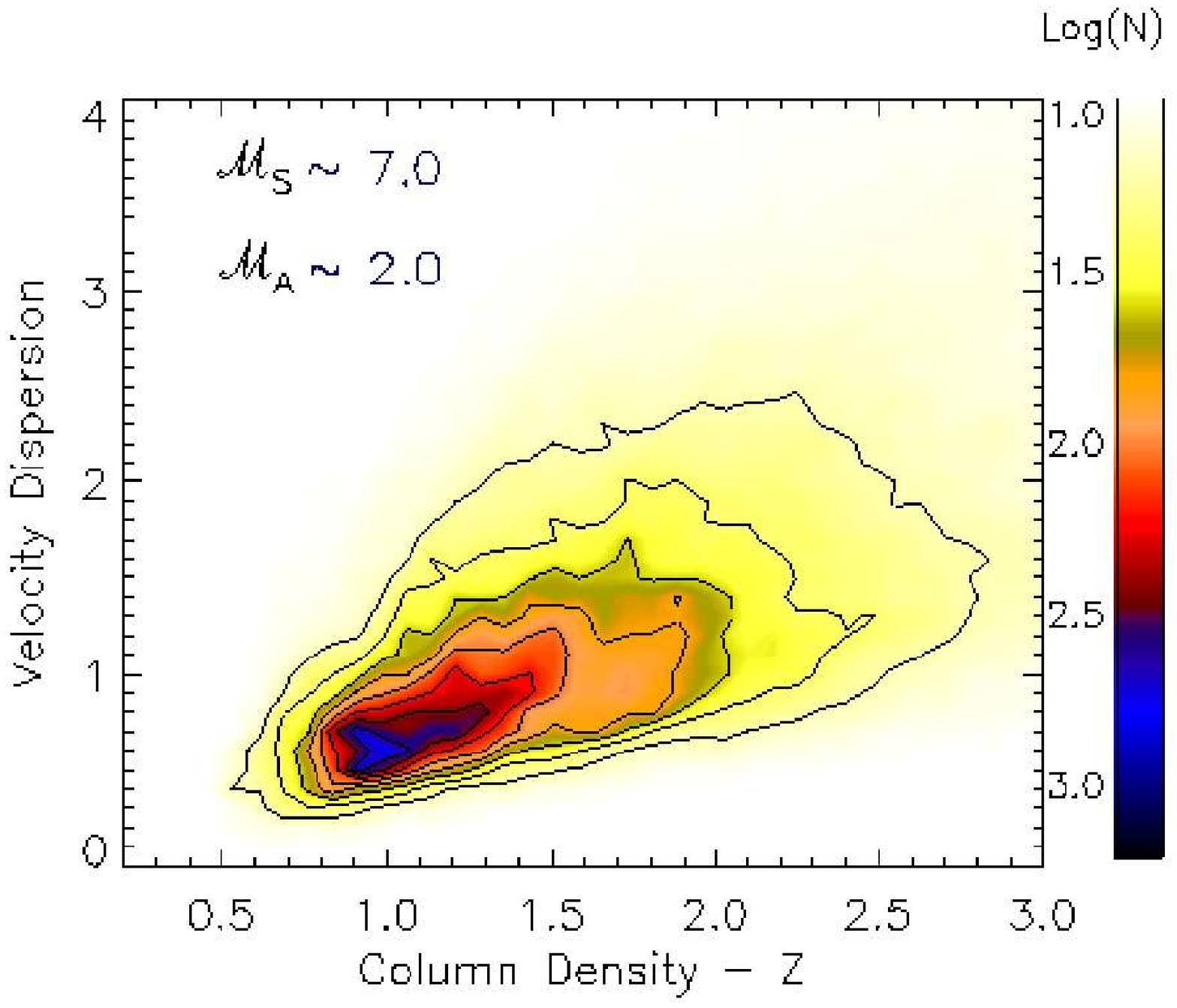} % \includegraphics[scale=.35]{apj/c512b.1p.01/paper/corr_coldens_vel_z}
\includegraphics[scale=.3]{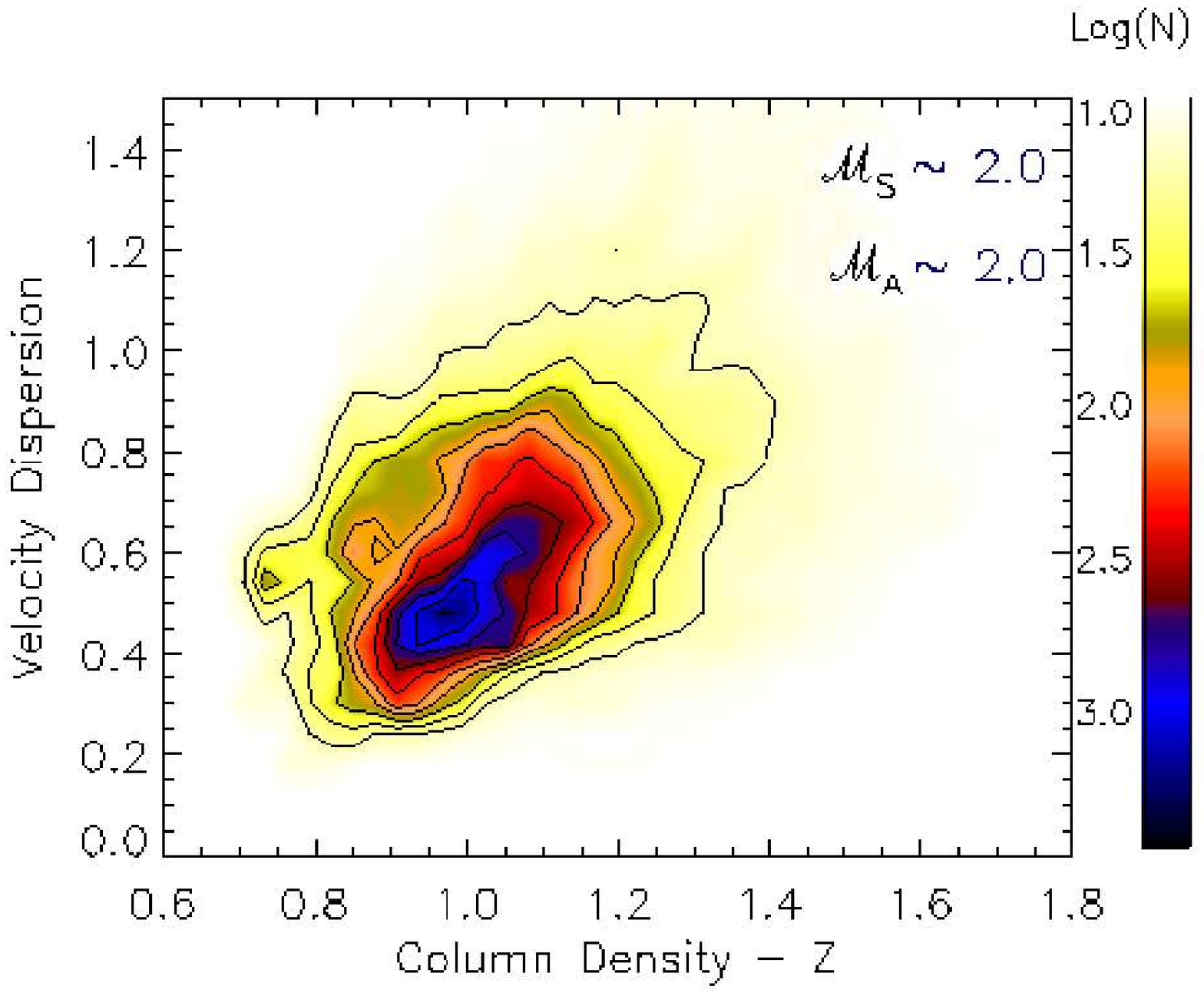} % \includegraphics[scale=.35]{apj/c512b.1p.1/paper/corr_coldens_vel_z}
\includegraphics[scale=.3]{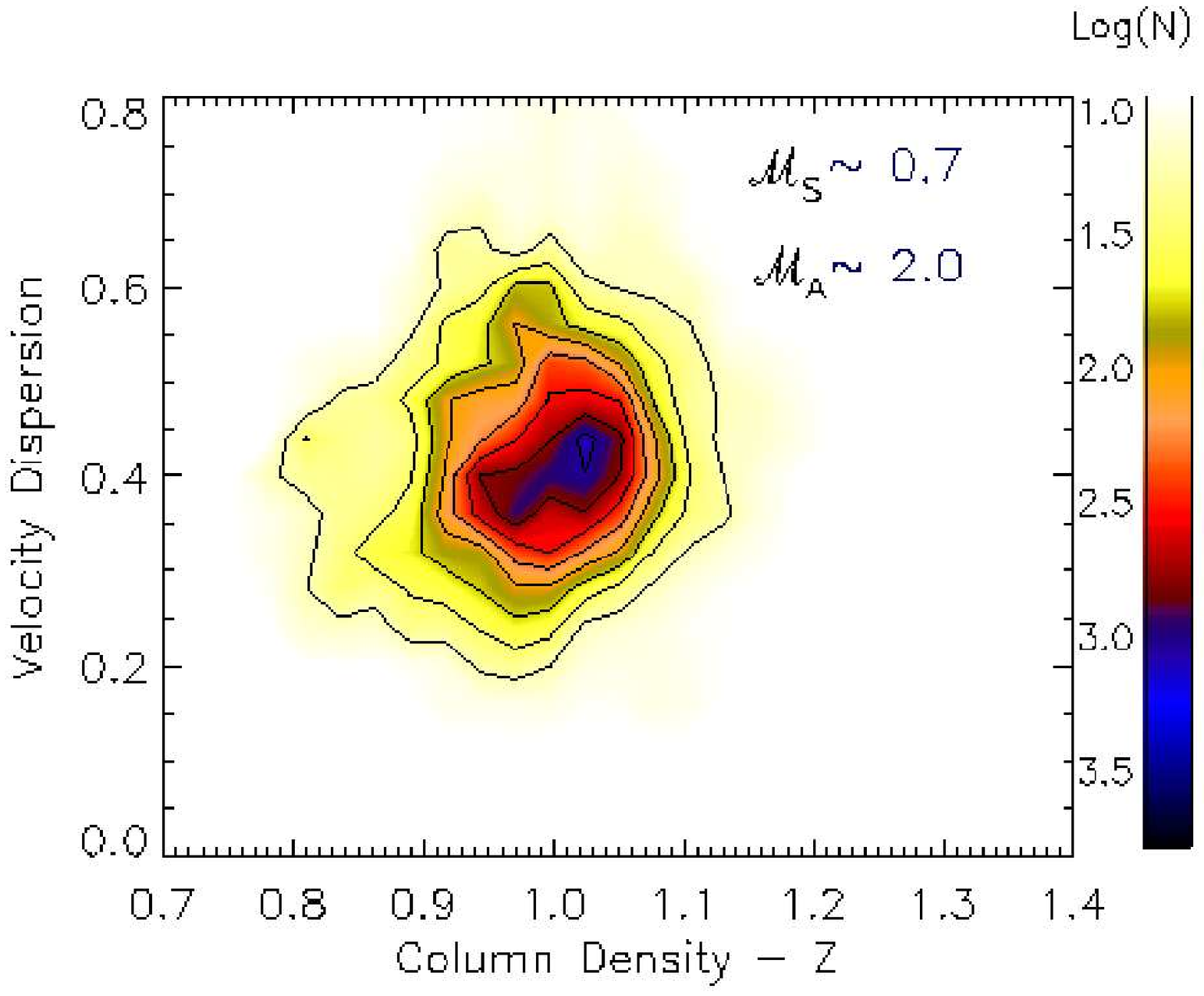} % \includegraphics[scale=.35]{apj/c512b.1p1/paper/corr_coldens_vel_z}
\includegraphics[scale=.3]{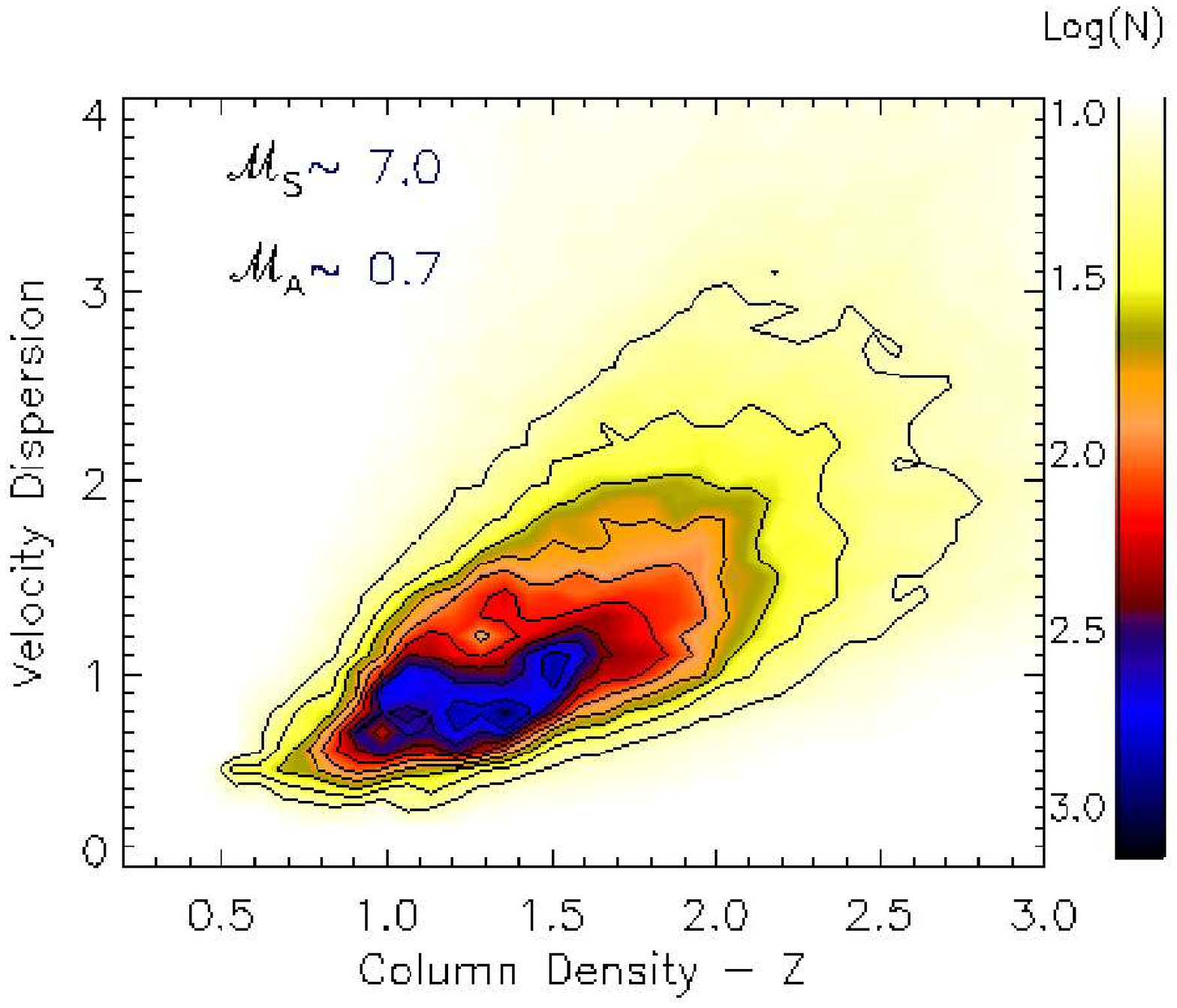} % \includegraphics[scale=.35]{apj/c512b1p.01/paper/corr_coldens_vel_z}
\includegraphics[scale=.3]{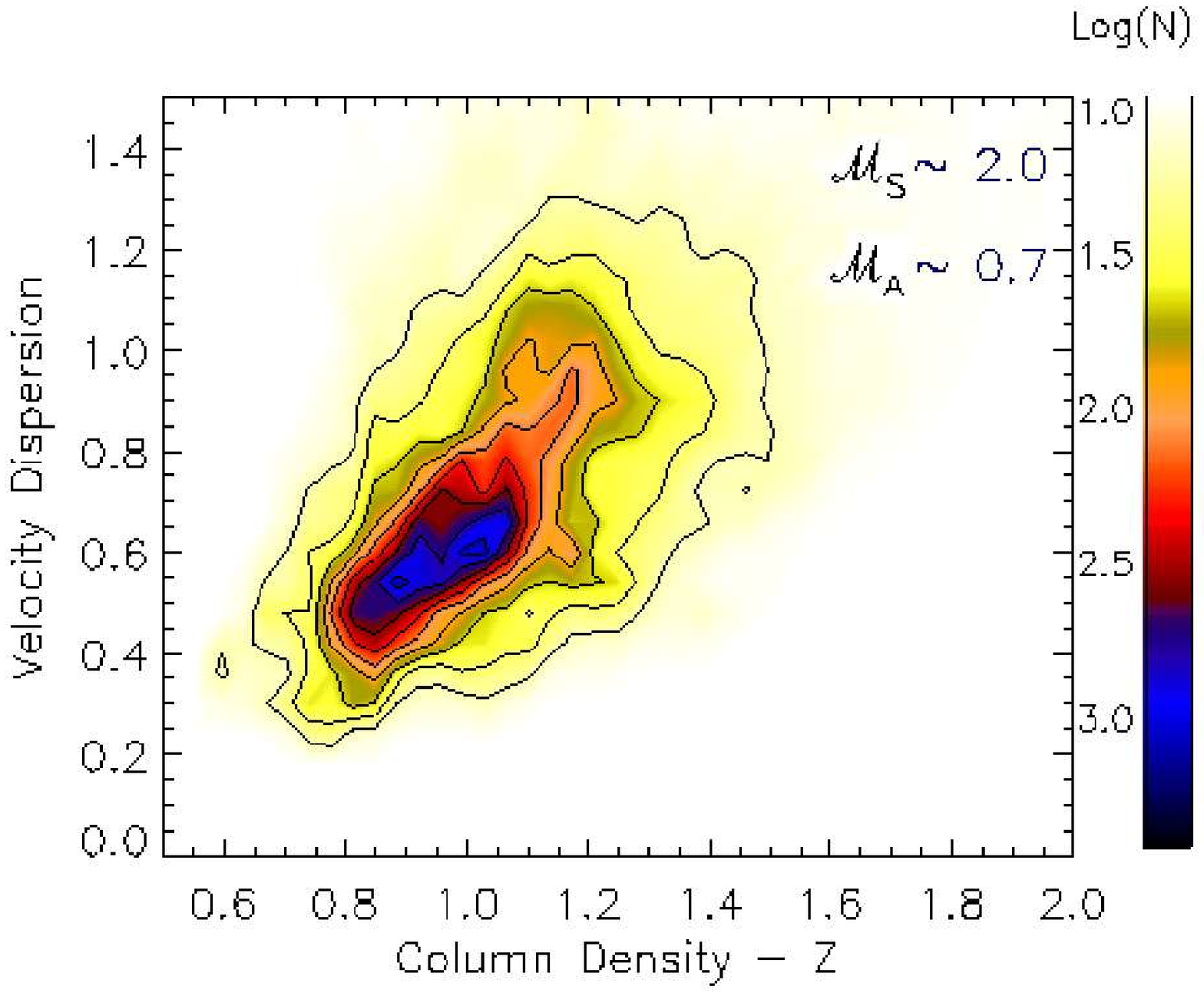} % \includegraphics[scale=.35]{apj/c512b1p.1/paper/corr_coldens_vel_z}
\includegraphics[scale=.3]{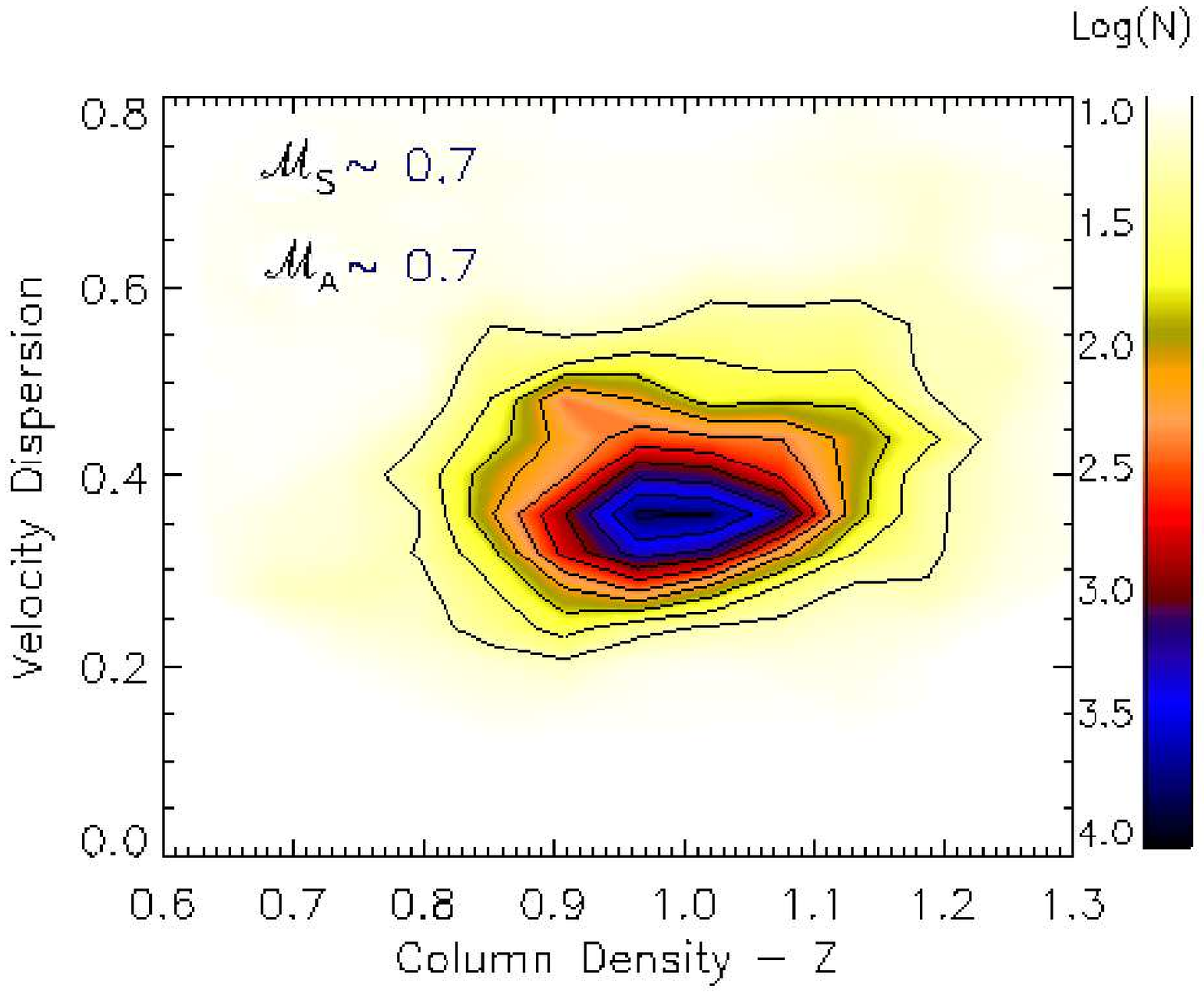} % \includegraphics[scale=.35]{apj/c512b1p1/paper/corr_coldens_vel_z}
\caption{The 2D correlation of velocity dispersion parallel to the line of sight (i.e. along the z direction) vs. column density in the z direction. The mean magentic field is $\perp$ to LOS.  The first row consists of super-Alfv\'enic models while the bottom row is sub-Alfv\'enic. Images are ordered left to right as supersonic to subsonic. Blue contours indicate regions of high data counts while red and yellow have lower counts.\label{fig:cden_vel_z}}
\end{figure*}

In order to compare the observational measures of velocities from spectral
lines with the synthetic velocity dispersions, we studied the correlations of column density and velocity dispersion parallel and perpendicular
to the line of sight. We present the figures for dispersion perpendicular to the LOS (shown in Figure~\ref{fig:cden_vel_z}). We calculate velocity dispersion by integrating $\sigma_{V \rho}$ along the LOS. Supersonic models present stronger correlation along the z LOS then along the x LOS due densities being perpendicular to the mean magentic field.  The effect of this orientation is explained in Figures ~\ref{fig:cden_bpar_z} and ~\ref{fig:cden_bper_z}.  For cases parallel to the magnetic field, matter is confined by the magnetic field and the flows tend to be along it, thus there is ultimately less dispersion seen in these cases.  When column density is perpendicular to the field, as shown in Figure~\ref{fig:cden_vel_z}, dispersion increases with density for supersonic cases.  The higher velocities give rise to higher compression of matter due to shock waves.  This is evident upon comparison between supersonic and subsonic models
Exactly how to relate the statistics of observed velocity line profiles to simulations is not straight forward (see Esquivel \& Lazarian 2005). The interaction between density fluctuations and velocity in synthetic MHD cubes must first be characterized before more complicated statistics can be applied and comparisons with observations are made.  We show the correlation between observable velocity dispersions in Figure ~\ref{fig:vdisp}. We study how the actual dispersion of velocity correlates with the measure available through spectral observations of the Doppler shifted lines. We understand the dispersion of velocity along the z axis to be  $V_{z}$ dispersion$=\sigma_{V_{z} \rho_{z}}$. Due to shock waves, we expect strong density fluctuations to be present for supersonic cases.  This will result in generally nonlinear correlations of velocity dispersions for constant vs. non constant $\rho$.  However, for subsonic cases, density fluctuations are less prevalent, and linear correlations should be present.  Considering dispersion, models with higher Mach numbers should reach higher values of velocity dispersion, by definition. All of these trends are clearly shown in Figure ~\ref{fig:vdisp}. Supersonic models show no correlation due to density fluctuations while subsonic models are linear.   The stronger the magnetic field, the tighter the correlation.  This trend is confirmed in Figure ~\ref{fig:cden_vel_z}.

\subsection{Correlation Coefficients}
\begin{table*}

\begin{center}
\caption{Table of log-log Correlation Coefficients \label{tab:coefficients1}}
\begin{tabular}{|c|c|c|c|c|} \hline \hline 
Model & Magnetic Energy vs. $\rho$ & Kinetic Energy vs. $\rho$ & $B_{\parallel}$ vs. $\Sigma_{X}$ & $B_{\perp}$ vs. $\Sigma_{X}$  \\ \tableline
1 &$-1.1700\pm{1x10^{-4}}$ & $-0.5810\pm{3x10^{-4}}$ & $0.020\pm{0.003}$ & $1.200\pm{0.004}$ \\
2 & $-0.10000\pm{4x10^{-5}}$& $-0.1500\pm{1x10^{-4}}$& $0.950\pm{0.001}$ & $1.240\pm{0.002} $ \\
3 & $0.00200\pm{2x10^{-5}}$& $-0.05000\pm{8x10^{-5}}$ & $0.990\pm{0.001}$& $1.240\pm{0.002}$\\
4 & $-1.0600\pm{4x10^{-4}}$ & $-0.6330\pm{3x10^{-4}}$&$0.720\pm{0.009}$ & $0.920\pm{0.003} $ \\
5 & $0.1500\pm{1x10^{-4}}$ &$-0.1300\pm{1x10^{-4}} $ & $2.270\pm{0.004}$ & $1.310\pm{0.001}$ \\
6 & $0.3400\pm{8x10^{-5}}$ &$-0.02000\pm{7x10^{-5}}$ &$2.110\pm{0.003}$ & $ 1.350\pm{0.001}$ \\
\hline
\end{tabular}
\end{center}
\end{table*}

\begin{table*}

\begin{center}
\caption{Table of log-log Correlation Coefficients \label{tab:coefficients2}}
\begin{tabular}{|c|c|c|c|c|c|} \hline \hline 
Model & $B_{\parallel}$ vs. $\Sigma_{Z}$ & $B_{\perp}$ vs. $\Sigma_{Z}$ & $M_{A}$ vs.$\rho$ & $V_{dispersion}$ vs. $\Sigma_{Z}$ & $V_{dispersion}$ vs. $\Sigma_{X}$\\ \tableline
1 &$-0.13\pm{0.01}$& $1.3600\pm{2x10^{-4}} $  & $ 0.7980\pm{2x10^{-4}}$ & $0.810\pm{0.006} $& $0.350\pm{0.007}$ \\
2 & $1.550\pm{0.004} $ & $0.9300\pm{6x10^{-4}} $ & $ 0.47300\pm{6x10^{-5}} $ & $1.180\pm{0.003}$&$-0.170\pm{0.004}$ \\
3 & $1.570\pm{0.003}$ & $ 0.9800\pm{5x10^{-4}} $ & $0.47100\pm{4x10^{-5}} $ & $ 1.190\pm{0.002}$& $-0.120\pm{0.004}$\\
4 & $2.130\pm{0.008}$ & $1.300\pm{0.003} $  & $0.7150\pm{3x10^{-4}} $ & $1.19\pm{0.005}$& $0.620\pm{0.005}$ \\
5 & $2.330\pm{0.004}$ & $1.320\pm{0.001} $ & $0.35700\pm{9x10^{-5}} $ & $1.1200\pm{2.7x10^{-4}} $& $0.230\pm{0.003}$   \\
6 & $2.020\pm{0.003}$ & $ 1.360\pm{0.001}$ & $0.32020\pm{6x10^{-5}} $ & $1.270\pm{0.002} $&  $0.616\pm{0.005}$ \\
\hline
\end{tabular}
\end{center}
\end{table*}

Finally, we present a table of select log-log correlation coefficients with error bars in Tables ~\ref{tab:coefficients1} and ~\ref{tab:coefficients2} to see quantitatively the degree of correlation in our plots. Ultimately, we seek to set an order on which models show the highest correlations. These correlation coefficients tell us the degree of the polynomial equation fitted  to quantities plotted.  In order to obtain these coefficients we computed the linear least-squared fit of the correlations, with error bars determined from the 1-sigma uncertainty estimates.  For magnetic energy vs. density, highly supersonic cases show positive correlations while subsonic present anti-correlation. However, it is of interest to note that the cases with the largest correlation differ in sonic Mach number, but both have ${\cal M}_{A}$=2.0. We see that from our results that both the ${\cal M}_{s}$ and ${\cal M}_{A}$ play a roll in the correlation of B and $\rho$, due to shocks creating density clumps in supersonic cases. The ${\cal M}_{A}$=2.0, ${\cal M}_{A}$=0.7 case shows the closest to Alfv\'enic perturbations ($B \approx \sqrt{\rho}$). In general, magnetic energy cannot be correlated with $\rho$ with a simple polynomial expression.  For kinetic energy vs. density, all show negative correlations with supersonic cases being correlated the greatest, showing that no strong correlation trend was obtained. For magnetic field parallel to LOS vs. x column density ($\Sigma_{x}$), the table shows that the most important factors in obtaining a strong correlation is the sonic Mach number and the strength of the field. Supersonic cases with large magnetic field are generally linear while supersonic cases with small magnetic field show an $x^2$ trend. Looking at magnetic field perpendicular to LOS, it is apparent all models are near linear. The orientation of the field perpendicular to the LOS creates correlations that are more nearly linear and this is evident in both the plots and the coefficients of $B_{\perp}$ vs. $\Sigma_Z$. We also show the correlation coefficients for $M_A$ vs. $\rho$.  No polynomial relationship is seen for any model. For velocity dispersion and $\Sigma_Z$, a strong linear relationship is established which is not seen for the models that are along $\Sigma_X$ (parallel to the magnetic field).  It is evident that the orientation of the field with respect to the LOS is very important in obtaining a strong correlation with dispersion.

\section{Bispectrum}
\label{sec:bispectrum}

\subsection{Definition and Calculation of bispectrum}

As turbulent vortexes evolve, they transfer energy from large to small scales. In this case, wave-wave interactions generate a hierarchical turbulent cascade as $k_1 + k_2 \rightarrow k_{3}$. For incompressible flows, under Kolmogorov's assumptions, we have $k_1 \simeq k_2 = k$ and $k_3 \simeq 2k$. For compressible and magnetized flows, this becomes more complicated and nonlinear wave-wave interactions may take place, mainly for $k_1 \neq k_2$. Also, 
we expect MHD turbulence to present more wave modes than purely hydrodynamical flows and, in this case, the energy cascade can be much more complicated. In order to study this phenomenon and characterize both amplitude and phase in  the turbulent cascade, the analysis of the three-point correlation function, or bispectrum, is required \cite[see][]{Bar02,Masa04}.

The bispectrum technique characterizes and searches for nonlinear interactions, which makes it a timely study for interstellar MHD turbulence. However, unlike the standard power spectrum, the bispectrum is a complex function and carries phase information. In order to compute either the power spectrum or bispectrum, one must first compute the Fourier transform of the signal.  In the case of either  a discrete or continuous Fourier transform we are left with information that includes both the magnitude  and the \textit{phase} of the signal at a given frequency.  The power spectrum measures solely the magnitude as a function of the frequency and discards the phase information. The power spectrum gives information about power distributions over all frequencies and looses information about wave-wave interactions and phase. Signals containing multiple frequencies and phases   are better described by the bispectrum.

The bispectrum is closely related to the power spectrum. The Fourier transform of the second-order cumulant, i.e. the autocorrelation function, is the power spectrum while the Fourier transform of the third order cumulant is known as the bispectrum. In a discrete system, the power spectrum is defined as:

\begin{equation}
P(\vec{k})=\sum_{\vec{k}=const.}\tilde{A}(\vec{k})\cdot\tilde{A}^{*}(\vec{k})
\end{equation} 

\noindent
In a similar way, the bispectrum can be defined as:

\begin{equation}
B(\vec{k_{1}},\vec{k_{2}})=\sum_{\vec{k_{1}}=const}\sum_{\vec{k_{2}}=const}\tilde{A}(\vec{k_{1}})\cdot\tilde{A}(\vec{k_{2}})\cdot\tilde{A}^{*}(\vec{k_{1}}+\vec{k_{2}})
\label{eq:bispectra}
\end{equation}

\noindent
where $k_{1}$ and $k_{2}$ are the wave numbers of two interacting waves, and $A(\vec{k})$ is the original discrete time series data with finite number of elements with $A^{*}(\vec{k})$ representing the complex conjugate of $A(\vec{k})$. As is shown in Equation~\ref{eq:bispectra} , the bispectrum is a complex quantity which will measure both phase and magnitude information between different wave modes. 

In practice, our calculation of the bispectrum involves performing a Fast Fourier transform of density or column density and the application of Equation~\ref{eq:bispectra}. Although we are primarily interested here in density fluctuations, the bispectrum can also be calculated for other fields as well.   We  randomly choose wavevectors and their directions, $k_{1}$ and $k_{2}$ and iterate over them, calculating $k_{3}$, which depends on  $k_{1}$ and $k_{2}$ since  $k_{1}+k_{2}=k_{3}$. We limit the maximum length of the wave vectors to half of the box size. We normalize direction vectors to unity, calculate positions in Fourier space, and finally, compute the bispectrum, which yields a complex number.  We then average this bispectrum over all frequencies and plot bispectral amplitudes. This gives us information on the degree of mode correlation in the system.

In this section we will explore the bispectrum of density and column density for the data cubes. Because this is the first application of the bispectrum to ISM related studies and we would like to have a comparison with observations, exploring fields other then density and column density is beyond the scope of this paper. Although the application of the bispectrum to velocity and magnetic fields would be interesting and provide a beneficial follow up to densities, we focus here only on density and column density maps in order to make our study comparable with observational column densities. Bispectral analysis can be applied to observational data in a similar way as the synthetic data presented in this paper utilizing Equation~\ref{eq:bispectra}. Our model does not include self-gravity or external gravity, and only considers the interplay between the gas pressure and magnetic pressure of isothermal gas. By examining the bispectrum of density and comparing it with column densities, we can characterize turbulent flows on resolutions that are realistically seen in observational data. We will look at different models of turbulence including a Gaussian, pure hydro, and different MHD regimes.  We will examine differences between the bispectrum of density and column density and discuss what information about the sonic Mach number and magnetic field that might be gained by applying the bispectrum to observable data.

\subsection{Bispectra of Density and Column Density}

\begin{figure*}[htb]
\centering
\includegraphics[scale=.25]{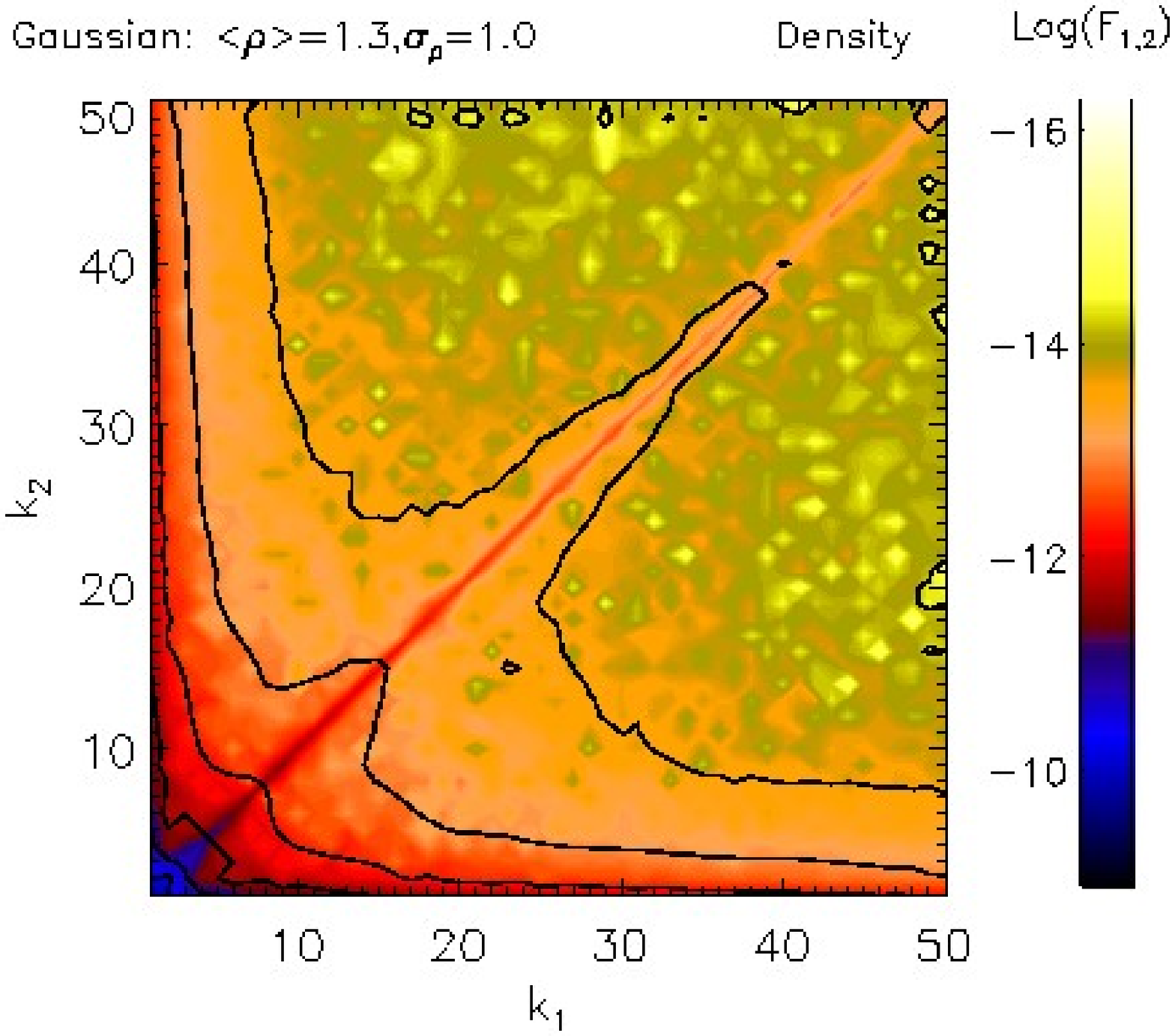} % \includegraphics[scale=.35]{apj/hydro/c512p.01/bispectrum_dens}
\includegraphics[scale=.25]{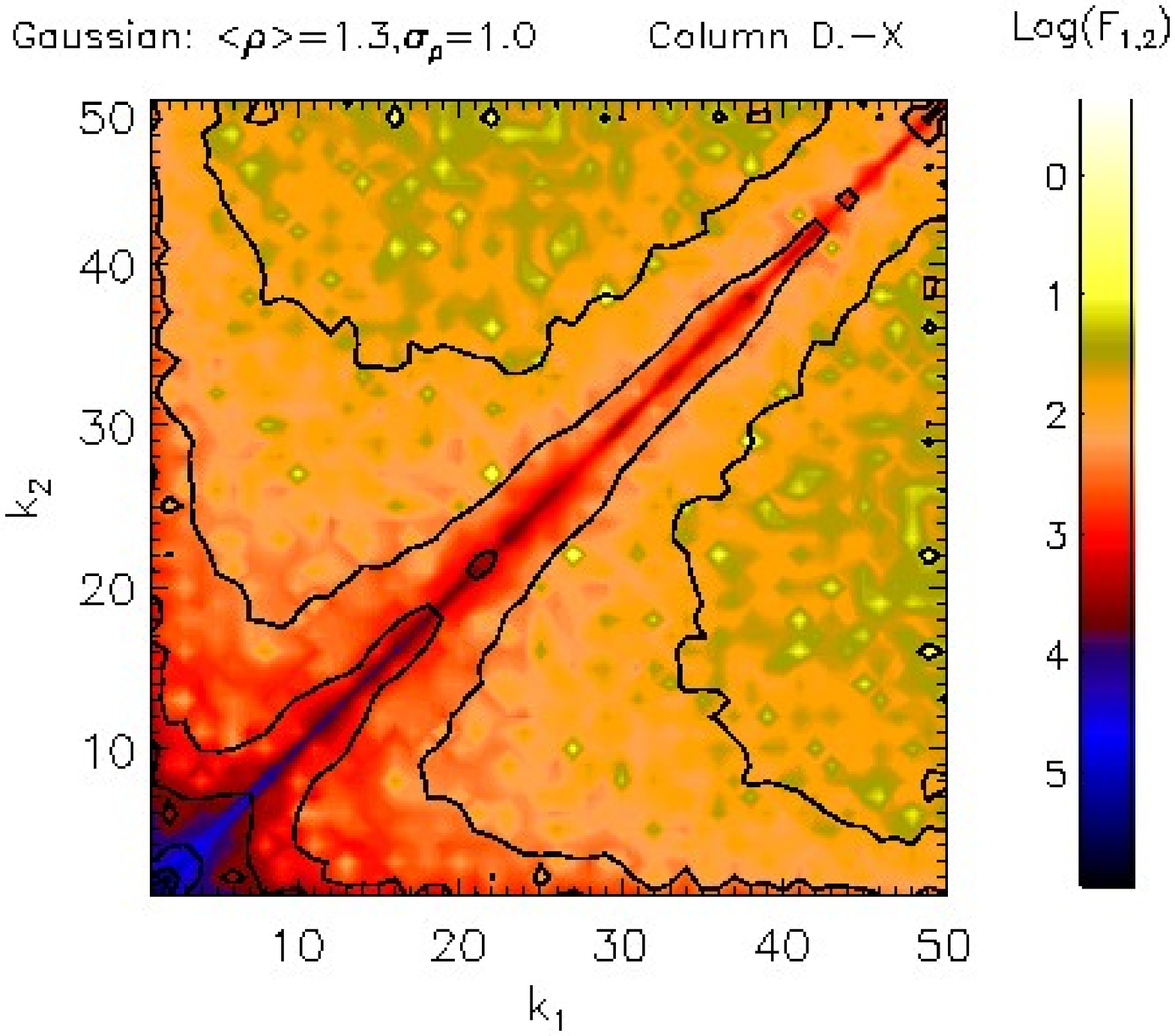} % \includegraphics[scale=.35]{apj/hydro/c512p.01/bispectrum_cd_x}
\includegraphics[scale=.25]{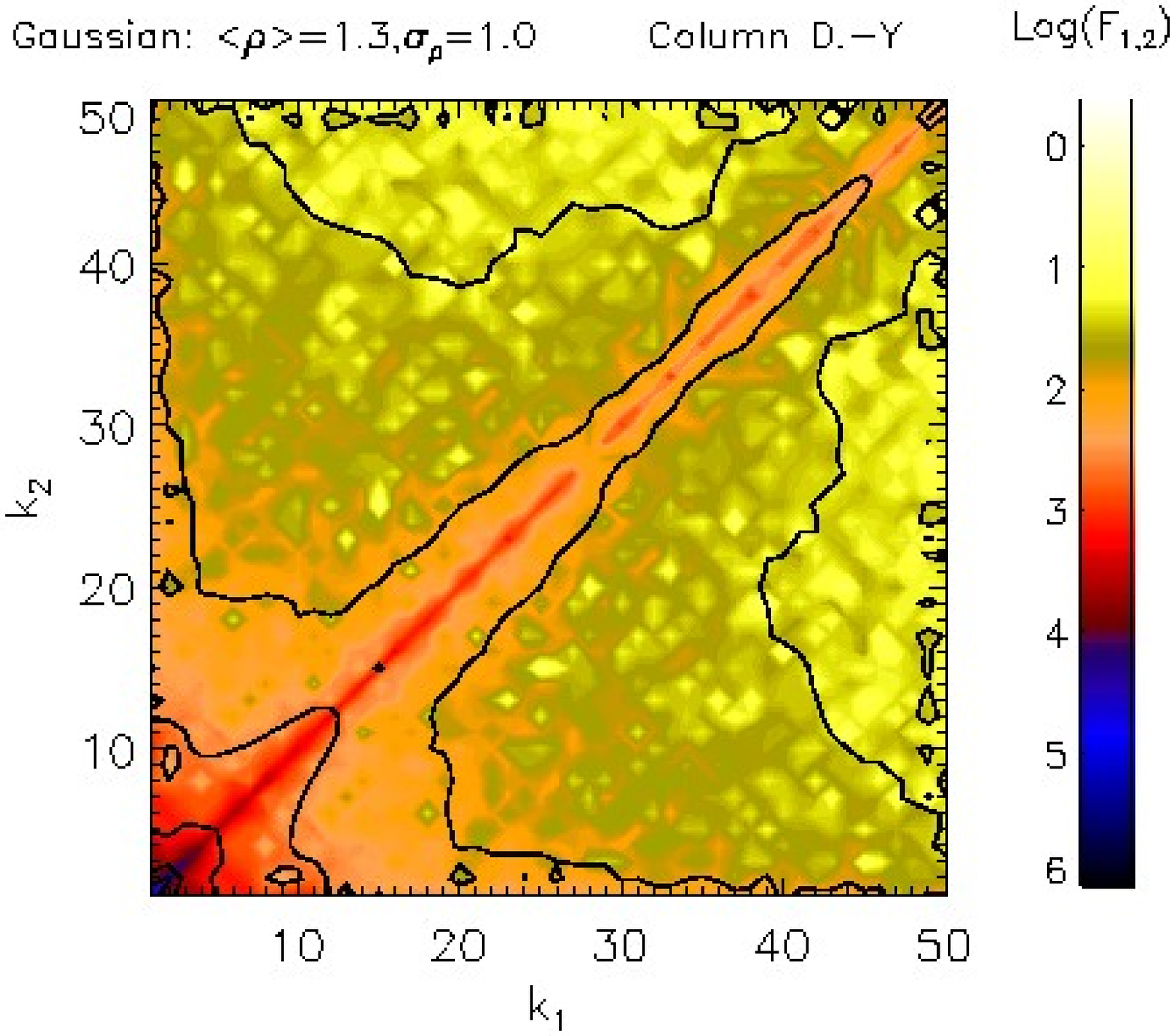} % \includegraphics[scale=.35]{apj/hydro/c512p.01/bispectrum_cd_y}
\includegraphics[scale=.25]{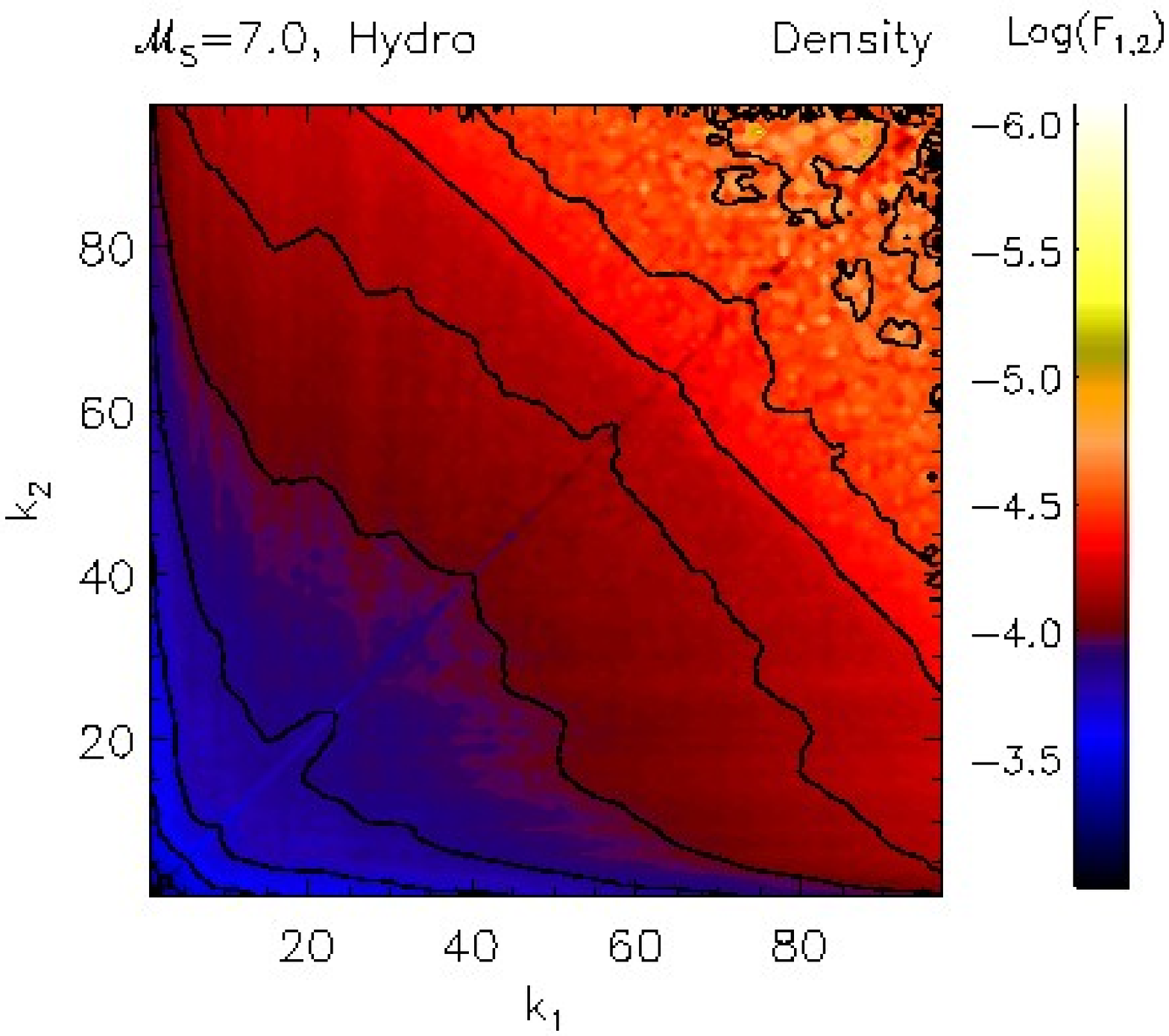} % \includegraphics[scale=.35]{apj/c512b.1p1/paper/bispectrum_dens}
\includegraphics[scale=.25]{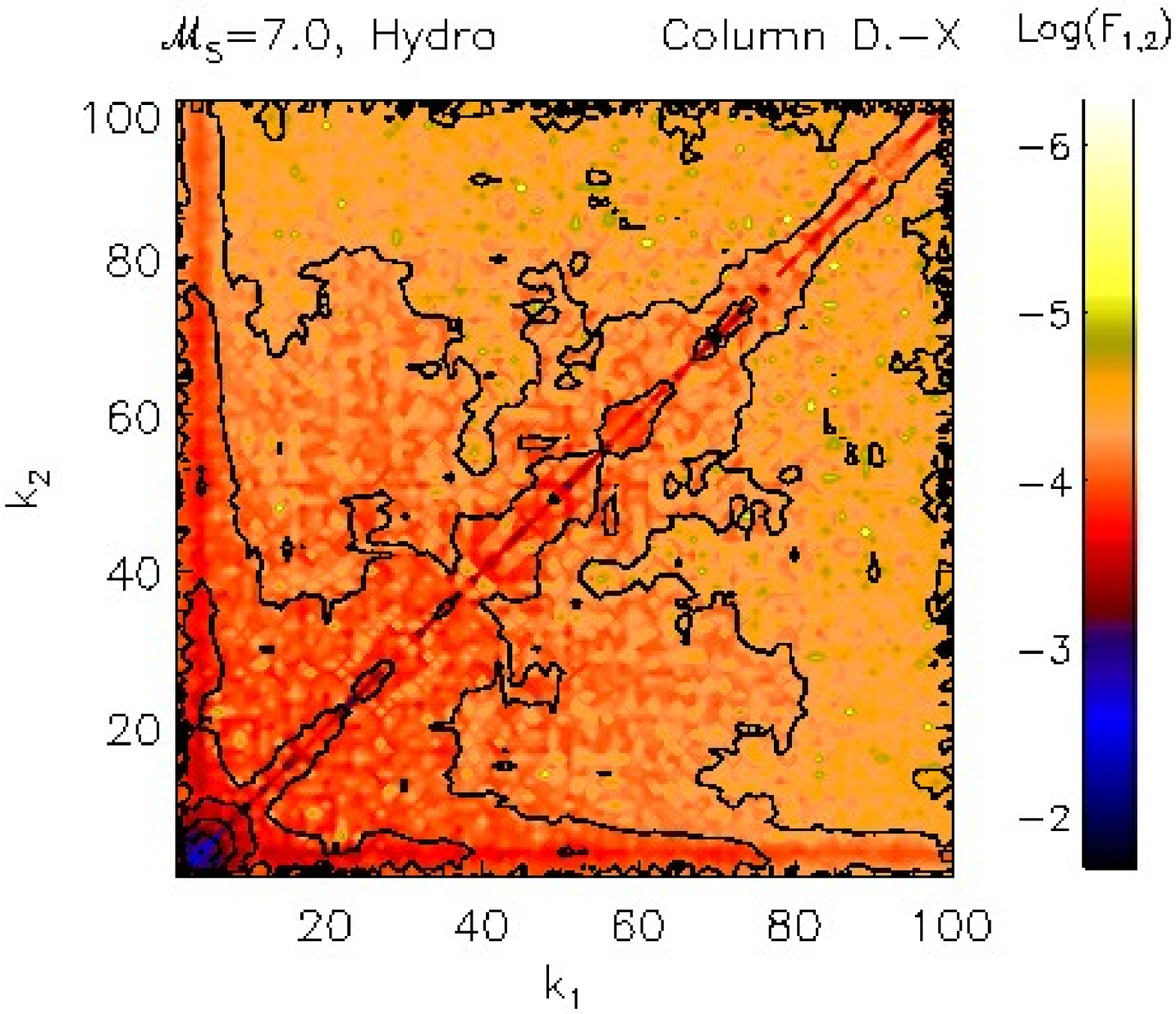} % \includegraphics[scale=.35]{apj/c512b.1p1/paper/bispectrum_cd_x}
\includegraphics[scale=.25]{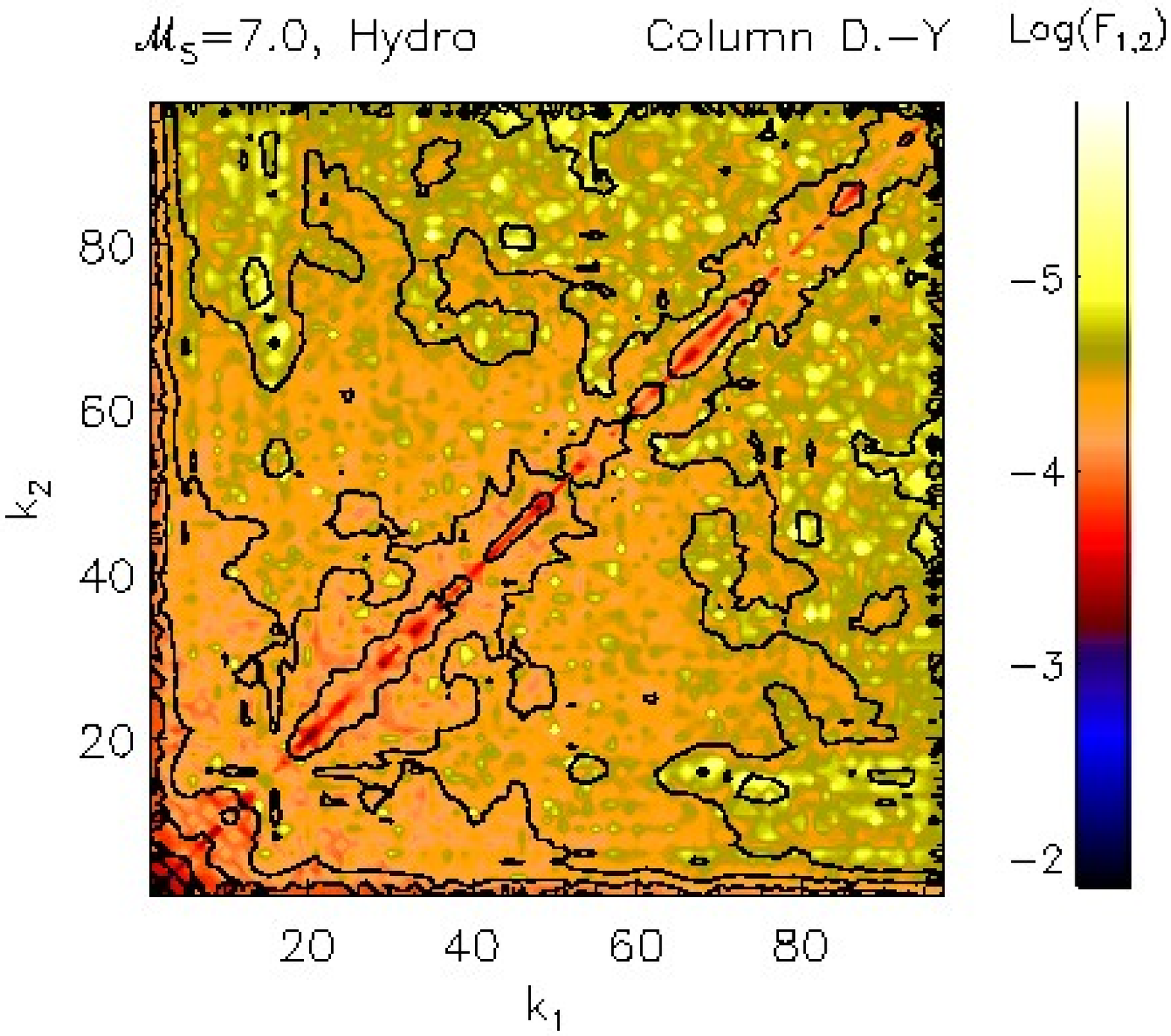} % \includegraphics[scale=.35]{apj/c512b.1p1/paper/bispectrum_cd_y}
\includegraphics[scale=.25]{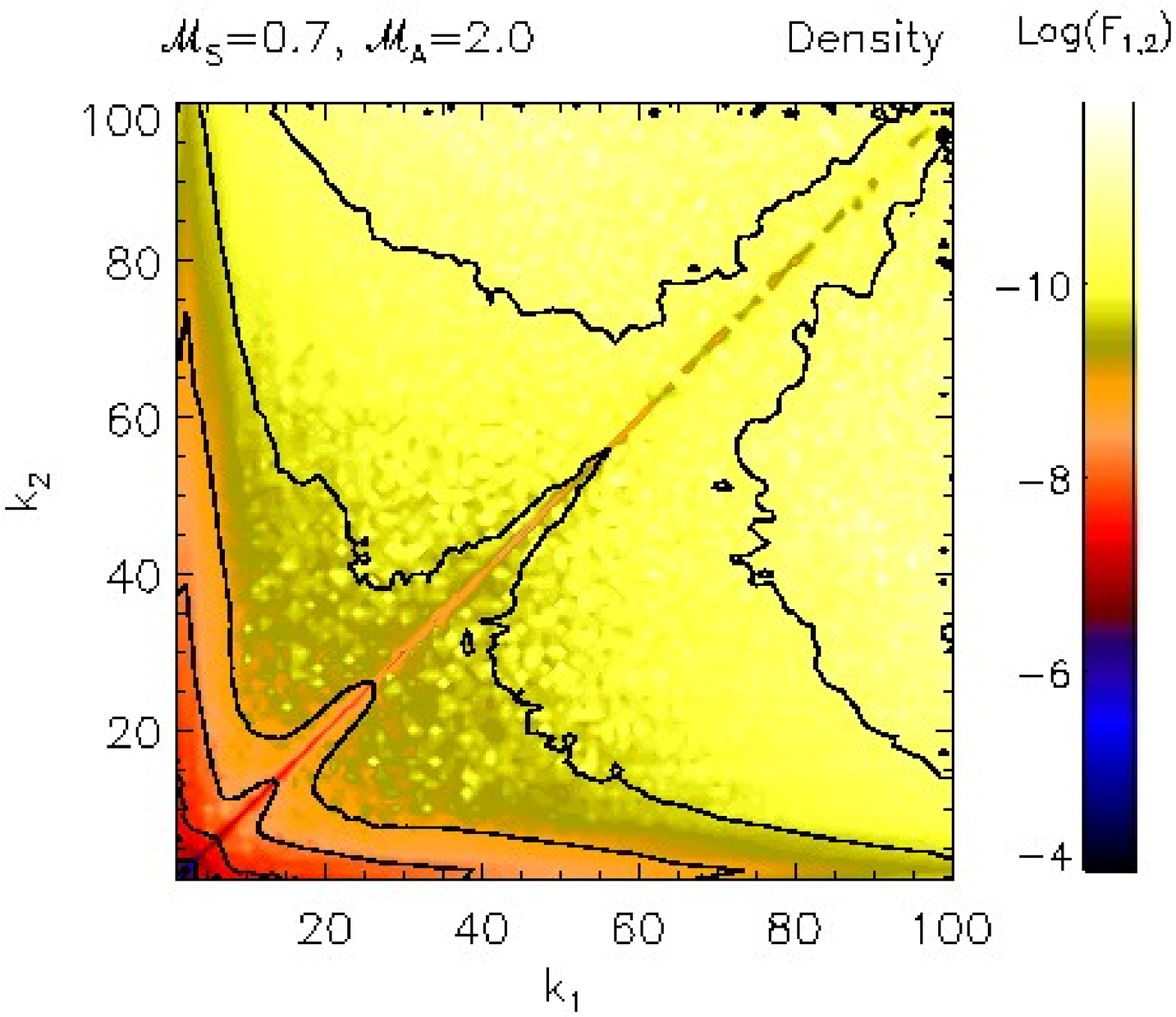} % \includegraphics[scale=.35]{apj/c512b1p1/paper/bispectrum_dens}
\includegraphics[scale=.25]{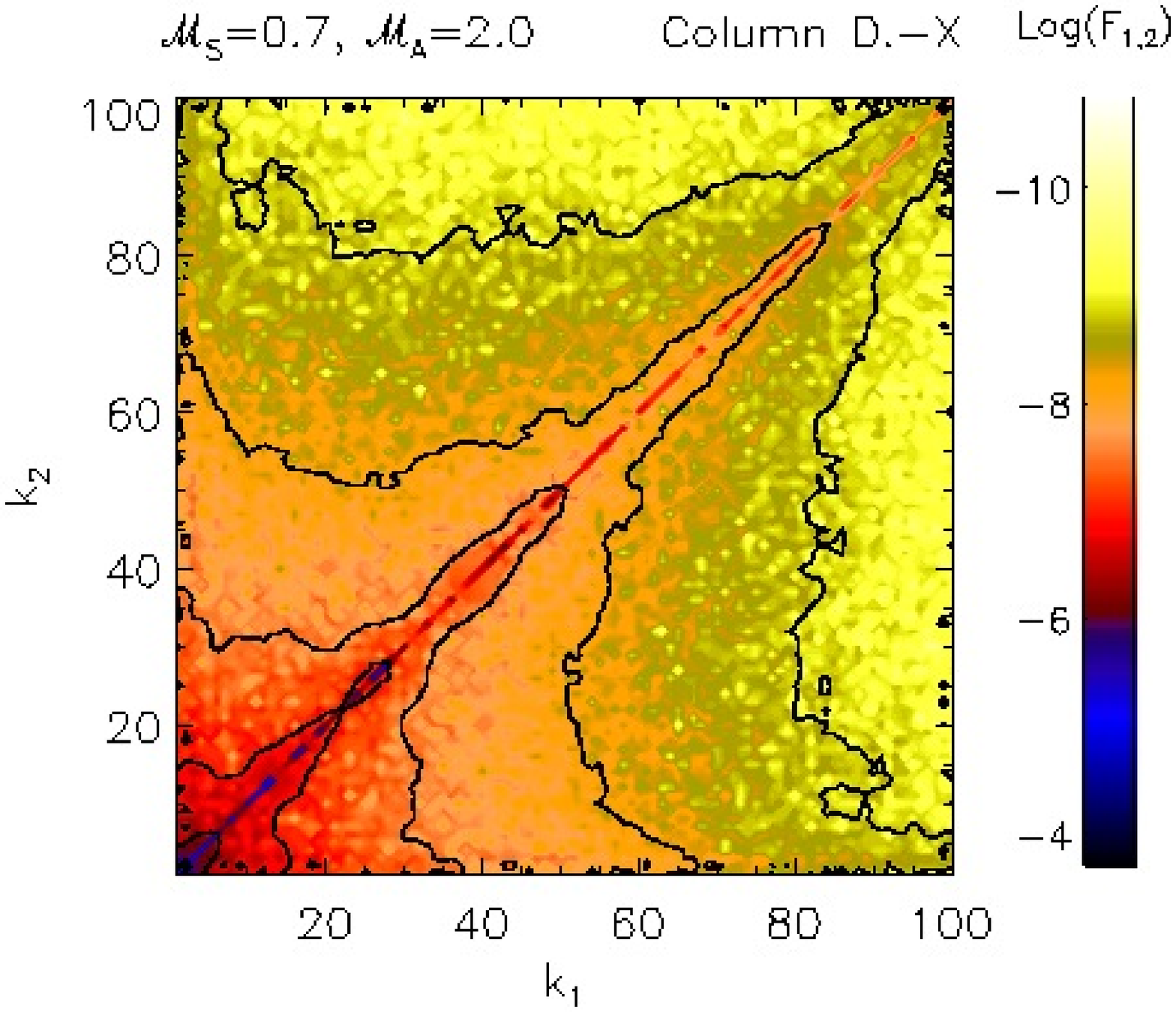} % \includegraphics[scale=.35]{apj/c512b1p1/paper/bispectrum_cd_x}
\includegraphics[scale=.25]{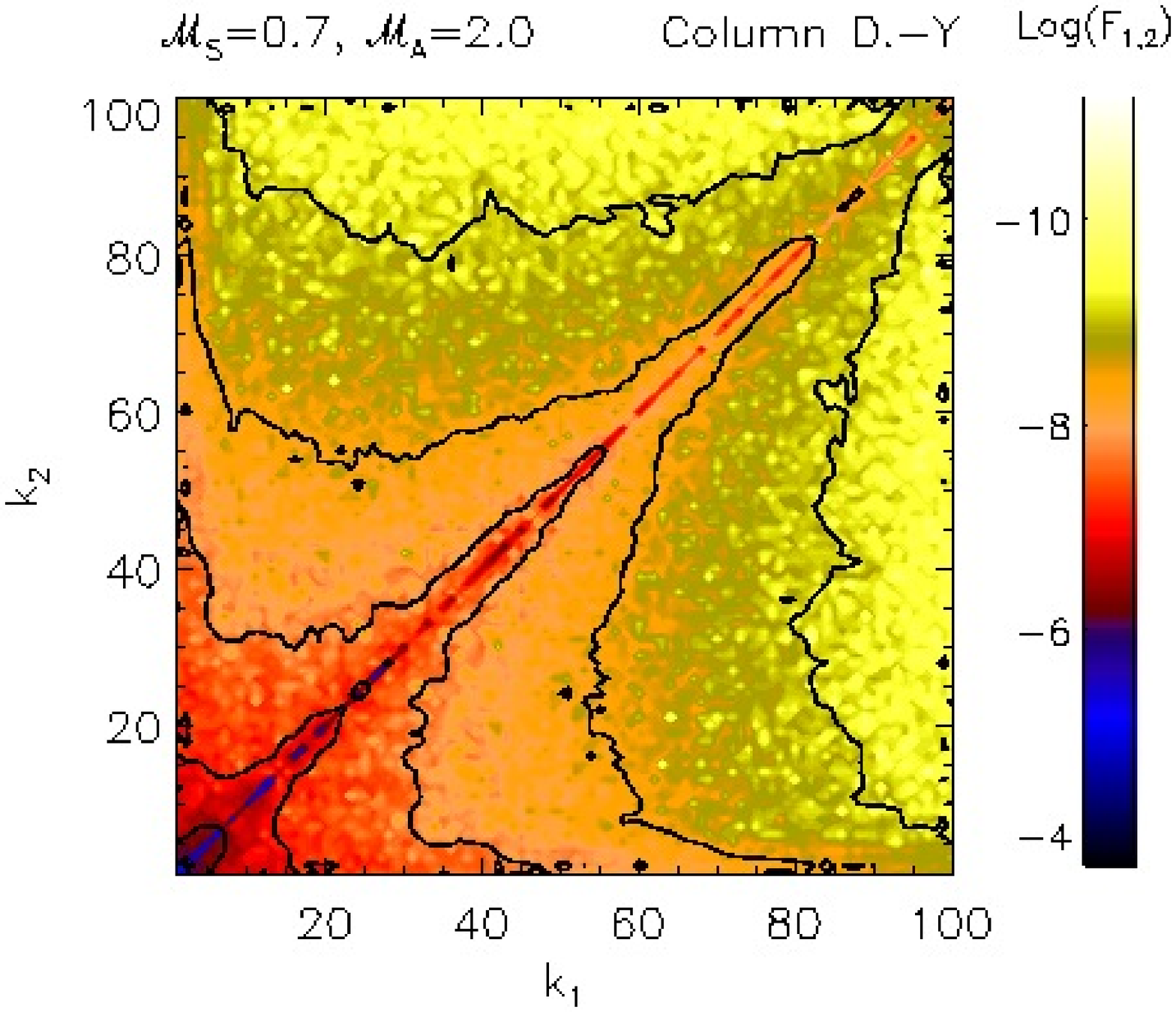} % \includegraphics[scale=.35]{apj/c512b1p1/paper/bispectrum_cd_y}
\includegraphics[scale=.25]{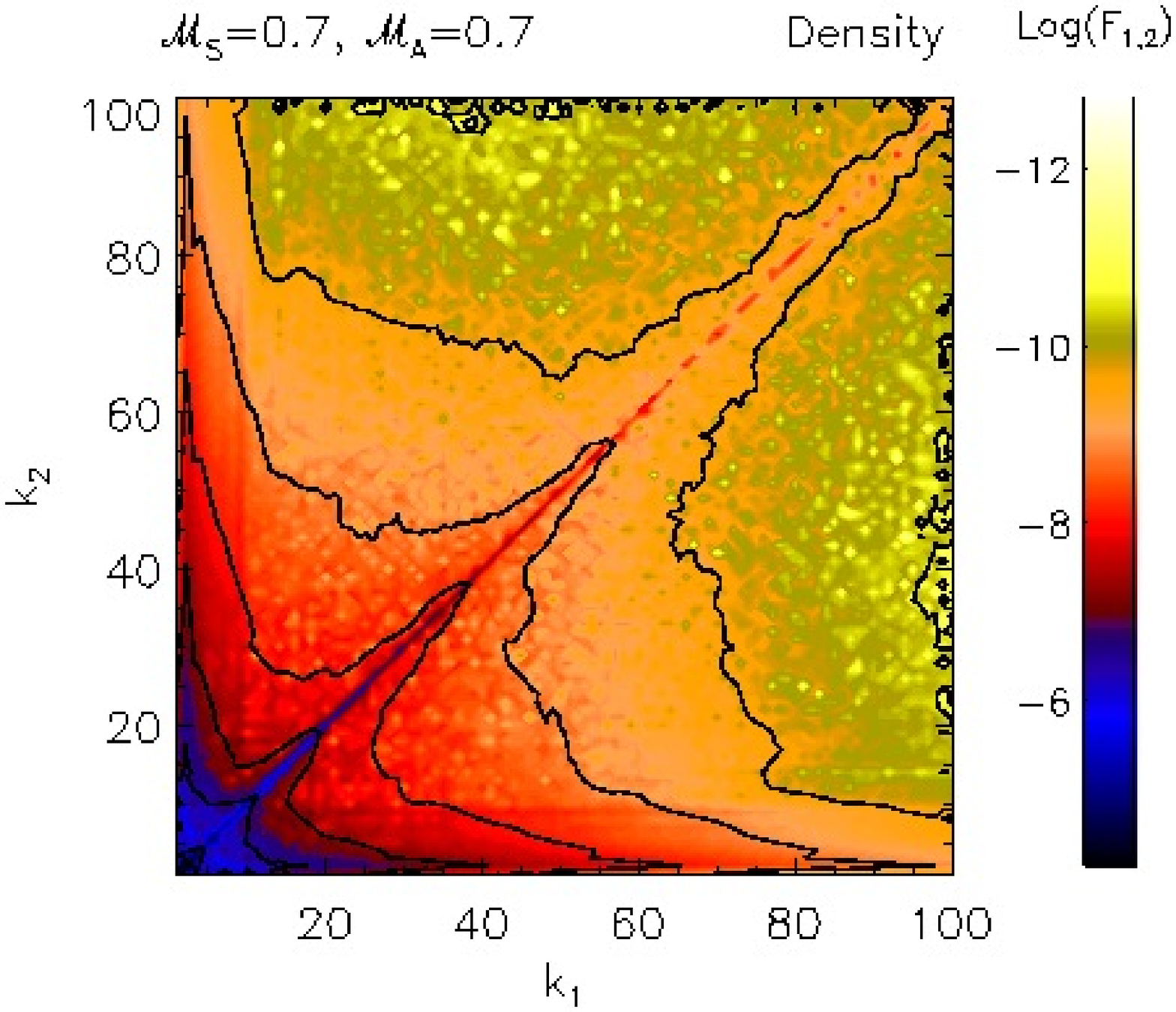} % \includegraphics[scale=.35]{apj/c512b.1p.01/paper/bispectrum_dens}
\includegraphics[scale=.25]{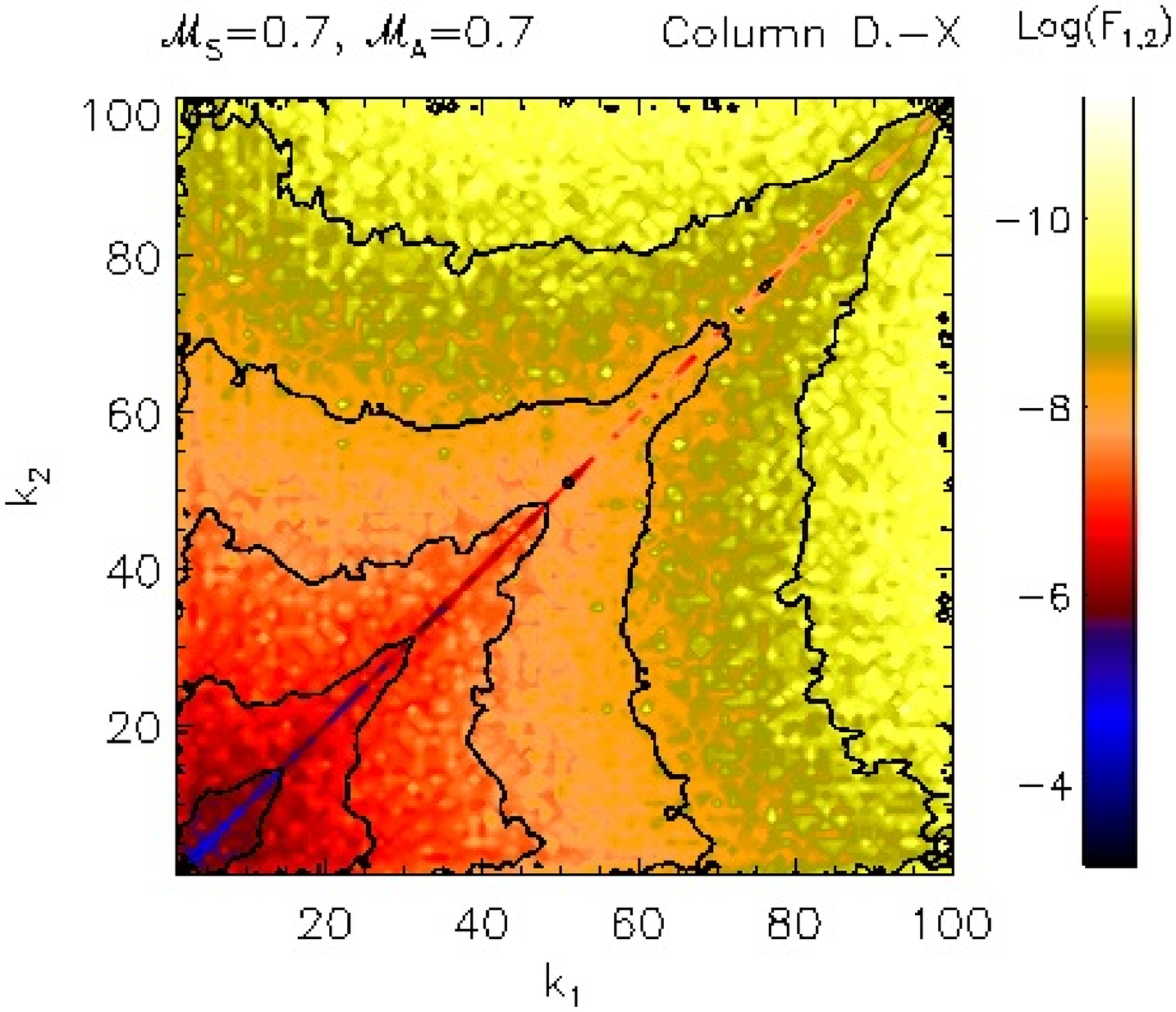} % \includegraphics[scale=.35]{apj/c512b.1p.01/paper/bispectrum_cd_x}
\includegraphics[scale=.25]{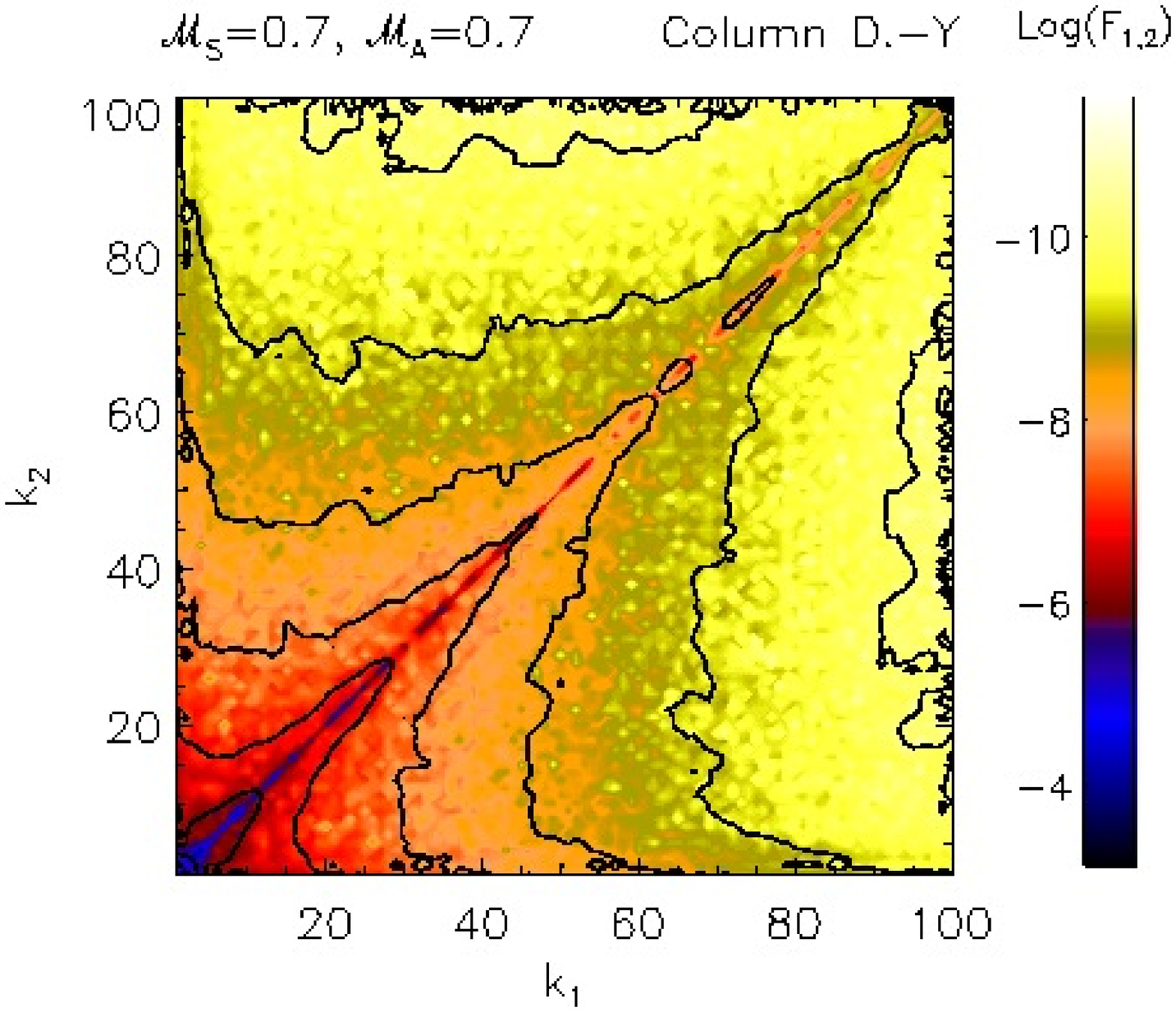} % \includegraphics[scale=.35]{apj/c512b.1p.01/paper/bispectrum_cd_y}
\includegraphics[scale=.25]{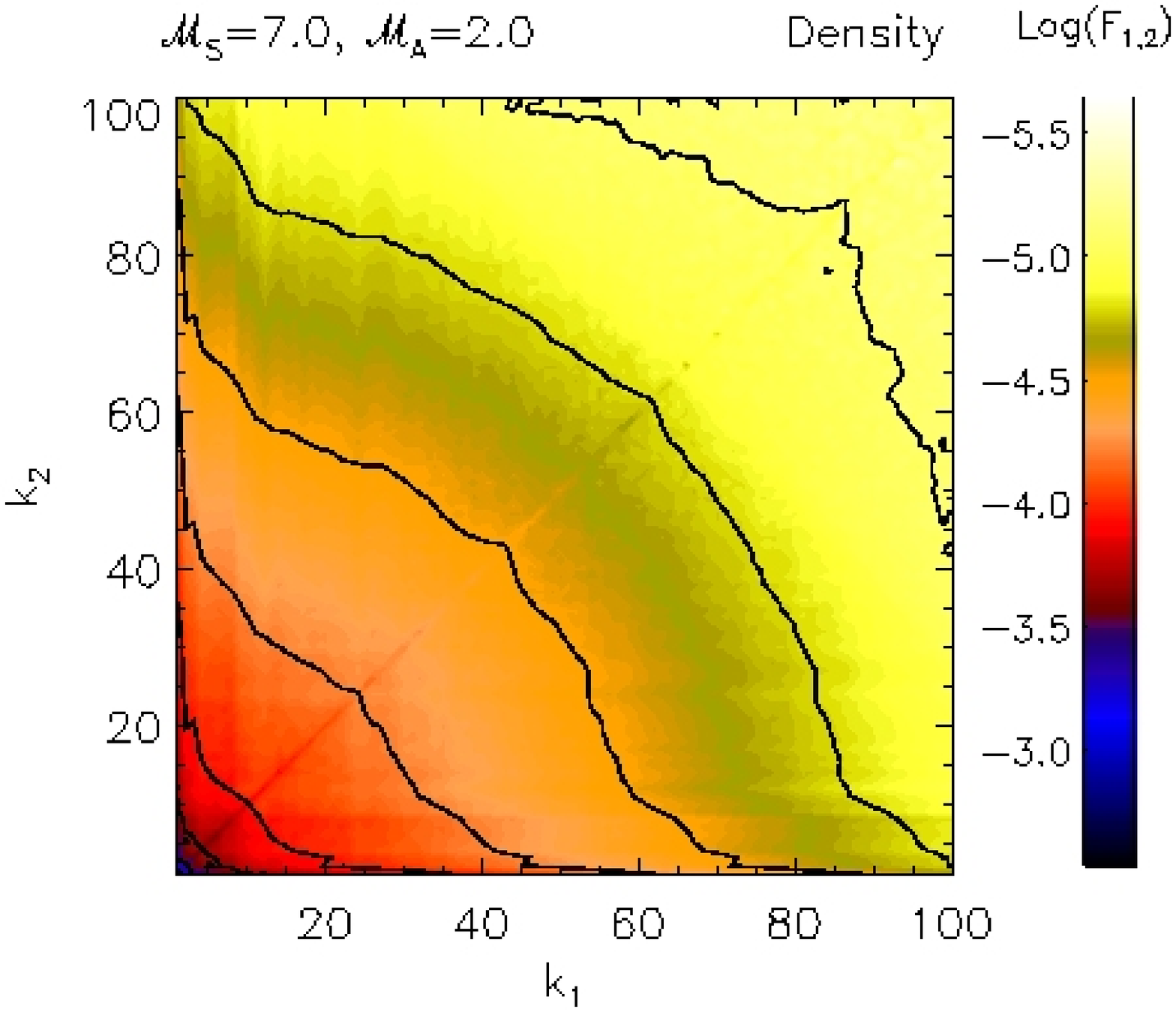} % \includegraphics[scale=.35]{apj/c512b1p.01/paper/bispectrum_dens}
\includegraphics[scale=.25]{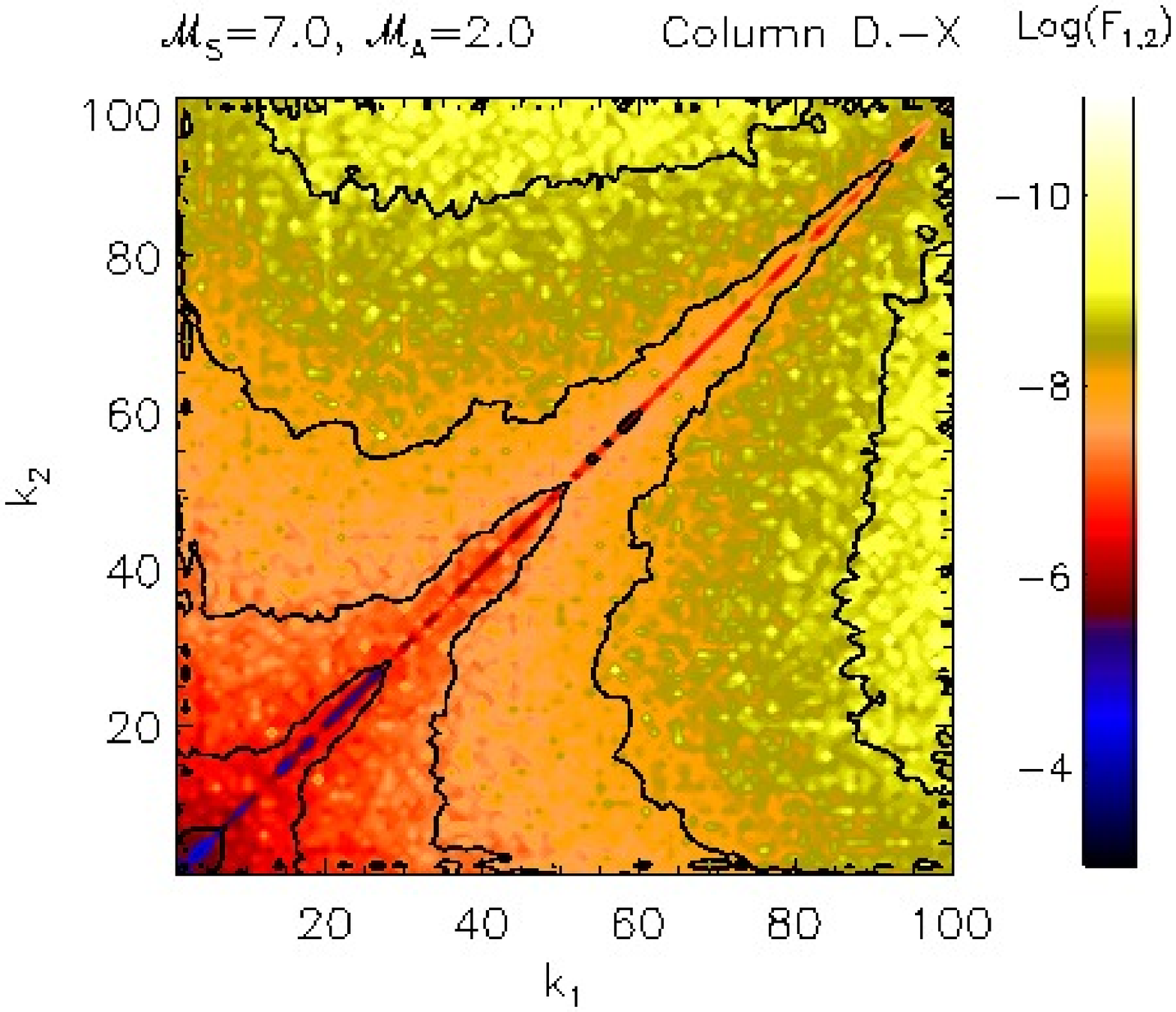} % \includegraphics[scale=.35]{apj/c512b1p.01/paper/bispectrum_cd_x}
\includegraphics[scale=.25]{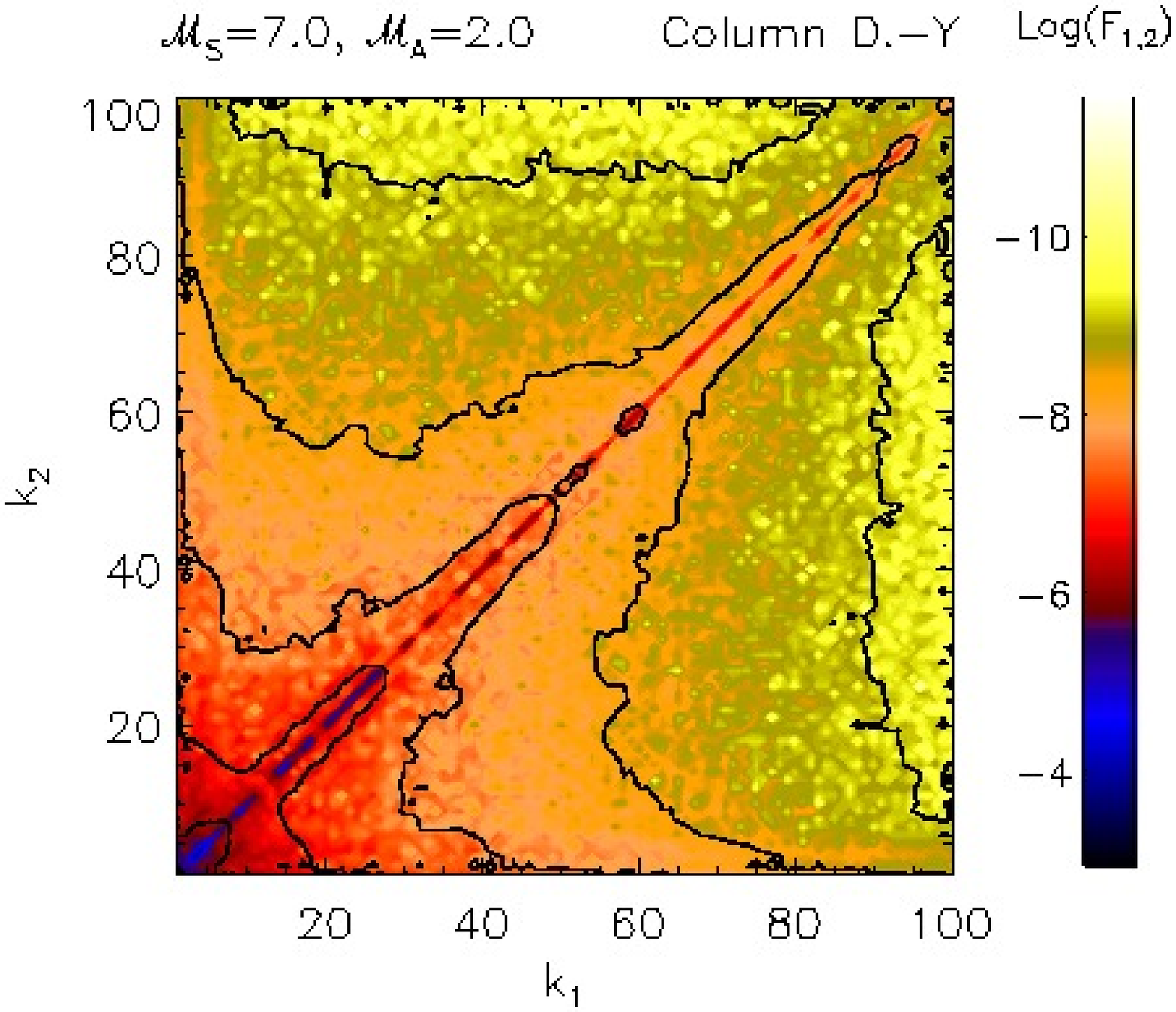} % \includegraphics[scale=.35]{apj/c512b1p.01/paper/bispectrum_cd_y}
\includegraphics[scale=.25]{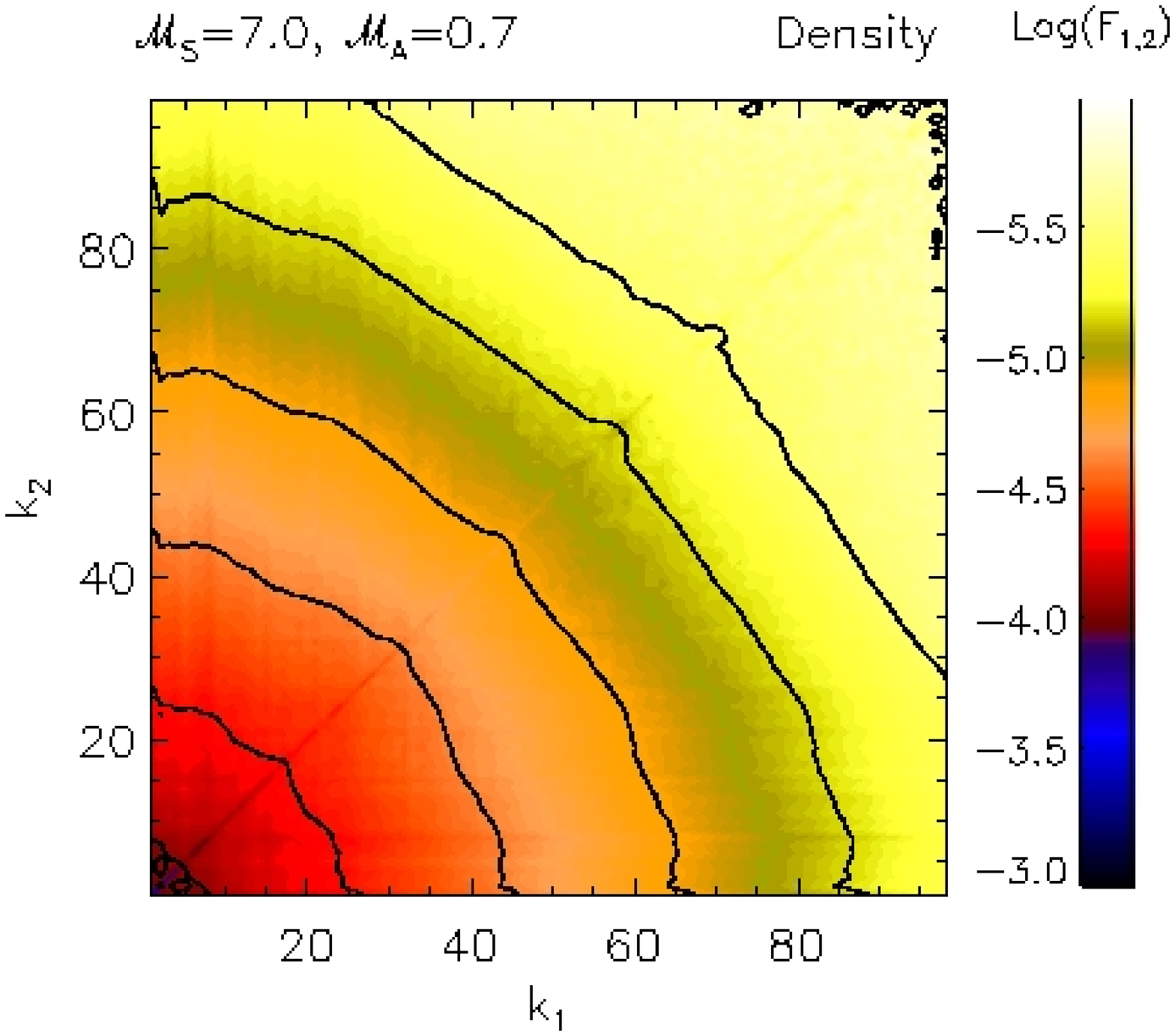} % \includegraphics[scale=.35]{apj/c512b1p.01/paper/bispectrum_dens}
\includegraphics[scale=.25]{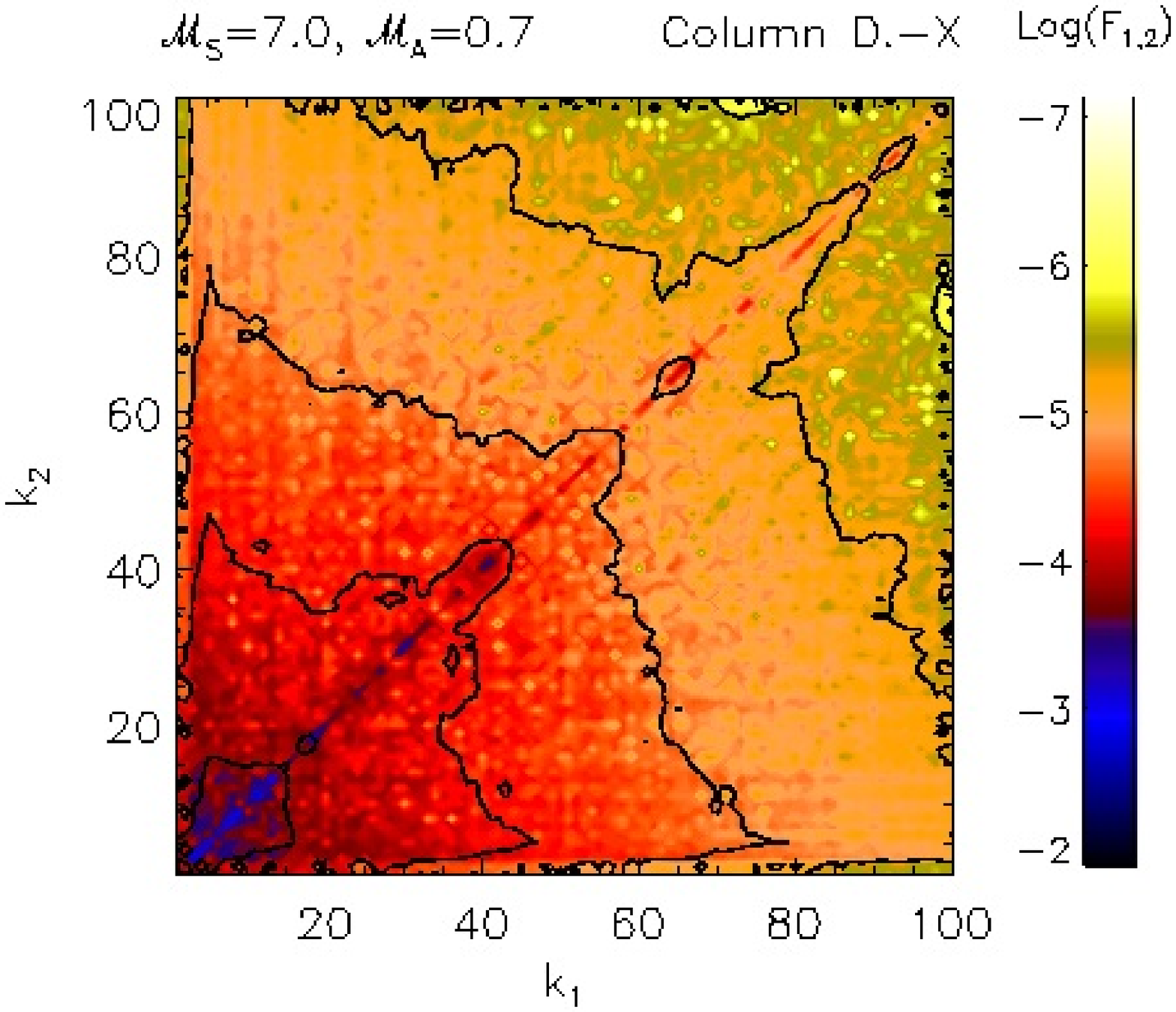} % \includegraphics[scale=.35]{apj/c512b1p.01/paper/bispectrum_cd_x}
\includegraphics[scale=.25]{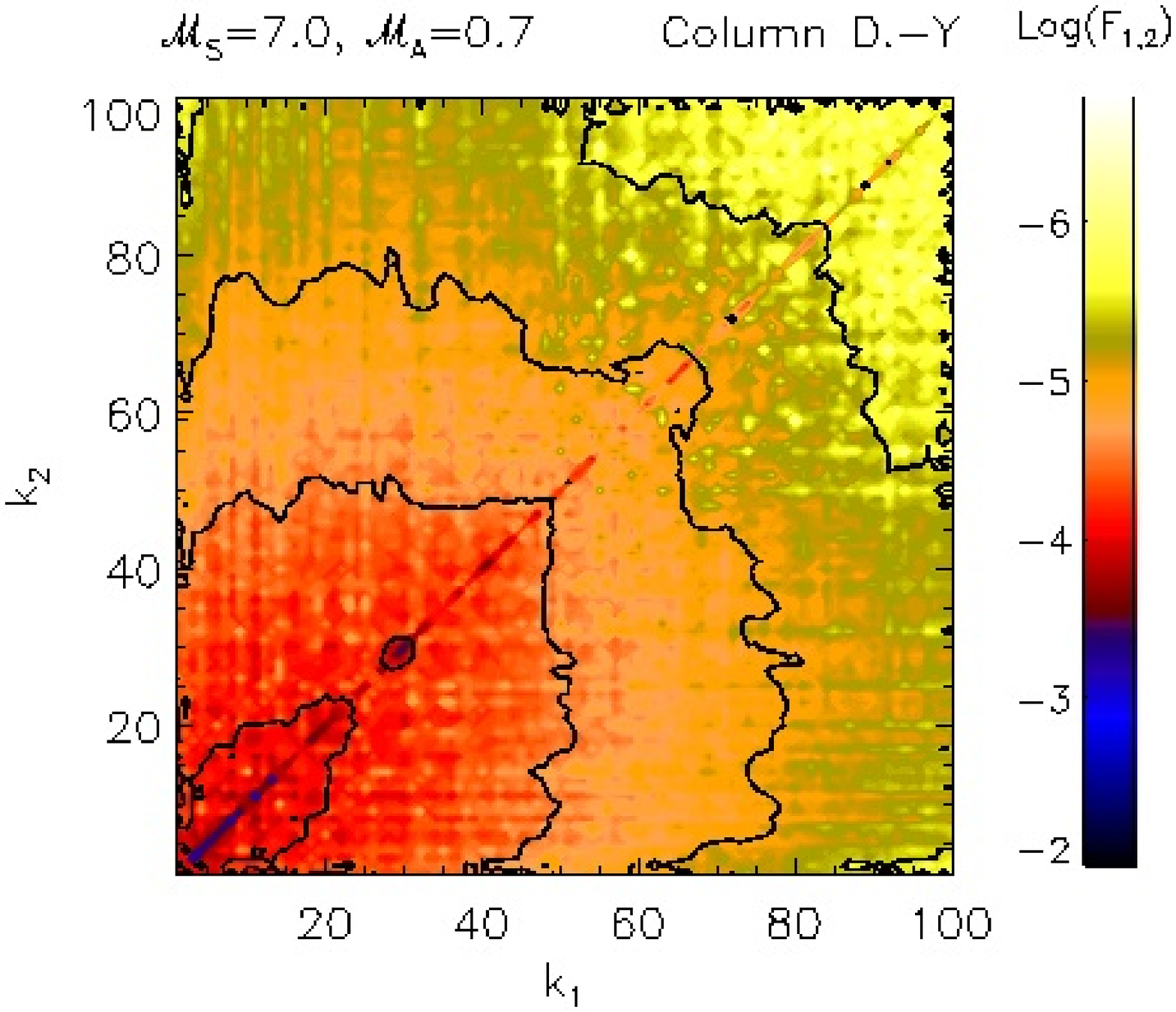} % \includegraphics[scale=.35]{apj/c512b1p.01/paper/bispectrum_cd_y}
\caption{\small Above is the contour analysis of the bispectrum for density and column density.  These figures show the degree of correlation between wavenumbers $k_{1}$ and $k_{2}$. The first column shows density, the second shows column density parallel to the magnetic field (x- column density) and the third shows column density perpendicular the the magnetic field (y- column density).  Here we compare different MHD sonic and Alfv\'enic regimes(third-bottom rows), a Gaussian model (top row) and pure hydro turbulence (second to top). Scales are slightly different due to flattening in the bispectrum for a single scale. Structure is clearly different for subsonic and supersonic cases. \label{fig:bispectra}}
\end{figure*}

We ran the bispectrum analysis for the last snapshots of the density and column density cubes on x and y-directions for all MHD models. We also performed the bispectrum calculation for two comparison models: a supersonic hydrodynamical simulation and a random synthetic Gaussian distribution of density. The density and column density integrated along x and y bispectra are shown in Figure~\ref{fig:bispectra}, on the left, center and right columns, respectively. In the the first row we present the results obtained for the Gaussian distribution, followed by the hydrodynamical case and then the MHD models. The models are labeled for the different turbulent regimes with varying values of ${\cal M}_{A}$ and ${\cal M}_{s}$. We plotted contours and their corresponding color amplitudes in order to see both differences in the shape and magnitude of the bispectrum.

Upon examination of Figure~\ref{fig:bispectra} one sees a general correspondence between the bispectrum of the column density and the underlying 3D bispectrum.  All cases show a prominent amplitude of the bispectrum and thus a high correlation of modes, at $k_1 = k_2$. This is to be expected since this is a trivial case where wave numbers will always show correlation. However, the amplitudes at $k_1 \neq k_2$ are different for each turbulent regime. The more circularly shaped the isocontours and stronger the amplitudes are, the more highly correlated the modes are. The Gaussian distribution shows almost no correlation (very weak amplitudes) for $k_1 \neq k_2$ and the distribution is mostly localized in the diagonal line (i.e. where $k_1 = k_2$).  This is because the distribution of density is entirely random and thus modes should lack association. The supersonic hydro model is similar to a super-Alfv\'enic supersonic MHD model, as expected, and is more strongly correlated then the Gaussian density distribution. The MHD models also present broader distributions over the {$k_1, k_2$} plane, thus showing high correlation. 

The bispectrum of MHD density is shown in the first column starting at the third map down. Larger wave numbers (representing larger frequencies and smaller wavelengths) are less correlated in all models, since higher wave numbers are present near the end of the energy cascade when more of the systems energy has dissipated. For the models with the same ${\cal M}_{A}$ but varying ${\cal M}_{S}$, we found an increase in the wave-wave coupling under the supersonic regime. This phenomenon is expected, as interactions between different modes generate small scale shocks. Shocks in supersonic models greatly increase wave-wave correlations over subsonic models, which are incompressible. The supersonic models of density (1st column, last and second to last rows) show such strong correlations that the strength of the $k_1 \neq k_2$ is comparable to the  $k_1 = k_2$ line.  Interestingly, comparing models with same sonic Mach number, the sub-Alfv\'{e}nic cases show increased wave-wave coupling for $k_1 \neq k_2$. Here, non-linear coupling is increased by the magnetic field.  Comparing density models with ${\cal M}_{s}=7.0$ we see that a stronger magnetic field shows slightly higher amplitudes and has a more circular isocontour profile.  Comparing ${\cal M}_{s}=0.7$ models, it is easy to see that stronger amplitudes exist for the cases with ${\cal M}_{A}=0.7$. Looking specifically at the second and third columns for the supersonic models, the amplitudes of the ${\cal M}_{A}=0.7$ models are much stronger then that of the ${\cal M}_{A}=2.0$ models. It is clear from these models that a stronger magnetic field causes enhanced correlation of modes. For super-Alfv\'{e}nic models, the large scale (low k) magnetic field configuration is destroyed and the MHD modes operate at smaller scales. Both effects reveal that the energy cascade, generated or amplified by $k_1 \neq k_2$ interactions, may work differently on the turbulent regimes, and Kolmogorov scalings may not well-characterize them. 

Regarding observable parameters, the column density bispectra do not show such prominent differences compared to the density analysis. However, the column density bispectra present larger noise because of worse statistics (512$^2$ maps compared to the $512^3$ density cube). However, the supersonic and sub-Alfv\'{e}nic model shows a notably different distribution, with larger amplitudes for $k_1 \neq k_2$, similarly to the density bispectrum. Smaller amplification is  seen in the subsonic cases, particularly the subsonic super-Alfv\'{e}nic model. It is clear from the bispectra of density and column density that supersonic models present the highest correlation of modes while models with a high magnetic field also enhances mode correlation.  If this is true for synthetic data then the bispectrum technique could be used to characterize magnetized turbulence in the ISM. Interestingly, the broadening of bispectrum distribution by the magnetic field is mostly independent on the orientation of the magnetic field lines regarding the line of sight, however differences are clear for supersonic sub-Alfv\'enic cases. The bispectrum is slightly broader for the x direction since the mean magnetic field is along this axis. The column density bispectra present larger noise because of worse statistics (512$^2$ maps compared to the $512^3$ density cube). From our studies it is obvious that the bispectrum requires very high resolution data in order to yield meaningful results.  Noise and other outside influences can completely mask a desired signal information in the bispectrum. Noise will produce a 'false' bispectral map, showing higher correlations for larger wavenumbers. Therefore, bispectral analysis of observational maps can be used to reveal the turbulent regimes operating in molecular clouds, but only if high resolution maps with minimal noise are available. 

\subsection{Bispectra and spectra}

Quantitative studies of compressible MHD turbulence in Cho \& Lazarian (2002, 2003), concentrated on decomposition of turbulence perturbations into slow, fast and Alfv\'{e}nic modes and studies of their spectra. Such a decomposition requires the knowledge of the local magnetic field, which is not the case for the data averaged along the line of sight. Therefore from observations the total spectra of turbulent perturbations are available. The density spectrum is readily available from column density measurements, while the velocity spectrum requires more sophisticated techniques, like Velocity Channel Analysis (VCA) or Velocity Coordinate Spectrum (VCS) to be used (see Lazarian \& Pogosyan 2000, 2006, 2008). Different numerical studies show that for MHD turbulence, the velocity spectrum gets steeper for velocity and shallow for density $\rho$ as the ${\cal M}_s$ increases\footnote{As a result the quantity $\rho^{1/3} v$ stays invariant to ${\cal M}_s$ both in hydro
(Kritsuk et al. 2006) and MHD (Kowal \& Lazarian 2006) turbulence.}. KLB showed that the dependences of the density spectrum on ${\cal M}_A$ is less prominent than on ${\cal M}_s$.

Above we studied the bispectrum of density and noticed the dependence of the measure of ${\cal M}_s$.  We see that the variations of the bispectrum with ${\cal M}_A$ for all $k_{1}$ and $k_{2}$, and find that the phases of the turbulent fluctuations are more correlated in the presence of the dynamically important magnetic fields.

\section{Discussion}
\label{dis}

\subsection{Comparison to KLB and other studies.}

KLB presented an extensive statistical analysis of MHD density for cubes of resolution $128^{3}$, $256^{3}$, and $512^{3}$. However, they only examined the skewness and kurtosis for the lower resolution model ($128^{3}$).  In this work we study cubes of $512^{3}$ resolution in order to compare with the statistics of lower resolution models. KLB examined a larger range of ${\cal M}_{s}$ values and noticed that the asymmetry grows with ${\cal M}_{s}$, but only if ${\cal M}_{s} \geq 0.5$. In this work, we study models with $ 0.7\leq{\cal M}_{s} \leq 7.0 $ and we find that statistical moments such as skewness and kurtosis are easy to obtain and are very useful for characterizing sonic Mach number in a magnetized turbulent system. We confirm the results of KLB in finding that the moments all have strong dependence on the sonic Mach number. As turbulence becomes increasingly supersonic, the skewness of density and column density increases, which is agreement with KLB. We also see a very rapid growth in kurtosis of density and column density with sonic Mach number. Both column density and density show increasing asymmetric distributions with increasing ${\cal M}_{s}$. If one observes a given skew and kurtosis of column density, the value of ${\cal M}_{s}$ can be inferred and related to density, which could be very useful when applied to observational data. It seems that in both our study and in KLB, ${\cal M}_{A}$ does not affect the moments of density.

Our study of magnetic fields and column density correlations in isothermal nonlinear MHD waves has similarities to that of Passot \& V\'azquez-Semadeni(2003) who found that the relation of magnetic field strength B on density $\rho$ depends on whether the MHD wave is a slow or fast mode. Their results confirmed that, except for cases of insufficient field fluctuations, the magnetic pressure behaves as: $B^{2} \simeq c_{1}-c_{2}\rho$ for slow modes (low ${\cal M}_{A}$) and $B^{2} \simeq \rho^{2}$ for fast modes. However, while this study incorporated a large variety of values for the Alfv\'en Mach number ($ 0.073\leq {\cal M}_{A} \leq 7.29)$, they lacked sampling of sonic Mach numbers, only focusing their analysis on supersonic values, $2.72 \leq {\cal M}_{s} \leq 7.28$. They also only investigated 2D models, while in this study we use 3D simulations of magnetic energy and density as well as 2D correlations of magnetic field and column density at higher resolutions. In this study we also explore subsonic as well as supersonic cases and find that the sonic Mach number is a critical parameter in determining how B and $\rho$ correlate, since it determines whether or not density clumps will develop. In agreement with Passot \& V\'azquez-Semadeni(2003), we also find that a larger value of ${\cal M}_{A}$ yields a tighter correlation between density and magnetic pressure, especially at higher density.  This effect can easily be seen by comparing the 3D correlations in Figure~\ref{fig:dens_emag}, the ${\cal M}_{s}=7.0$ cases. However, Figures~\ref{fig:cden_bpar_x}-\ref{fig:cden_bper_z} confirm a  more linear relationship for slow modes ( i.e. low ${\cal M}_{A}$) over larger ${\cal M}_{A}$, which contain both slow and fast modes. We also see that the orientation of the field along the LOS can effect these results and we confirm  Passot \& V\'azquez-Semadeni's result that the angle between the magnetic field and the wave propagation will play a role in determining the importance of the modes. Subsonic models generally show little correlations, yet in the 2D correlations we show that when the LOS is perpendicular to the magnetic field and mean field (Figure~\ref{fig:cden_bper_z}), subsonic correlations begin to show a degree of linearity.

While higher order moments show virtually no dependence on ${\cal M}_{A}$, correlations, on the other hand, can yield information regarding the dynamical importance of the magnetic field in the evolution of densities. According to Padoan and Norlund(1999), cloud dynamics only become strongly affected by magnetic field in very dense regions and on small scales.  It was  also further stated in Padoan et al.(2004) that, while the volume averaged magnetic field strength in a molecular cloud has never directly been measured, there is evidence from the power spectrum of gas densities that super-Alfv\'enic turbulence is dominate in these star forming regions. From the 3D correlation in Figure~\ref{fig:dens_emag}  it is clear that magnetic fields play an important role in shaping density.  This is especially true of supersonic models where a correlation of magnetic energy and density exists.  Higher density regions are good candidates for star formation, thus it may seem promising that higher magnetic field strength could indicate more star forming regions.  Also, from Figure~\ref{fig:dens_malf} it is clear that even though turbulence may be globally super or sub-Alfv\'enic, dense \textit{clumps} influence magnetic field to a point where the local Alfv\'en Mach number can change greatly.  In the case of super-Alfv\'enic turbulence, high density results in smaller values ${\cal M}_{A}$ (although they generally remain super-Alfv\'enic) while in sub-Alfv\'enic turbulence, ${\cal M}_{A}$ steadily increases with density to the point where turbulence becomes super-Alfv\'enic. However, even though clumps in globally sub-Alfv\'enic clouds are able to become locally super-Alfv\'enic, this does not mean that the cloud as a whole becomes super-Alfv\'enic. We do not see any evidence that the cloud as a whole is unaffected by a strong magnetic field, although we do see that indeed a magentic field has a strong impact on small scale turbulence. Therefore, our globally super-Alfv\'enic models support Padoan and Norlund(1999)'s claim that magnetic field plays a leading role in the dynamics of very dense clumpy regions (i.e. as density increases, $M_{A} decreases$) , however there is no evidence for this in our sub-Alfv\'enic models, which show that on small scales density clumps are increasingly super-Alfv\'enic, and thus have a lower B field.  Overall, these correlations only give us information on how the magnetic field effects local density clumps and not whole cloud. However, it is perhaps promising that our results tend to be locally super-Alfv\'enic for high densities, despite the global magnetic field.  More studies will be needed in order to confirm Padoan and Norlund(1999) results.

While density statistics are improtant to understanding turbulent processes, the observational velocity information provided by radio spectroscopic data should be utilized in order to gain a broader study of turbulent interactions.  The difficulty in directly comparing synthetic velocity dispersions with observational spectral velocities lies in the fact that the emissivity of a spectral line has dependence on velocity and density. In order to better understand this, we calculate correlations of velocity dispersion vs. column density and also investigate how dispersion of velocity correlates with  the measures available through Doppler shifted spectral lines.  Lazarian et al.(2001) studied emissivity statistics and the density-velocity correlation to better understand how density and velocity interact in a turbulent media. In this study, we confirm that a moderate correlation between column density and velocity exists, specifically seeing that higher velocities indicate higher compression of matter. However, we further find here that these correlations gain strength when the LOS is $\perp$ to the magnetic field. The Lazarian et al. study used  lower resolution cubes and only examined 3D cases, while here we investigate 2D cases at higher resolutions.  In general, further studies must be made in order to understand how best to relate observational velocity line profiles to simulated velocity dispersions.

\subsection{Density Correlations and Relation to Observables: 2D vs. 3D}
	
	 In order to compare correlations of 3D densities to quantities more similar to what is avaliable  through observations (i.e. column density, velocity dispersions), we provide 2D column density correlations.  We do indeed see several similarities between 3D correlation maps and their 2D counterparts.  For instance, when comparing the correlation of density and magnetic energy (Figure~\ref{fig:dens_emag}) with correlations of magnetic field and column density (Figures~\ref{fig:cden_bpar_x}-\ref{fig:cden_bper_z}), we see that for supersonic cases, magnetic energy ($B^2$) and magnetic field both increase with density and column density. This implies that for 2D and 3D cases, magnetic fields become trapped in high density clumps. However, because Figure ~\ref{fig:dens_emag} is a volume plot and we plot magnetic energy instead of field, we see differences.  Most notably, we see a $B \approx \rho $ for 2D maps and $B^2$ increases more exponentially with $\rho$ for supersonic cases. We can see this trend from Table~\ref{tab:coefficients2}, with  2D correlations being linear  with cases with the field perpendicular to the LOS being the most linear. Therefore, although the magnetic field and energy  both  see a similar growth with density for supersonic turbulence, the rate of growth is different. Subsonic cases for both 3D and 2D correlations both showed the least correlation. 

  Specific kinetic energy (${\cal E}_k$) vs. density showed no special relationship. However, the kinetic energy weighted by density ($E_k$) shows a nearly linear correlation for supersonic cases. We see a similar behavior for velocity dispersions vs. column density.  These dispersions are also weighted by density, since observers must also account for the column density when a measure of velocity dispersion is made. Because of density clumps, both 2D and 3D quantities increase with  dispersion and kinetic energy, respectively.  For both correlations, subsonic models show less linear correlations.  However, despite many similarities, it is not necessarily straightforward to compare the two, since the velocity dispersion seems to be effected by the orientation of the magnetic field with the LOS. We have not tested this effect on kinetic energy density.  It is obvious from the 2D correlations that the field orientation effects how correlated dispersions are with column density.  When the field is perpendicular to the LOS, the most linear correlations are obtained, as is evident from Table~\ref{tab:coefficients2}.  Dense clumps can drag the field lines and, therefore, present a more isotropic velocity field. This effect results in an increase in the dispersion of the velocity measured along the LOS.

\subsection{2D and 3D Bispectrum}
	The bispectrum can give us valuable information regarding how the modes of a nonlinear system correlate. The power spectrum has been used in numerous papers to analyze observations, \citep[see][]{Arm95,Stan99,Stan01}. In Section~\ref{sec:bispectrum} we analyzed the bispectrum for densities and column densities and we have found several interesting features which may provide new ways of analyzing data. Different turbulent parameters display very different bispectra, and thus the bispectrum maybe be used to characterize Mach number and magnetic field in observational data.
	
	For instance, the bispectrum of density and column density  gives information as to how shocks and magnetic fields affect turbulence. It has been shown by KLB and Beresnyak et al.(2005) that in supersonic turbulence, shocks produce compressed density and a shallower spectrum. Looking at Figure~\ref{fig:bispectra} for density (first column) it is clear that these shocks play a crucial role in the correlation of modes. The subsonic cases show little correlation between any points except the case of $k_{1}=k_{2}$. The compressed densities from supersonic turbulence are critical for correlations between wavenumbers, since waves will be compressed together and therefore have  more interactions. Also interesting to consider is the differences in sub-Alfv\'{e}nic and super-Alfv\'{e}nic cases for density. We find excellent agreement with KLB in that density structures are well correlated to the presence of magnetic fields. For super-Alfv\'enic cases, it is clear from Figure~\ref{fig:bispectra} that the system lacks the stronger correlations that are a characteristic of the sub-Alfv\'{e}nic models. For super-Alfv\'{e}nic simulations, correlations in frequency are not as readily made due to large dispersion of density structure. The bispectrum of supersonic hydro models is similar to the super-Alfv\'{e}nic, supersonic cases. It is clear from these cases that the presence of a magnetic field assists in the correlation of modes. 

	In order to relate to observations, the bispectrum of column densities is provided in order to compare with density. It should be noted that while our model does not include self or external gravity, it has been shown by \cite[]{Elmegreen04} that self-gravity partitions the gas into clouds which contributes to scale-free motions generated by turbulence. Thus, self-gravity enhances small scale structure and our results should be relatively unaffected, as we are primarily concerned with large scale turbulent cascades. For both density and column density, strongest correlations are found in the case of supersonic sub-Alfv\'{e}nic turbulence. Differences in image quality between density and column density maps arise from the fact that density cubes have higher resolution ($N^{3}$ vs. $N^{2}$). Because the magnetic field enhances correlations, the bispectrum could be used to characterize the magnetic field in studies similar to Goodman et al.(1995); Padoan \& Nordlund (1999).
	
	The bispectrum has proved to be an important addition to the tools of statistical studies of MHD turbulence. This new tool could be used to characterize the Mach number of gas in the interstellar medium as well as the magnetic field.  However, it should be noted that very high quality data is required since the multipoint statistics are known to increase noise. For the interstellar medium it has been suggested in \cite[]{Lazarian99} to compare various regions of the sky using the bispectrum to search for like signals. These studies are to be done elsewhere.

\section{Conclusions}
In this paper we investigated several different statistics of density structure of compressible MHD turbulence.  We examined the skewness and kurtosis of the data as well as studied several different measures of correlations.  In particular, we examined the bispectrum of compressible MHD turbulence.  We found:

\begin{itemize}
\item  We confirm KLB's results with higher resolution cubes that sonic Mach number has dependence on the 3rd and 4th order moments.
\item Correlations of kinetic energy density vs. density are linear for supersonic cases and show no correlation for subsonic cases 
\item  Magnetic energy increases with density clumps for supersonic turbulence. 
\item For the super-Alfv\'{e}nic cases, the local Alfv\'en Mach number, i.e. $M_{A}\equiv \frac{\delta V}{V_{A}}$, of clumps increases up to the mean density then falls of as B increases with increasing density clumps. 
\item For sub-Alfv\'{e}nic cases, the local $M_{A}$ increases with density for both supersonic and subsonic cases.
\item Correlations of magnetic field vs. column density are linear for supersonic turbulence and are strongest when the field is perpendicular to the LOS. 
\item Correlations of velocity dispersion vs. column density show a degree of linearity and are strongest when the mean magnetic field is perpendicular to the LOS.
\item Subsonic velocity dispersions are not strongly effected by density fluctuations.
\item Applying the bispectrum to density fields of MHD turbulence we find that: 

\begin{enumerate}
\item There are strong correlations for cases where $k_{1}=k_{2}$
\item There are virtually no correlations for $k_{1}\neq k_{2}$ for subsonic cases. 
\item Supersonic models show the strongest $k_{1}\neq k_{2}$ correlations.
\item The introduction of a magnetic field enhances correlations for all $k_{1}$ and $k_{2}$ .\\
\end{enumerate}
\end{itemize}

\acknowledgments
B.B. is supported by the NSF funded Research Experience for Undergraduates (REU) program through NSF award
AST-0453442. B.B. is thankful for valuable discussions with Dr. James T. Lauroesch and Thiem Hoang. D.F.G. thanks the Brazilian agencies 
FAPESP (No.\ 06/57824-1) and CAPES (No.\ 4141067). We also 
thank the financial support of the NSF (No.\ AST0307869) and the Center for Magnetic Self-Organization 
in Astrophysical and Laboratory Plasmas.

\end{document}